\documentclass[english]{svjour3}
\usepackage[T1]{fontenc}
\usepackage[latin9]{inputenc}
\usepackage{wrapfig}
\usepackage{url}
\usepackage{amsmath}
\usepackage{amssymb}
\usepackage{graphicx}

\makeatletter

\providecommand{\tabularnewline}{\\}

\RequirePackage{fix-cm}

\smartqed  % flush right qed marks, e.g. at end of proof
\usepackage{physics}

\@ifundefined{showcaptionsetup}{}{%
 \PassOptionsToPackage{caption=false}{subfig}}
\usepackage{subfig}
\makeatother

\usepackage{babel}
\begin{document}

\title{Performance of adiabatic melting as a method to pursue the lowest
possible temperature in \textsuperscript{3}He and \textsuperscript{3}He\textendash \textsuperscript{4}He
mixture at the \textsuperscript{4}He crystallization pressure\thanks{}}

\titlerunning{Performance of an adiabatic melting method}

\author{T. S. Riekki \and A. P. Sebedash \and J. T. Tuoriniemi}

\authorrunning{T. S. Riekki \and  A. P. Sebedash \and  J. T. Tuoriniemi}

\institute{T. S. Riekki\and  J. T. Tuoriniemi\at Aalto University\\
School of Science\\
Low Temperature Laboratory\\
P.O. BOX 15100 FI-00076 Aalto, Finland\\
\email{tapio.riekki@aalto.fi}\\
A. P. Sebedash \at P. L. Kapitza Institute for Physical Problems
RAS\\
Kosygina 2, 119334 Moscow, Russia\\
}

\date{Received: date / Accepted: date}
\maketitle
\begin{abstract}
We studied a novel cooling method, in which \textsuperscript{3}He
and \textsuperscript{4}He are mixed at the \textsuperscript{4}He
crystallization pressure at temperatures below $0.5\,\mathrm{mK}$.
We describe the experimental setup in detail, and present an analysis
of its performance under varying isotope contents, temperatures, and
operational modes. Further, we developed a computational model of
the system, which was required to determine the lowest temperatures
obtained, since our mechanical oscillator thermometers already became
insensitive at the low end of the temperature range, extending down
to $\left(90\pm20\right)\,\mathrm{\mu K\approx}\frac{T_{c}}{\left(29\pm5\right)}$
($T_{c}$ of pure \textsuperscript{3}He). We did not observe any
indication of superfluidity of the \textsuperscript{3}He component
in the isotope mixture. The performance of the setup was limited by
the background heat leak of the order of $30\,\mathrm{pW}$ at low
melting rates, and by the heat leak caused by the flow of \textsuperscript{4}He
in the superleak line at high melting rates up to $500\,\mathrm{\mu mol/s}$.
The optimal mixing rate between \textsuperscript{3}He and \textsuperscript{4}He,
with the heat leak taken into account, was found to be about $100..150\,\mathrm{\mu mol/s}$.
We suggest improvements to the experimental design to reduce the ultimate
achievable temperature further. 

\end{abstract}

\section{Introduction\label{sec:Introduction}}

Strong motivation for pursuing ever lower temperatures in helium fluids
is the anticipated superfluid transition of the \textsuperscript{3}He
component in dilute \textsuperscript{3}He\textendash \textsuperscript{4}He
mixtures. Pure liquid \textsuperscript{3}He undergoes superfluid
transition when fermionic \textsuperscript{3}He atoms start to form
BCS-like pairs \cite{BCS}. The phenomenon occurs only at sufficiently
low temperatures, which for pure \textsuperscript{3}He is about 1
mK at saturated vapor pressure, and just below 3 mK at \textsuperscript{3}He
crystallization pressure ($\sim3.4\,\mathrm{MPa}$). We, however,
are interested in systems where \textsuperscript{3}He is diluted
by \textsuperscript{4}He. The \textsuperscript{4}He component of
the mixture becomes superfluid already at around $2\,\mathrm{K}$,
and at millikelvin regime it acts as a thermally inert background
contributing to the interactions between the\textsuperscript{ 3}He
atoms. The requirement for the BCS-pairing is an attractive interaction
between the particles, and since a very weak attraction is still present
in the mixture systems, the \textsuperscript{3}He component superfluid
transition is expected to occur at some ultra-low temperature \cite{Ebner_thesis,Bardeen1967,AL-SUGHEIR2006,Sandouqa2011,Effective_3He_interactions}.
Rysti \emph{et al.} \cite{Effective_3He_interactions} calculated the
highest transition temperature $\sim100\,\mathrm{\mu K}$ to occur
at $\sim10\,\mathrm{bar}$ in saturated mixture, while at the crystallization
pressure of \textsuperscript{4}He ($\sim2.6\,\mathrm{MPa}$), it
was estimated to be about $40\,\mathrm{\mu K}$. 

Superfluid mixture of \textsuperscript{3}He and \textsuperscript{4}He
would be a dense mixture of fermionic and bosonic superfluids, and
thus a completely unique system. Mixture superfluidity has been studied
in rare quantum gases \cite{Kinnunen2015,Chevy2015,Delehaye2015,Onofrio2016},
where it has been observed both in mixtures of \textsuperscript{6}Li
and \textsuperscript{7}Li \cite{Bose_Fermi_superfluids}, and \textsuperscript{6}Li
and \textsuperscript{174}Yb \cite{Roy2017}. However, the interactions
between Fermi and Bose superfluids are significantly weaker there
than they would be in liquid helium of $10^{4}$ times higher density,
making the superfluid helium isotope mixture a fascinating system
to study. Furthermore, since the melting method can be used to cool
pure \textsuperscript{3}He phase as well, it could be used to study
the exotic f-wave pairing state of superfluid \textsuperscript{3}He,
that has been anticipated to take place below $50\,\mathrm{\mu K}$ \cite{Nguyen2019}.
The majorana quasiparticle surface states should also manifest themselves
at low enough temperature \cite{Bunkov2013,Bunkov2014}.

To reach for such extreme conditions, cooling techniques need to be
perfected. The situation was similar in the 1960s during the search
for \textsuperscript{3}He superfluidity \cite{Osheroff1972,Osheroff1972a},
that saw, for example, the development of the Pomeranchuk cooling
method \cite{pomeranchuk1950theory,anufriev1965soviet}. Oh \emph{et
al.} \cite{Oh1994} used a two-stage nuclear demagnetization refrigerator
to cool a small mixture sample, with a $4000\,\mathrm{m^{2}}$ heat-exchanger
surface area, to $97\,\mathrm{\mu K}$ at 1 MPa. The major problem
with an external cooling method, such as that, is the rapidly increasing
thermal boundary resistance, or Kapitza resistance, between liquid
helium and metallic coolant. To fight against it, one needs to increase
the surface area of the experimental volume, but eventually, it will
become practically unviable.

The adiabatic melting method \cite{Sebedash1997,Sebedash2000,Tuorinemi_Martikainen_Pentti,Adiabatic_Melting}
overcomes the Kapitza bottleneck by relying on internal cooling that
takes place directly in helium fluid. In this setup, the nuclear demagnetization
refrigerator provides only precooling conditions, and thus the surface
area of the cell will no longer be the ultimate limiting factor. The
physical origin of cooling is similar to that of a conventional dilution
refrigerator, except that the melting method operates cyclically at
an elevated pressure. The phase separation between helium isotopes
is achieved by increasing the pressure in the system to the crystallization
pressure of \textsuperscript{4}He $2.564\,\mathrm{MPa}$ \cite{Pentti_Thermometry}.
When \textsuperscript{4}He solidifies at sufficiently low temperature,
it expels all the \textsuperscript{3}He component \cite{Balibar2000,Balibar2002,Pantalei2010},
and ideally in the end we have a system consisting of pure solid \textsuperscript{4}He,
with negligibly small entropy, and pure liquid \textsuperscript{3}He.
If the system is then cooled to far below the superfluid transition
temperature of pure \textsuperscript{3}He $T_{c}=2.6\,\mathrm{mK}$ \cite{Pentti_Thermometry},
the entropy of the liquid component can also be reduced dramatically.

Good initial temperature would be of order 0.5 mK, which is straightforward
to bring about by using adiabatic nuclear demagnetization of copper.
Next, the solid phase is melted, releasing liquid \textsuperscript{4}He
allowing \textsuperscript{3}He to mix with it again forming a saturated
mixture with $8.1\%$ \cite{Pentti_etal_solubility} molar \textsuperscript{3}He concentration.
Per mole, \textsuperscript{3}He\textendash \textsuperscript{4}He
mixture contains a large amount of entropy compared to superfluid
\textsuperscript{3}He, and going adiabatically from the system of
solid \textsuperscript{4}He\textendash superfluid \textsuperscript{3}He
to mixture is only possible if the temperature of the system decreases.
Theoretically, the cooling factor in the melting process can exceed
1000, but in practice, things like remnant mixture in the initial
state, and external heat leak can severely limit it. To repeat the
process, solid then needs to be regrown and the heat released from
the phase separation absorbed to the precooling stage. A more thorough
discussion about the thermodynamics of the adiabatic melting method
can be found in Ref.~\cite{Riekki2019}.

Our experiment takes advantage of the lessons learned from the earlier
run \cite{Adiabatic_Melting}, and places the sinter needed for precooling
into a separate volume to reduce the heat load from the precooler
to the melting cell at the coldest stages of the experiment. We have
also improved the design of the superleak line. Superleak is a capillary
filled with tightly packed powder, that allows only superfluid \textsuperscript{4}He
to flow through it. The performance of the superleak is essential
to the success of the experiment, as the solid \textsuperscript{4}He
phase is grown, or melted, by transferring superfluid \textsuperscript{4}He
to, or from, the cell. Further, the cooling power of the melting process
is directly proportional to the melting rate, whence the superleak
needs to be able to sustain large enough flow.

The present paper is a complete recollection of our recent melting
experiment run. Our earlier publications \cite{Riekki2019,Sebedash_QFS,Riekki2019a,Riekki_kappa}
laid the groundwork for the results presented here, and will be frequently
referred to. We start by describing the experimental setup, briefly
summarizing the entire cooling system, but focusing on the low temperature
parts, as well as give a rundown of a typical melting run. Then we
will build upon the computational thermal model of the system, first
introduced in Ref.~\cite{Riekki_kappa}. The computational model was
needed, because at the lowest temperatures, the quartz tuning fork
oscillators we used for thermometry had become insensitive \cite{Riekki2019a}.
To complete the model, we first determine the Kapitza
resistance coefficients of our system that determine the thermal connection
between the melting cell and the demagnetization precooler. Then we
will describe the determination of distribution of the helium isotopes
between the three phases present in the system at different stages
of the experiment, before moving on to the next topic, the heat leak
during the melting process. We will show that it consists of two components:
generic background heat leak, and melting-rate dependent contribution.
Once the computational model is completed, we use it to determine
the lowest achieved temperatures. We conclude that there was no observation
of the superfluidity of the mixture phase. We will also show examples
of how the setup behaved at higher temperatures ($>0.5\,\mathrm{mK}$),
under varying conditions, as well as simulate how altering certain
parameters would have affected the lowest possible temperature achievable
in the system. Finally, we will also suggest improvements for the
next iteration of the experiment.\vfill\clearpage

\section{Experimental Setup\label{sec:Experimental-Setup}}

\subsection{Cooling system}

The cooling system that allowed us to reach for the record-low temperatures
in helium fluids essentially consisted of five stages. The cryostat
was submerged in liquid \textsuperscript{4}He bath, which provided
starting temperature of 4 K. Liquid from the bath was also used to
operate a \textsuperscript{4}He evaporation cooler, or pot, to provide
1 K base temperature to the vacuum insulated inner parts of the cryostat.
The pot was needed to liquefy the incoming \textsuperscript{3}He
to the closed-cycle \textsuperscript{3}He\textendash \textsuperscript{4}He
dilution refrigerator, in which\textsuperscript{ 3}He was continuously
mixed with \textsuperscript{4}He to decrease the temperature to about
10 mK. In fact, we had two pots connected together with one providing
the general cooling to 1 K and condensing of \textsuperscript{3}He,
while the secondary pot was used to thermalize the capillaries connecting
to the melting cell.

The dilution refrigerator was, in turn, used to precool the adiabatic
nuclear demagnetization stage \cite{Pobell,Lounasmaa}. There the nuclear spins of copper were
first aligned in a large magnetic field, and the heat of magnetization
was absorbed by the dilution refrigerator, after which the two stages
were thermally disconnected by an aluminum heat switch. Then, the
magnetic field was slowly lowered, while maintaining alignment of
the nuclear spins, which cooled the system further, dropping the copper
electron temperature to below 0.5 mK. The electron temperature was
monitored by a pulsed \textsuperscript{195}Pt NMR thermometer (PLM).
The nuclear stage cannot be operated continuously, as increasing the
magnetic field heats the system to about 50 mK. Attached to the nuclear
stage was the fifth and final cooling stage: the melting cell.

\subsection{Cell\label{subsec:Cell}}

\begin{figure}[t]
\includegraphics[width=0.9\columnwidth]{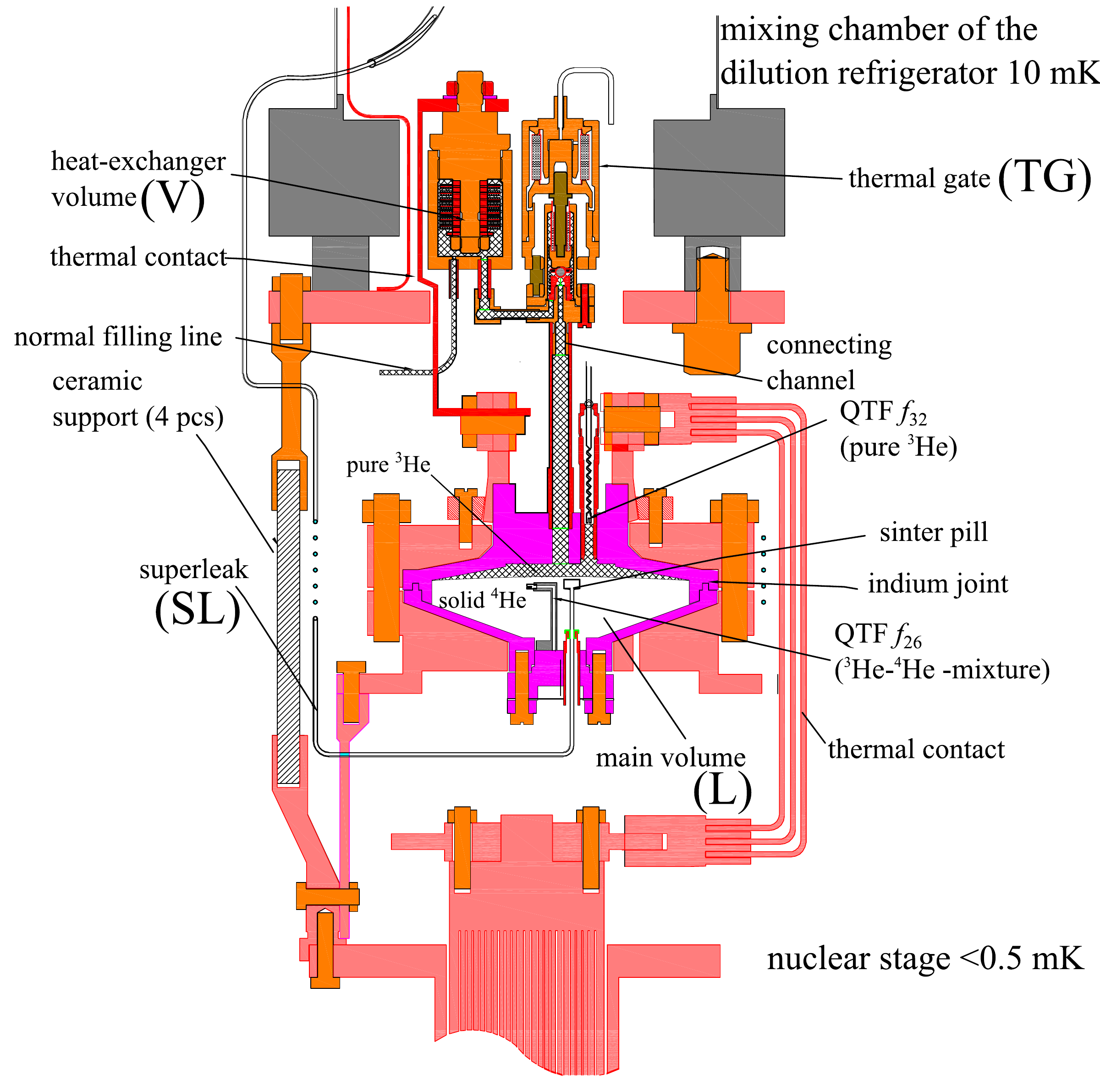}\caption{(color online) Schematic drawing of the low temperature parts of the
experimental setup \cite{Sebedash_QFS}. The cell consists of a main
volume (L) connected to a separate heat-exchanger volume (V) with
a thermal gate (TG) in between. The superleak line (SL) connects to
the bellows system shown in Fig.~\ref{fig:Bellows-schematic} \label{fig:Cell}}
\end{figure}
The total volume of the experimental cell, shown in Fig.~\ref{fig:Cell},
was $\left(82\pm2\right)\,\mathrm{cm^{3}}$ and it consisted of two
separate parts: a large main volume ($77\,\mathrm{cm^{3}}$) and a
small sinter-filled heat-exchanger volume ($5\,\mathrm{cm^{3}}$)
connected together by a channel that could be restricted by a pressure-operated
cold-valve, dubbed as the thermal gate. The cooling process occurred
in the main volume that housed liquid \textsuperscript{3}He, \textsuperscript{3}He\textendash \textsuperscript{4}He
mixture, and solid \textsuperscript{4}He at varying proportions depending
on the stage of the cooling cycle. At most, about 90\% of the main
volume was filled by solid. The connecting channel and the heat-exchanger
were filled with liquid \textsuperscript{3}He, and as the name suggests,
its purpose was to provide thermal connection between the liquid in
the cell and the nuclear stage during precooling periods.

The body of the main volume was made of two high-purity copper shells
that were encased in between thick copper flanges to provide rigidity
to sustain high pressures. The shells were sealed by an indium joint
and tightened by 16 bolts through the copper flanges. The bottom surface
of the cell had a grafoil strip on it to act as a nucleation site
for \textsuperscript{4}He crystal \cite{Balibar2005}.

The heat-exchanger volume was also made of copper with a stack of
8 sintered discs attached to the top half of the volume. Each of them
was a silver disc covered by silver sinter on both sides with silver-plated
copper spacers in between to provide good thermal contact throughout
the entire stack. We determined the surface area of the stack to be
about $10\,\mathrm{m^{2}}$, while the plain walls of the main cell
volume had about $0.12\,\mathrm{m^{2}}$ surface area.

The setup had two filling lines to transport liquid helium in the
system. An ordinary capillary line attached to the heat-exchanger
volume was used to fill the cell with \textsuperscript{3}He, while
a superleak line connected to the main volume of the cell was used
to transfer superfluid \textsuperscript{4}He to and from the cell.
The \textsuperscript{4}He crystallization pressure in porous superleak
is higher than in bulk and thus that line remained open when the normal
capillary was already blocked by solid helium. The cell-side end of
the superleak line was placed in the middle of the main volume to
allow the crystal to grow to large enough size, and had a cylindrical
pill of sintered powder attached to it to prevent solid from blocking
it prematurely. The other two feedthroughs were for the quartz tuning
fork oscillators, discussed further in Section \ref{subsec:Quartz-tuning-fork}.

\subsubsection{Thermal gate\label{subsec:Thermal-gate}}

Ultimately, the main volume of the cell was the coldest part of the
experiment. The purpose of the thermal gate was to isolate the cell
main volume from the heat-exchanger volume at times like this to minimize
the heat flow coming from the nuclear stage. Thermal gate was a pressure
operated needle valve, where the ``needle'' was a stainless steel
ball at the end of a Vespel rod pressed against a conical copper saddle
(see Fig.~\ref{fig:Cell}). The valve was operated by a miniature
stainless steel bellows system with brass framework and a copper bottom
flange. Vespel was used to provide heat insulation both between the
upper and the lower part of the bellows as well as between the frame
and the bottom flange. The upper bellows was fitted with both normal
and superleak lines with the normal line allowing us to introduce
\textsuperscript{3}He to the system, while the superleak was used
to operate the bellows via transfer of superfluid \textsuperscript{4}He.

As the pressure in upper bellows is increased, the ball is pressed
against the saddle, restricting the width of the channel. At 0.1 MPa
the gate was fully open, while at 0.3 MPa  it was closed. The setup
was not intended to be superfluid \textsuperscript{3}He tight, but
rather it was supposed to sufficiently limit the flow of normal-fluid
\textsuperscript{3}He that is responsible of the entropy transfer
between the volumes. Further information about the thermal gate can
be found in Ref.~\cite{Sebedash_QFS}.

\subsubsection{Superleak line and the bellows system\label{subsec:Bellows-system}}

\begin{figure}[t]
\center\includegraphics[width=0.75\columnwidth]{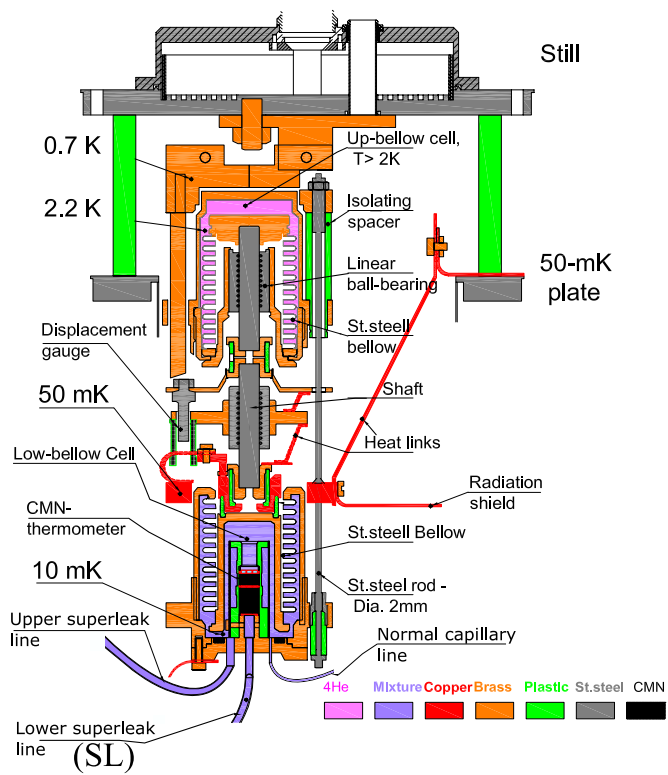}

\caption{(color online) Schematic drawing of the bellows system placed within
the dilution refrigerator \cite{Sebedash_QFS}. The lower superleak
line (SL) connects to the melting cell (cf. Fig.~\ref{fig:Cell})\label{fig:Bellows-schematic}}
\end{figure}
The cell superleak mentioned in Section \ref{subsec:Cell} consisted
of two pieces. The first one started from the main volume of the cell
and ended up in the lower of the two bellows attached to the dilution
refrigerator of the cryostat. This bellows system, shown in Fig.~\ref{fig:Bellows-schematic},
was similar to that of the bellows in the thermal gate, but at larger
size \cite{Sebedash_QFS}. From the lower bellows, the second superleak
piece continued to the still plate of the dilution refrigerator at
$\sim0.7\,\mathrm{K}$. From there on, an ordinary capillary line
continued towards room temperature, thermalized to $1\,\mathrm{K}$
and $4\,\mathrm{K}$ along the way (not shown in Fig.~\ref{fig:Bellows-schematic})

The still plate thermalization was made weak on purpose to allow us
to warm-up the upper end of the superleak if needed. The melting curve
of \textsuperscript{4}He is flat $2.5\,\mathrm{MPa}$ up to about
$1.5\,\mathrm{K}$, above which the pressure starts to increase. As
the solid in the cell fixes the pressure in the system to a value
slightly higher than this, the open upper-end of the superleak had
to be warm enough to keep it free from solid, and available for \textsuperscript{4}He
transport. It turned out that no additional heating was required,
but the heat link itself was weak enough to keep the temperature sufficiently
high. However, that also meant that we could not block the superleak
easily at will, but to do so we had to increase the pressure enough
to force crystallization at the upper end. The original intention
was to allow the superleak to become blocked during the precooling
of the cell to decrease heat load coming through it, but achieving
it easily was thus not possible.

The purpose of the two-part superleak design was to isolate the experimental
cell from any $>\text{1}\,\mathrm{K}$ parts of the cryostat. Of special
concern was preventing the fourth sound modes potentially generated
at the high-temperature end of the superleak line from reaching the
cell. Such sound modes can be generated in porous materials, at temperatures
near \textsuperscript{4}He superfluid transition temperature $\sim2\,\mathrm{K}$,
where there still is a finite amount of normal fluid component along
with the superfluid. With this design, their propagation should terminate
at the lower bellows and the possible heat generated would be absorbed
to the dilution refrigerator at no detriment to the melting cell.

To grow the solid phase, \textsuperscript{4}He was introduced from
a gas bottle at room temperature, and then pushed through a liquid
nitrogen trap to the superleak and to the cell. The melting was performed
by pumping the line with a scroll pump, or simply to an empty volume.
The flow of \textsuperscript{4}He was measured using\emph{ Bronkhorst
F-111C-HA-33-V EL-FLOW} flowmeter, calibrated against helium flow
from a known volume storage tank at various pressure gradients. Accurate
flow measurement was important as the amount of transferred \textsuperscript{4}He
is directly proportional to the change in the amount of solid in the
cell main volume.

The \textsuperscript{4}He transport carried out this way inherently
had a connection from room temperature to the lowest temperature parts
of the experiment. Even if there were several thermalizations and
buffer volumes along the way, there were concerns about the heat leak
this direct procedure could cause to the experiment. Hence, the actual
purpose of the bellows system was to provide an alternate cell operation
method without this direct room temperature path.

The lower bellows (Fig.~\ref{fig:Bellows-schematic}) was filled
with saturated \textsuperscript{3}He\textendash \textsuperscript{4}He
mixture at 10 mK, and was monitored by a CMN-susceptibility thermometer,
while the upper bellows had pure \textsuperscript{4}He kept at above
1 K. By changing the pressure in the upper bellows, the lower bellows
could be compressed or depressed to provide flow to, or from, the
main cell volume. This way the solid growth and melt would be isolated
from room temperature by the isolation between the lower and upper
bellows, and the flow in the superleak would only involve parts at
temperatures at 10 mK or below. The areas of the bellows system were
designed so that changing the upper bellows pressure between $1-2.5\,\mathrm{MPa}$
would utilize its entire range of motion without risking the formation
of solid \textsuperscript{4}He there. A more detailed description
of the bellows is also in Ref.~\cite{Sebedash_QFS}.

\subsection{Quartz tuning fork\label{subsec:Quartz-tuning-fork}}

\begin{figure}[b]
\subfloat{\includegraphics[width=0.5\columnwidth]{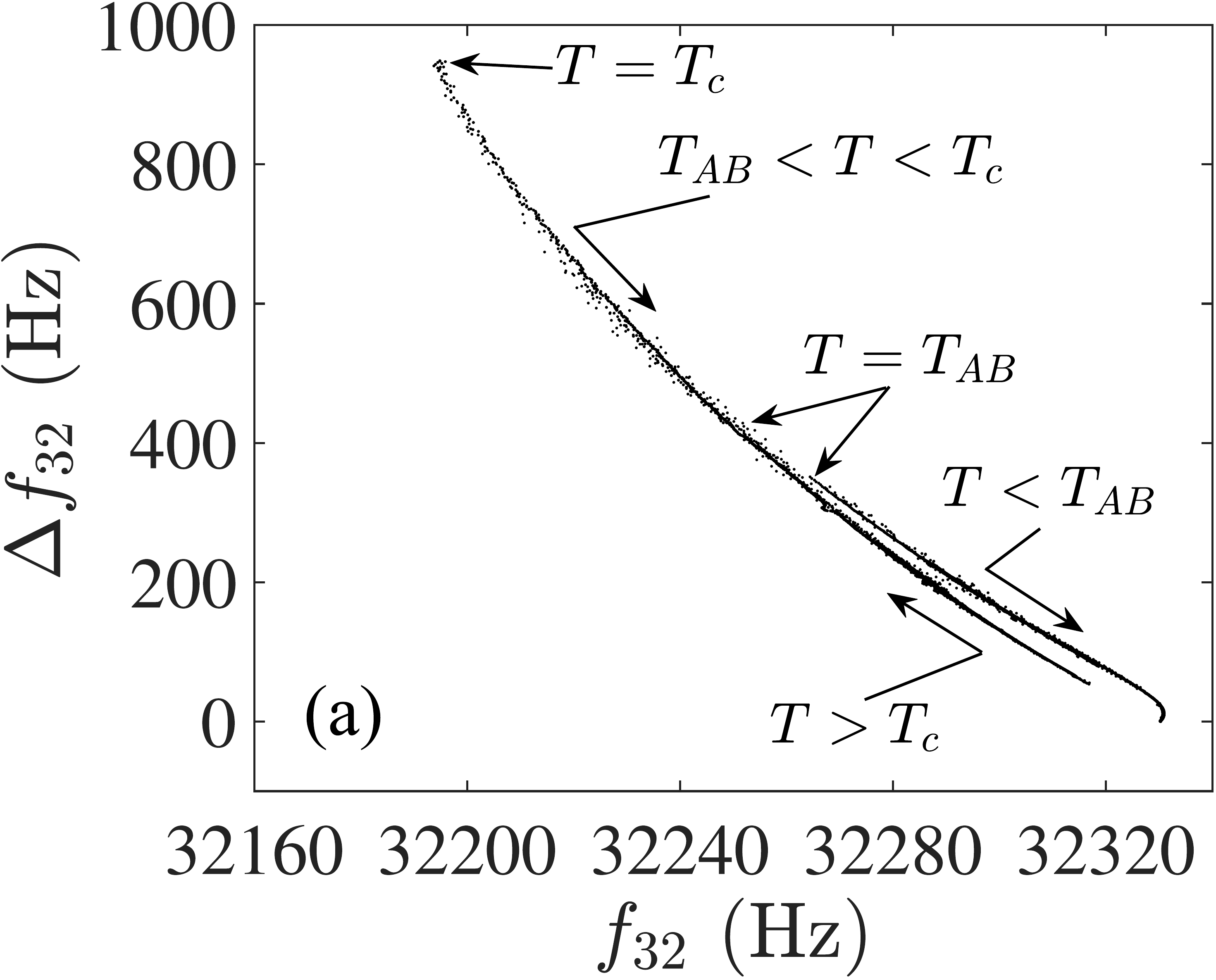}}\quad{}\subfloat{\includegraphics[width=0.5\columnwidth]{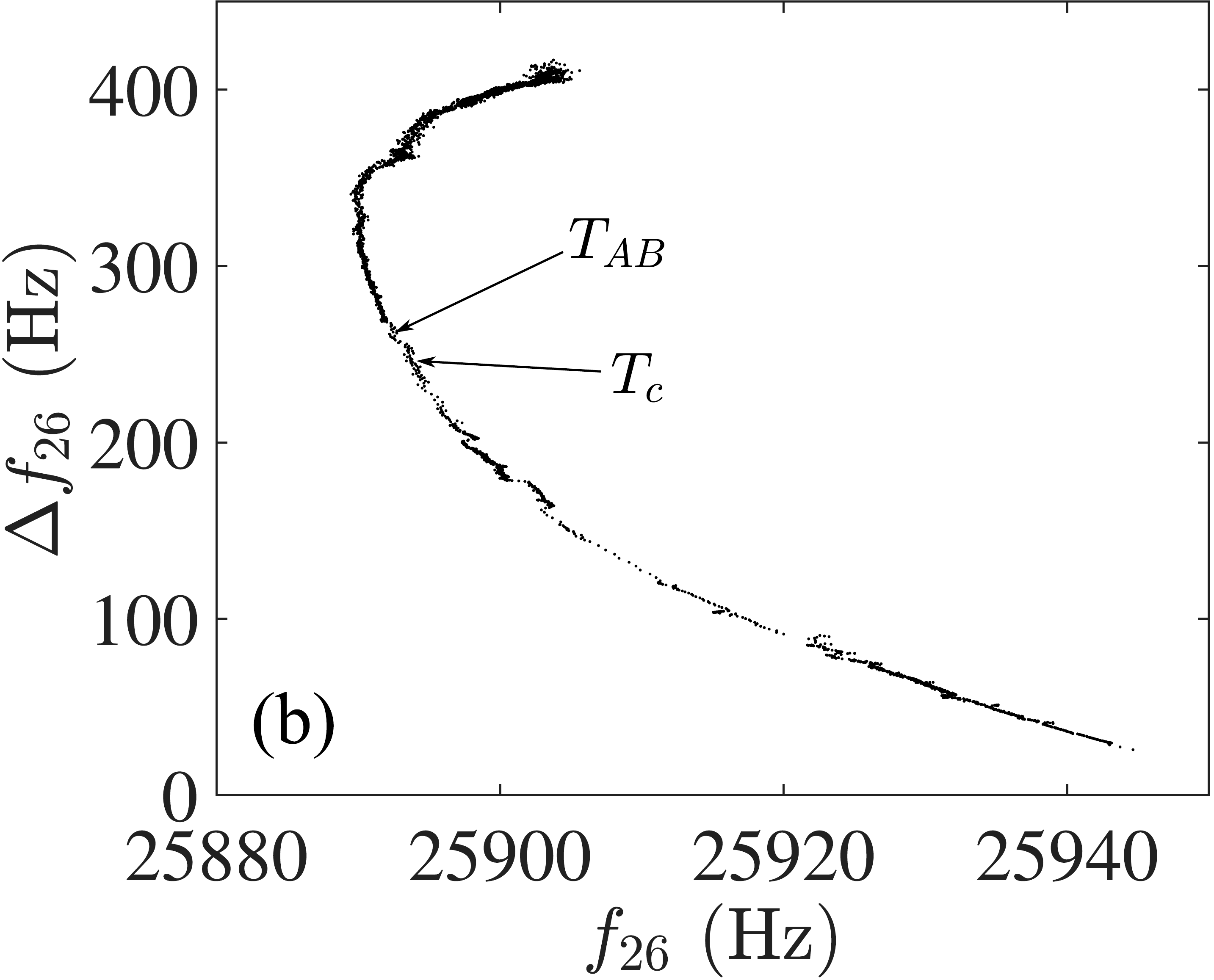}}\caption{Resonance width against resonance frequency for both \protect\textsuperscript{3}He
QTF $f_{32}$ (a), and \protect\textsuperscript{3}He\textendash \protect\textsuperscript{4}He
mixture QTF $f_{26}$ (b), at the \protect\textsuperscript{4}He crystallization
pressure. Significant points of temperature are also shown\label{fig:Resonace-width-freq}}
\end{figure}
The main volume of the experimental cell was monitored by two quartz
tuning fork resonators (QTFs): one in a tubular extension on the top
half of the cell structure, and the other in the middle of the volume.
The upper QTF was $32\,\mathrm{kHz}$ resonance frequency \emph{ECS
Inc.} \emph{ECS-.327-8-14X} oscillator placed so that it would always
be in the pure \textsuperscript{3}He phase acting as our main thermometer.
The second QTF was $26\,\mathrm{kHz}$ resonance frequency \emph{EPSON}
\emph{C-2 26.6670K-P:PBFREE} oscillator situated in the middle of
the main volume and thus either in liquid \textsuperscript{3}He\textendash \textsuperscript{4}He
mixture or frozen in solid \textsuperscript{4}He (and thus inoperable),
depending on the amount of solid. The forks had different resonance
frequencies to ensure that they did not interfere with each other.

The forks were operated by two separate circuits with excitation provided
by a signal generator and the signal read with a combination of a
pre-amplifier and a lock-in amplifier. The two parameters determined
from the readout were the resonance frequency and its width (full-width
at half-maximum). The circuits could be operated either in a full-frequency
sweep mode, or a single-point tracking mode \cite{Pentti_Rysti_Salmela}.
The tracking mode assumes a Lorentzian lineshape and conservation
of energy resulting in a constant resonance curve area. It is useful
at small resonance widths to increase the data collection rate, because
it circumvents the necessity to wait a time proportional to the inverse
of the width after changing the frequency.

The response of each fork to temperature is illustrated in Fig.~\ref{fig:Resonace-width-freq}.
As we approach the $T_{c}$ from above, the viscosity of normal fluid
\textsuperscript{3}He increases with decreasing temperature (Fig.~\ref{fig:Resonace-width-freq}a),
observed as increased width of the $32\,\mathrm{kHz}$ QTF. At $T=T_{c}=2.6\,\mathrm{mK}$
the width is $\Delta f_{32}\approx926\,\mathrm{Hz}$, after which
the dissipation of the fork starts to decrease due to the now-superfluid
nature of \textsuperscript{3}He. When we cooled to below the $T_{c}$,
liquid \textsuperscript{3}He tended to undercool before transitioning
to superfluid state, and the width-at-$T_{c}$ value was determined
during the warm-up instead. \textsuperscript{3}He is first an A-phase
superfluid with the transition to B-phase occurring at $T=T_{AB}=0.917T_{c}$ \cite{Greywall1986}$\approx2.4\,\mathrm{mK}$,
indicated by a jump in the resonance width from $424\,\mathrm{Hz}$
to $350\,\mathrm{Hz}$. As the temperature in the superfluid \textsuperscript{3}He
is decreased, the number of quasiparticles decreases, and the mean-free
path of the remaining particles increases. Eventually, it exceeds
the dimensions of the experimental cell and we reach the ballistic
regime. This occurs at $\Delta f_{32}\approx20\,\mathrm{Hz}$, corresponding
to about $0.25T_{c}\approx0.7\,\mathrm{mK}$ \cite{Todoshchenko2014},
below which the resonance frequency no longer changes significantly
but the width still has temperature resolution down to about 0.3 mK,
where it saturates to our minimum observed width $\sim0.14\,\mathrm{Hz}$.

The caveat regarding the $32\,\mathrm{kHz}$ QTF is that it was not
in pure bulk \textsuperscript{3}He. Rather, since there was \textsuperscript{4}He
present in the system, all available surface was covered by a superfluid
\textsuperscript{4}He film. This included the surfaces of the ``\textsuperscript{3}He
QTF'' as well. In principle, deep in bulk superfluid \textsuperscript{3}He
phase, we should be able to reach the vacuum resonance width of the
QTF of order 10 mHz, as the superfluid-induced dissipation disappears.
But, as said, the lowest observed resonance width was of order 100
mHz, leading us to believe that the \textsuperscript{4}He film coverage
is responsible for the additional dissipation. The effect of the film
starts to be meaningful when the width drops below $\sim1\,\mathrm{Hz}$
. More discussion about such film influence on mechanical oscillators
can be found in Refs.~\cite{Riekki2019a,Peshkov1975,Kim1993,Boldarev2011,Murakawa2012}.

\begin{figure}[b]
\center\includegraphics[width=0.75\columnwidth]{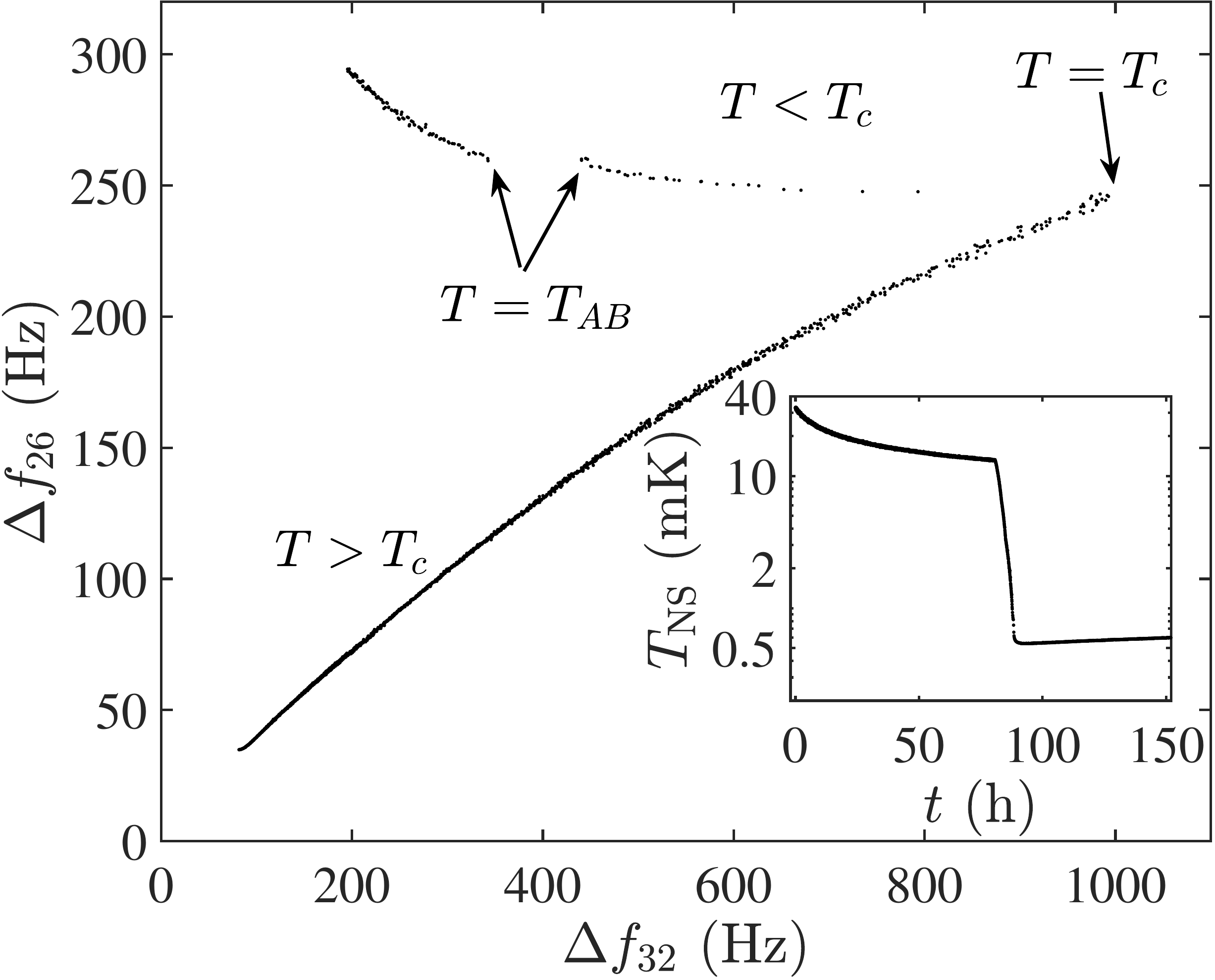}\caption{Resonance width of mixture QTF $f_{26}$ against resonance width of
the \protect\textsuperscript{3}He QTF $f_{32}$. Inset shows the
nuclear stage temperature during the run\label{fig:f32vsf26}}
\end{figure}
The second QTF was immersed in the mixture phase, where the \textsuperscript{3}He
component remained always in the normal state. Thus, the measured
resonance width increased monotonically with decreasing temperature
(Fig.~\ref{fig:Resonace-width-freq}b), saturating to 405 Hz at about
$1\,\mathrm{mK}$. The main purpose of this QTF was to show the anticipated superfluid
transition in mixture that would have resulted in sudden decrease
in the resonance width. It could also be used as another thermometer
down to its saturation. We performed one precool with small enough
crystal to keep both forks available the entire time to cross-check
their response to temperature. This is shown in Fig.~\ref{fig:f32vsf26}
demonstrating congruent temperature response between them, and that
there were no discernible temperature gradients within \textsuperscript{3}He
and mixture phase down to 1 mK during this particular run.

The width of the \textsuperscript{3}He QTF was converted to temperature
with a phenomenological formula defined piecewise in \textsuperscript{3}He-B,
\textsuperscript{3}He-A, and normal fluid. For the normal fluid region,
we combined the hydrodynamic tuning fork equations from Ref.~\cite{Blaauwgeers},
and the bulk \textsuperscript{3}He viscosity from Ref.~\cite{Pentti_Rysti_Salmela}
to give
\begin{equation}
\left(\frac{T}{\mathrm{mK}}\right)^{2}=\frac{1}{6.65}\left\{ \left[\left(\frac{\Delta f_{32}}{\mathcal{A}}\right)^{2}\left(\frac{f_{\mathrm{vac}}}{f_{32}}\right)^{4}\frac{4\pi}{\rho f_{32}}\,\mathrm{\frac{kg\,Hz}{m^{3}}}-10^{-5}\right]^{-1}-12.8\right\} ,\label{eq:calib-normal}
\end{equation}
where $f_{32}$ and $\Delta f_{32}$ are the measured resonance frequency
and width, respectively, $f_{\mathrm{vac}}=32765.9\,\mathrm{Hz}$
is the vacuum resonance frequency, $\rho=112.7\,\mathrm{g/cm^{3}}$ \cite{Kollar2000a}
is the density of liquid \textsuperscript{3}He at \textsuperscript{4}He
crystallization pressure, and $\mathcal{A}=0.429\,\mathrm{Hz}$ is
a fitting parameter, determined from $\Delta f_{32}\left(T=T_{c}\right)=926\,\mathrm{Hz}$,
and $f_{32}\left(T=T_{c}\right)=32187\,\mathrm{Hz}$.
\begin{figure}[b]
\center\includegraphics[width=0.75\columnwidth]{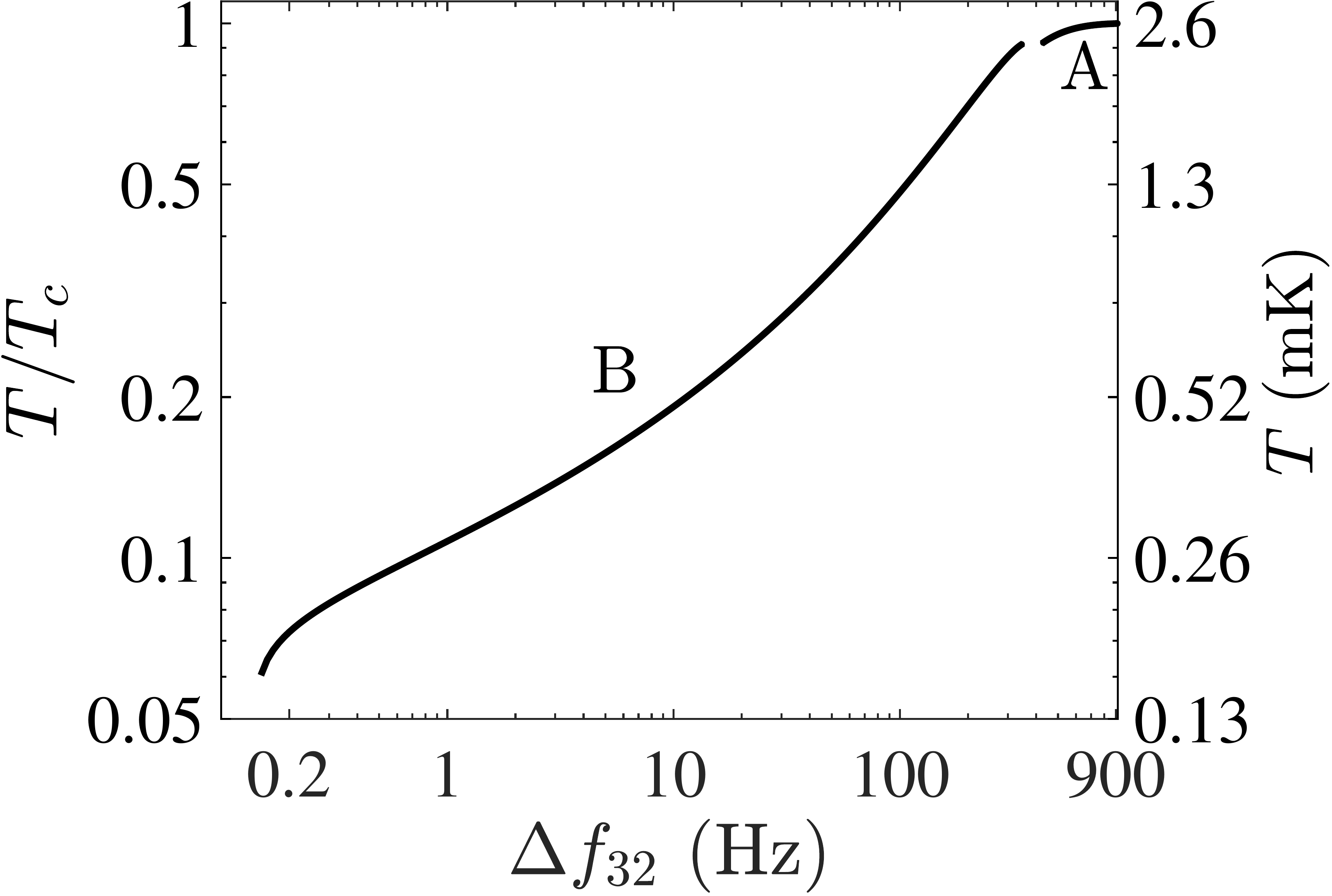}

\caption{Conversion of the width of the \protect\textsuperscript{3}He QTF
to temperature below the $T_{c}$, at \protect\textsuperscript{4}He
crystallization pressure, in superfluid A and B-phases\label{fig:calibration}}
\end{figure}

In the B-phase, we used $T_{AB}$ as a fixed point, and the ballistic
cross-over temperature as a ``semi-free'' point with temperature
fixed to $0.25T_{c}$ but with the width value adjustable between
$15$ and $30\,\mathrm{Hz}$. The calibration is least accurate below
1 Hz widths, when the \textsuperscript{4}He film covering the QTF
started to affect the measurement. To the narrow region of A-phase,
we fitted a simple exponential function that was continuous with the
B-phase formula both in value and in the first derivative. We ended
up with
\begin{align}
\frac{T}{T_{c}}= & \left[1+\frac{\Delta f_{32}^{0.3}-\left(\Delta f_{b}-\Delta f_{0}\right)^{0.3}}{\mathcal{B}^{0.3}}-\frac{\Delta f_{32}^{1.4}-\left(\Delta f_{b}-\Delta f_{0}\right)^{1.4}}{\mathcal{C}^{1.4}}\right]\cdot\label{eq:calib-B}\\
 & \left[4-\mathcal{D}\ln\left(\frac{\Delta f_{32}-\Delta f_{0}}{\Delta f_{b}-\Delta f_{0}}\right)\right]^{-1}\nonumber 
\end{align}
for \textsuperscript{3}He-B, and
\begin{equation}
\frac{T}{T_{c}}=\left[1-\mathcal{E}\exp\left(-\frac{\Delta f-\Delta f_{AB}}{\mathcal{F}}\right)\right]\label{eq:calib-A}
\end{equation}
for \textsuperscript{3}He-A, where $\Delta f_{b}=22\,\mathrm{Hz}$
is the ballistic cross-over width, $\Delta f_{0}=0.14\,\mathrm{Hz}$
the residual width, $\Delta f_{AB}=424\,\mathrm{Hz}$ the width at
the AB-transition, while $\mathbb{\mathcal{B}}-\mathcal{F}$ are fitting
parameters whose values are listed in Table~\ref{tab:fittingparams}.
The exponents 0.3 and 1.4 in Eq.~\eqref{eq:calib-B} were determined
empirically to give credible behavior across the whole span of the
B-phase. The conversion from resonance width to temperature below
the $T_{c}$ is shown in Fig.~\ref{fig:calibration}.

\subsection{Measurement procedure\label{subsec:Measurement-procedure}}

\begin{wraptable}{L}{0.25\columnwidth}%
\centering%
\begin{tabular*}{0.25\columnwidth}{@{\extracolsep{\fill}}cc}
variable & \multicolumn{1}{c}{value}\tabularnewline
\hline 
$\mathcal{A}$ & 0.429 Hz\tabularnewline
$\mathcal{B}$ & $2.138$ kHz\tabularnewline
$\mathcal{C}$ & 391.0 Hz\tabularnewline
$\mathcal{D}$ & 1.250\tabularnewline
$\mathcal{E}$ & 0.085\tabularnewline
$\mathcal{F}$ & 129.5 Hz\tabularnewline
\end{tabular*}\caption{Values of the fitting parameters of Eqs.~\eqref{eq:calib-normal}-\eqref{eq:calib-A}.\label{tab:fittingparams}}
\end{wraptable}%
A successful melting run requires sufficient amount of good quality
solid \textsuperscript{4}He, low enough precooling temperature, and
a reasonable melting rate of the solid. In order to have solid phase
of absolutely pure \textsuperscript{4}He, the crystal must always
be kept below 50 mK, as temperatures above that allow \textsuperscript{3}He
to start to dissolve into it \cite{Balibar2000,Balibar2002,Pantalei2010}.
To ensure the quality of the solid, we performed the initial nucleation
and growth to maximal size below 20 mK, and usually the crystals intended
to be used to pursue the lowest possible temperatures were grown entirely
below the \textsuperscript{3}He $T_{c}$. At no point we observed
any indication of, or had a reason to suspect, \textsuperscript{3}He
inclusions in the solid phase.

Nucleation of the solid \textsuperscript{4}He phase in the main cell
volume was not always straightforward, as it often tended to occur
in the bellows volume, and sometimes even at the upper end of the
superleak line. This was counterintuitive as the main cell resided
below the bellows volume to have even gravity favor the nucleation
there. During the experiment, we had to warm-up the setup to liquid
nitrogen temperature three times, due to trouble with the 1 K pot.
We used that as an opportunity to change the amount of \textsuperscript{3}He
in the main cell volume, as the cell could be emptied at such a high
temperature. We noted, that after each such thermal cycle, the nucleation
of \textsuperscript{4}He in the cell became more and more difficult
with no apparent reason. Eventually, to entice the nucleation in the
cell, we had to warm up the bellows volume to above 100 mK, while
the main volume was below the $T_{c}$. In the end, we resolved to
nucleate a new crystal as few times as possible. This resulted in
increased uncertainty in the determined amount of solid, since the
error in the \textsuperscript{4}He flow measurement accumulated when
the solid was melted and grown repeatedly without a fresh start from
zero.

The precooling of the main cell consisted of two stages: first the
cooling after magnetization of the nuclear stage to the dilution refrigerator
temperature, and then precooling by the demagnetization of the nuclear
stage. After the initial nucleation and growth near the $T_{c}$,
the nuclear stage was magnetized, which resulted the temperature to
increase near 50 mK followed by precool to dilution refrigerator temperature
around 12 mK. Then, after demagnetization, it took approximately 24
hours for the main volume to reach the pure \textsuperscript{3}He
$T_{c}$.

When the temperature of the cell dropped below the $T_{AB}$, we melted
the solid in the cell almost completely, leaving only a small nugget
of a crystal to not have to perform a new nucleation. This procedure
was deemed necessary as at the beginning of precool the crystal had
been close to the temperatures where\textsuperscript{ 3}He would
have been able to dissolve into it. When the solid phase was regrown
to the maximal size below the $T_{AB}$, we demagnetized the nuclear
stage further to reach precooling temperatures below 0.5 mK.

As the precooling proceeded, we allowed a small amount of flow out
of the superleak line to ensure that it was fully open. We had observed
that if the cell-side end of the superleak was blocked by solid, its
removal would result a harmful heat pulse to the cell. Furthermore,
at the ultimate precooling temperature, we would switch the pure \textsuperscript{3}He
QTF from sweep mode to the tracking mode to enable us to receive datapoints
more rapidly, with as small excitation as possible. Then, we would
start to remove \textsuperscript{4}He from the cell via the superleak
line, slowly increasing flow from zero to the desired value making
sure that no heat spikes would occur on the way. Towards the end,
as the solid was running out, we slowly decreased the flow back towards
zero, but leaving a small outward flow in place ($<1\,\mathrm{\mu mol/s}$)
for couple of hours to ensure that there would be no flow into the
cell during the post-melting warm-up period. Any flow into the cell
would regrow the crystal, which would cause heating as pure \textsuperscript{3}He
and mixture phases would separate. The melting process took anything
from a few minutes to a couple of hours, depending on the melting
rate. The warm-up period was observed for several hours, or at least
until the quartz tuning fork reading became more reliable indicator
of temperature again, i.e. the width was 1 Hz or above. Next, if the
nuclear stage temperature was still low enough, crystal was regrown
and precooled again for a new melt, and if not, a new magnetization
was commenced.

\section{Thermal model\label{sec:Thermal-model}}

\begin{figure}[b]
\center\includegraphics[width=0.75\columnwidth]{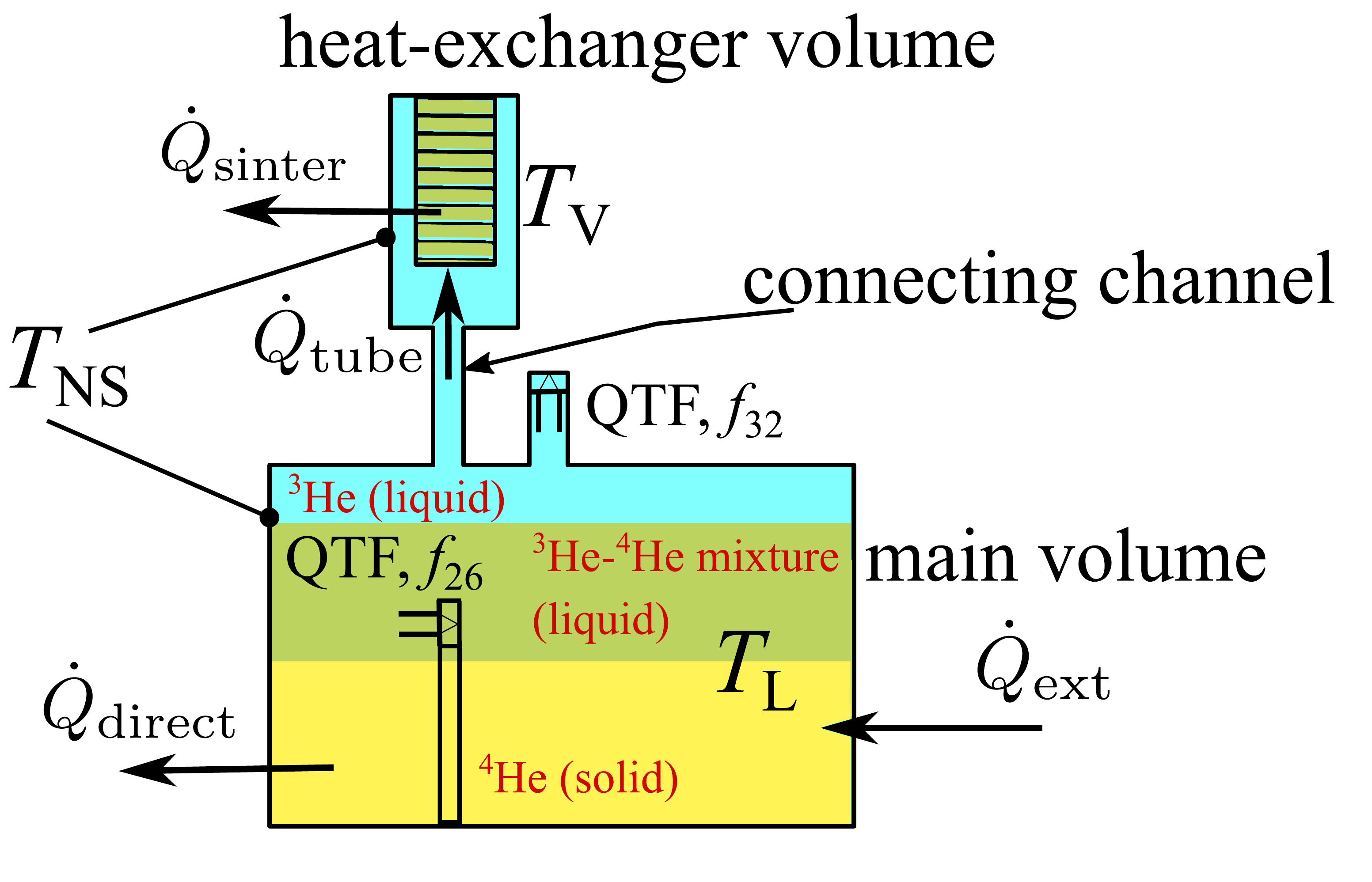}\caption{(color online) Simplified drawing of the experimental cell, showing
heat flows, temperatures, and phases in the system \cite{Riekki_kappa}.
Thermal gate (TG) is omitted here\label{fig:cell-heatflows} }

\end{figure}
Here, we will shift the focus to the computational model of the system,
which was required for temperature evaluations in the regimes where
the QTFs were no longer sensitive. The thermal model of our experimental
setup has already been discussed in Ref.~\cite{Riekki_kappa}, but
we will repeat the key considerations here. 

The heat balance equation for the main cell volume is
\begin{equation}
C_{\mathrm{L}}\dot{T}_{\mathrm{L}}=\dot{Q}_{\mathrm{direct}}+\dot{Q}_{\mathrm{tube}}+\dot{Q}_{\mathrm{melt}}+\dot{Q}_{f}+\dot{Q}_{\mathrm{ext}}.\label{eq:L-heat}
\end{equation}
Here $T_{\mathrm{L}}$ is the temperature of the liquid helium in
the main volume, $C_{\mathrm{L}}$ its heat capacity containing both
pure \textsuperscript{3}He and mixture contributions. A dot above
a symbol indicates time derivative. Different heat contributions $\dot{Q}_{\{\}}$
are evaluated as follows. We assume, that the Kapitza resistance can
be treated using the power law $R_{K}=R_{0}/\left(AT^{p}\right)$,
where $T$ is temperature, $A$ the surface area, and $R_{0}$ and
$p$ are constants to be determined. In our analysis we have combined
$A$ and $R_{0}$ into one constant $r=A/R_{0}$ to give $\dot{Q}_{\mathrm{direct}}=\frac{r_{\mathrm{L}}}{p_{\mathrm{L}}+1}\left(T_{\mathrm{NS}}^{p_{\mathrm{L}}+1}-T_{\mathrm{L}}^{p_{\mathrm{L}}+1}\right)$,
which is the heat transmitted between the cell liquid and the nuclear
stage through the plain cell wall. Then 
\begin{equation}
\dot{Q}_{\mathrm{tube}}=D\intop_{T_{\mathrm{V}}}^{T_{\mathrm{L}}}\kappa\left(T'\right)\mathrm{d}T'\label{eq:tube-heat}
\end{equation}
is the heat flowing between the main volume and sinter volume through
the connecting channel, where $D$ is a parameter depending on the
tube dimensions, and $\kappa$ is the \textsuperscript{3}He
thermal conductivity \cite{Riekki_kappa,Greywall1984}. When solid
\textsuperscript{4}He is grown, or melted, \textsuperscript{3}He
is transferred between pure \textsuperscript{3}He and mixture phase
with associated latent heat $\dot{Q}_{\mathrm{melt}}=T_{\mathrm{L}}\dot{n}_{\mathrm{3}}\left(S_{\mathrm{m,3}}-S_{3}\right)$,
where $\dot{n}_{\mathrm{3}}$ is the phase-transfer rate, and $S_{3}$
and $S_{\mathrm{m,3}}$ are the entropies of pure and mixture phase
per mole of \textsuperscript{3}He \cite{Riekki2019}, respectively.
Below about $0.2T_{c}$, we can use the low temperature approximation
$\dot{Q}_{\mathrm{melt}}=109\,\mathrm{\frac{J}{mol\,K^{2}}}\dot{n}_{3}T^{2}$.
$\dot{Q}_{f}$ represents the flow-dependent heat leaks in the system,
while $\dot{Q}_{\mathrm{ext}}$ is the generic background heat leak
to the cell main volume ($20-300\,\mathrm{pW}$). 

Next, the heat balance equation for the heat-exchanger volume reads
\begin{equation}
C_{\mathrm{V}}\dot{T}_{\mathrm{V}}=\dot{Q}_{\mathrm{sinter}}-\dot{Q}_{\mathrm{tube}}+\dot{Q}_{\mathrm{extV}},\label{eq:V-heat}
\end{equation}
where $T_{\mathrm{V}}$ is the temperature of the liquid in the sinter
volume, and $C_{\mathrm{V}}$ its heat capacity, while $\dot{Q}_{\mathrm{sinter}}=\frac{r_{\mathrm{V}}}{p_{\mathrm{V}}+1}\left(T_{\mathrm{NS}}^{p_{\mathrm{V}}+1}-T_{\mathrm{V}}^{p_{\mathrm{V}}+1}\right)$
is the heat transferred between the liquid and the nuclear stage through
the sinter Kapitza resistance and $\dot{Q}_{\mathrm{tube}}$ is from
Eq.~\eqref{eq:tube-heat}. Lastly, $\dot{Q}_{\mathrm{extV}}$ is
the background heat leak arriving directly to the heat-exchanger volume,
but it was immeasurably small and is omitted in the following. The
heat flows and temperatures are illustrated in the simplified schematic of the cell
in Fig.~\ref{fig:cell-heatflows}.

To finalize the computational model, we still need to determine the
Kapitza resistance coefficients of the system ($r_{\mathrm{L}}$,
$p_{\mathrm{L}}$, $r_{\mathrm{V}}$ and $p_{\mathrm{V}}$), the flow-rate
dependent heating $\dot{Q}_{f}$, as well as the background heat leak
$\dot{Q}_{\mathrm{ext}}$.

\section{Kapitza resistances\label{subsec:Kapitza-resistances}}

\subsection{Plain cell wall\label{subsec:cell-wall}}

\begin{figure}[b]
\includegraphics[width=1\columnwidth]{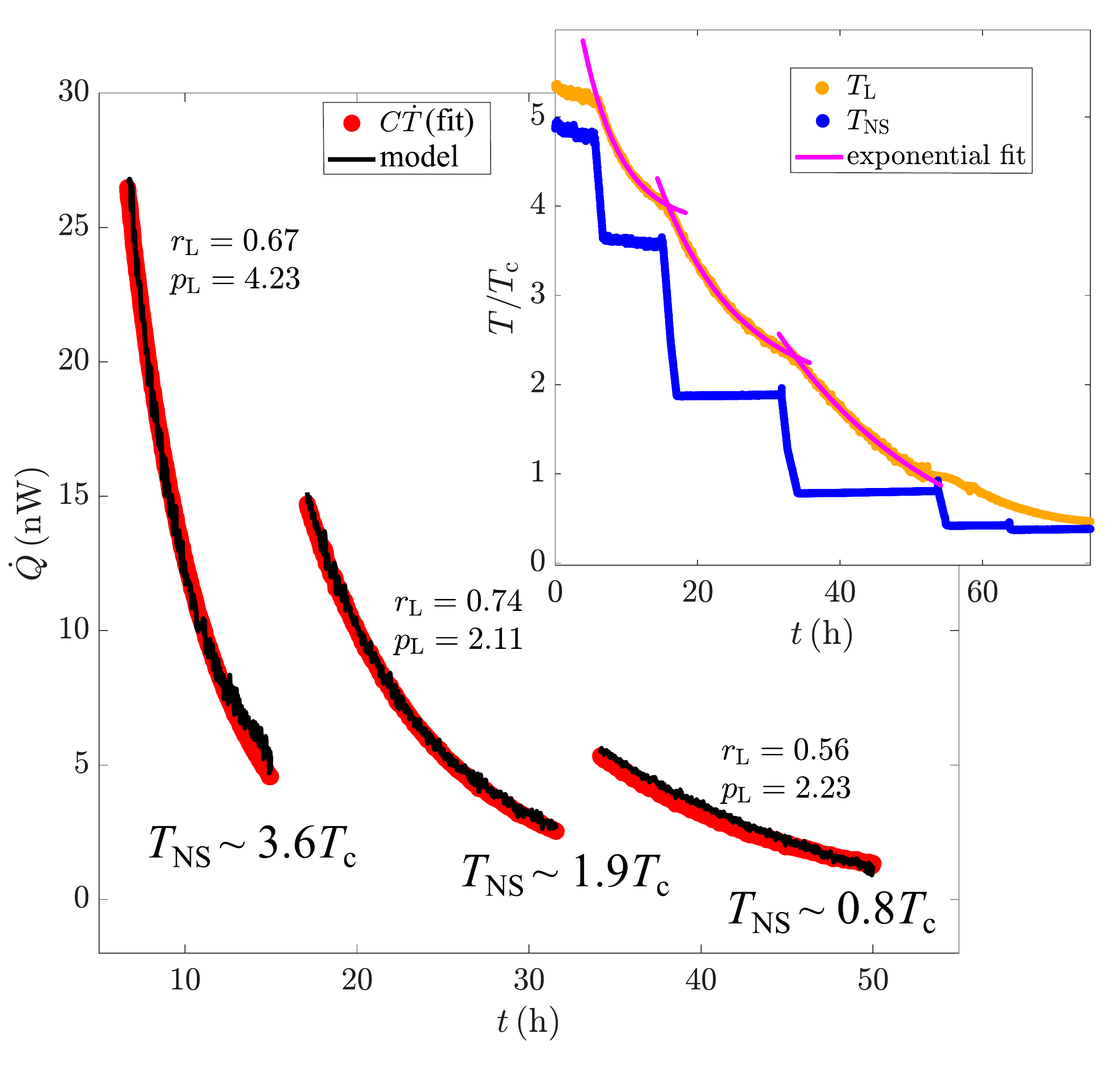}

\caption{(color online) Heat transferred between the main volume of the cell
and the nuclear stage directly through the plain cell wall at different
nuclear stage temperatures. Inset shows the measured cell main volume
temperature $T_{\mathrm{L}}$ and the nuclear stage temperature $T_{\mathrm{NS}}$,
as well as exponential fits to the $T_{\mathrm{L}}$ data at each
cooling step \label{fig:direct-Kapitza}}
\end{figure}
The Kapitza resistance of the plain cell wall plays important role
to the heat flow from the experimental cell to the nuclear stage at
the beginning of the precooling process, whereas below the $T_{c}$
its contribution rapidly becomes negligibly small. At temperatures
above the $T_{c}$, we can simplify the cell main volume heat balance
equation Eq.~\eqref{eq:L-heat} to
\begin{equation}
C_{\mathrm{L}}\dot{T}_{\mathrm{L}}=\frac{r_{\mathrm{L}}}{p_{\mathrm{L}}+1}\left(T_{\mathrm{NS}}^{p_{\mathrm{L}}+1}-T_{\mathrm{L}}^{p_{\mathrm{L}}+1}\right)-D\kappa_{0}\ln\left(\frac{T_{\mathrm{L}}}{T_{\mathrm{NS}}}\right),\label{eq:HB-directKapitza}
\end{equation}
since now it is safe to assume that the sinter volume is at the same
temperature as the nuclear stage, \textsuperscript{3}He is in normal
state everywhere, and $\dot{Q}_{\mathrm{melt}}=\dot{Q}_{f}=0$ because
the amount of solid \textsuperscript{4}He usually does not change
during precooling. Furthermore, we can ignore the background heat
leak $\dot{Q}_{\mathrm{ext}}$ of order $0.1\,\mathrm{nW}$ at these
temperatures. Here $\kappa_{0}=9.69\cdot10^{-5}$~W/(K$\,$m) \cite{Greywall1984} is the normal fluid \textsuperscript{3}He thermal conductivity.

We decreased the nuclear stage temperature stepwise and then observed
the cooling of the cell main volume. To produce smoother derivative
$\dot{T}_{\mathrm{L}}$, we fitted an exponential ($\propto\exp\left(-1/T\right)$)
function to cell temperature data at each cooling step. Measured temperatures
and these fits are shown in the inset of Fig.~\ref{fig:direct-Kapitza}.
Now, since we also know the properties of the connecting tube in the
last term of Eq.~\eqref{eq:HB-directKapitza}, we are left with two
parameters to be fitted: $r_{\mathrm{L}}$ and $p_{\mathrm{L}}$.
Comparison between the smoothed $C_{\mathrm{L}}\dot{T}_{\mathrm{L}}$
data and the data reproduced using the obtained Kapitza parameter
values are shown in the main panel of Fig.~\ref{fig:direct-Kapitza},
where the model calculation was performed using the original measured
$T_{\mathrm{L}}$ data (not the exponential fits). 
\begin{figure}[b]
\center\includegraphics[width=1\columnwidth]{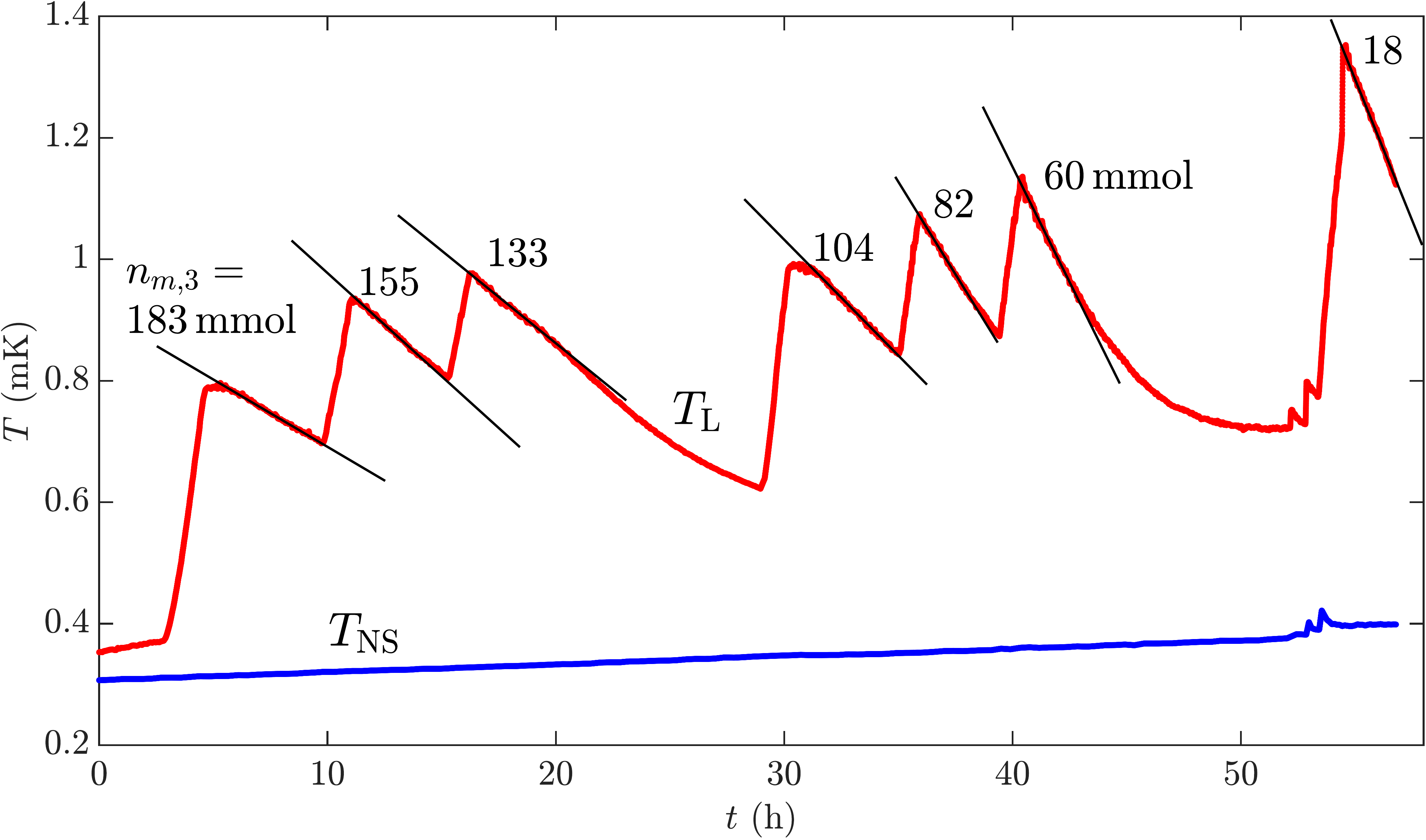}

\caption{(color online) Nuclear stage temperature $T_{\mathrm{NS}}$ and cell
main volume temperature $T_{\mathrm{L}}$ while growing solid \protect\textsuperscript{4}He
stepwise. The amount of \protect\textsuperscript{3}He in the mixture
phase after each growth is shown next to the $T_{\mathrm{L}}$ graph,
while the black lines indicate linear fits to each step with slopes
$-6.2\cdot10^{-6}$, $-9.5\cdot10^{-6}$, $-8.5\cdot10^{-6}$, $-1.1\cdot10^{-5}$,
$-1.7\cdot10^{-5}$, $-2.1\cdot10^{-5}$, and $-2.6\cdot10^{-5}\,\mathrm{mK/s}$
from left to right\label{fig:stepwise-growth}}
\end{figure}

Before the demagnetization begins, the independently measured $T_{\mathrm{L}}$
and $T_{\mathrm{NS}}$ deviate from each other more than our thermal
analysis suggests, which may be in part a result of inaccuracy in
our nuclear stage PLM-thermometer calibration, or fork calibration,
or both, at these high temperatures. If we scale up $T_{\mathrm{NS}}$
by 5\% to match the readings, the recomputed Kapitza parameter values
from left to right in Fig.~\ref{fig:direct-Kapitza} will read ($r_{\mathrm{L}}=0.93$,
$p_{\mathrm{L}}=2.46$), ($0.68$, $1.69$) and ($0.56$, $2.06$).
It is evident that the unadjusted $T_{\mathrm{NS}}$ overestimates
the Kapitza resistance while the constant scaling across the entire
temperature range likely gives too low values. Therefore we took an
average value of those as our final parameters. Furthermore, our approximations
are no longer quite as good when we get closer to the $T_{c}$, making
the fit to the last precooling step the most unreliable of the three.
After analyzing five more similar datasets as in Fig.~\ref{fig:direct-Kapitza},
we ended up with average values $p_{\mathrm{L}}=\left(2.6\pm0.2\right)$
and $r_{\mathrm{L}}=\left(0.7\pm0.2\right)\,\mathrm{WK}^{-p_{\mathrm{L}}-1}$,
where the confidence bounds were determined as the standard error
of the fitted parameter values. The estimated cell wall area is $0.12\,\mathrm{m^{2}}$
giving us $R_{0,\mathrm{L}}=\left(0.17\pm0.05\right)\,\mathrm{m^{2}}\mathrm{K}^{p_{\mathrm{L}}+1}\mathrm{W^{-1}}$
as the area-scaled constant $R_{0}=A/r$. The exponent $p_{\mathrm{L}}$
is close to the theoretical value 3 from the acoustic mismatch model,
while the prefactor $R_{0,\mathrm{L}}$ is about 1000 times larger
than typically found in sintered heat-exchangers \cite{Voncken_sinter}.

\subsection{Sinter\label{subsec:sinter}}

To study the Kapitza resistance parameters of the sinter ($r_{\mathrm{V}}$
and $p_{\mathrm{V}}$), we grew, or melted, a small amount of solid
\textsuperscript{4}He periodically below the $T_{c}$ and observed
the relaxation of the system across a certain temperature span at
an approximately constant nuclear stage temperature $T_{\mathrm{NS}}$,
example of which is shown in Fig.~\ref{fig:stepwise-growth}. By
changing the amount of solid \textsuperscript{4}He, we altered the
amount of \textsuperscript{3}He\textendash \textsuperscript{4}He
mixture in the main volume of the cell. Since mixture is the main
contributor to the total heat capacity of the cell well below the
$T_{c}$, this gives us an opportunity to map the Kapitza parameters
over a large span of thermal loads.
\begin{figure}[b]
\includegraphics[width=1\columnwidth]{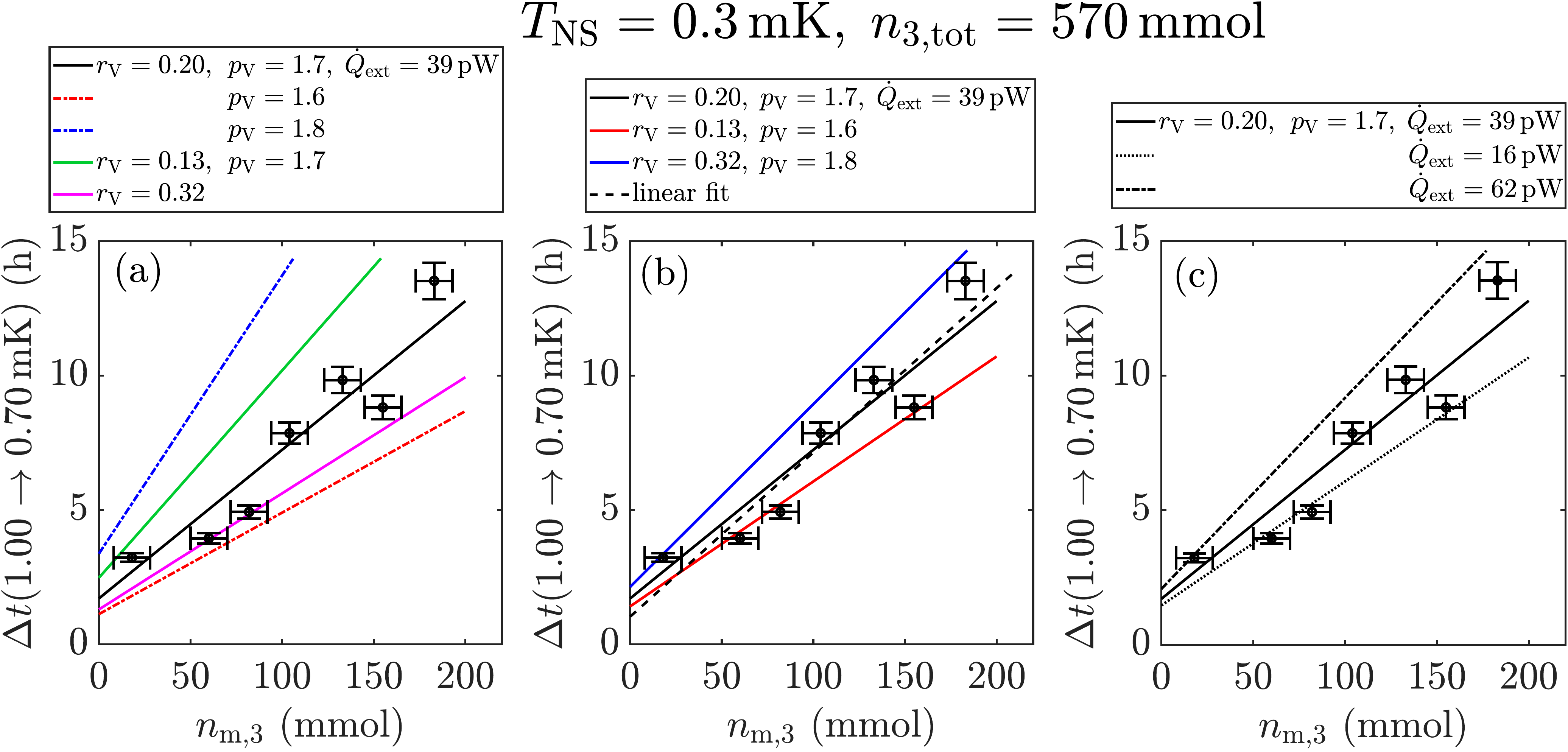}\caption{(color online) Relaxation time of the cell main volume temperature
from $T_{\mathrm{L}}=1.00\,\mathrm{mK}$ to $0.70\,\mathrm{mK}$ as
a function of \protect\textsuperscript{3}He in the mixture phase
(cf. Fig.~\ref{fig:stepwise-growth}). Various lines were obtained
by modeling the system using equations of Section~\ref{sec:Thermal-model}:
(a) constant $r_{\mathrm{V}}$/changing $p_{\mathrm{V}}$, and changing
$r_{\mathrm{V}}$/constant $p_{\mathrm{V}}$, with a constant background
external heat leak $\dot{Q}_{\mathrm{ext}}$, (b) adjusted $r_{\mathrm{V}}$
for each exponent with a constant heat leak (dashed line shows a linear
fit to the datapoints, for comparison), and (c) constant $r_{\mathrm{V}}$
and $p_{\mathrm{V}}$ at various heat leaks\label{fig:delta_T-1}}
\end{figure}

\begin{figure}[t]
\includegraphics[width=1\columnwidth]{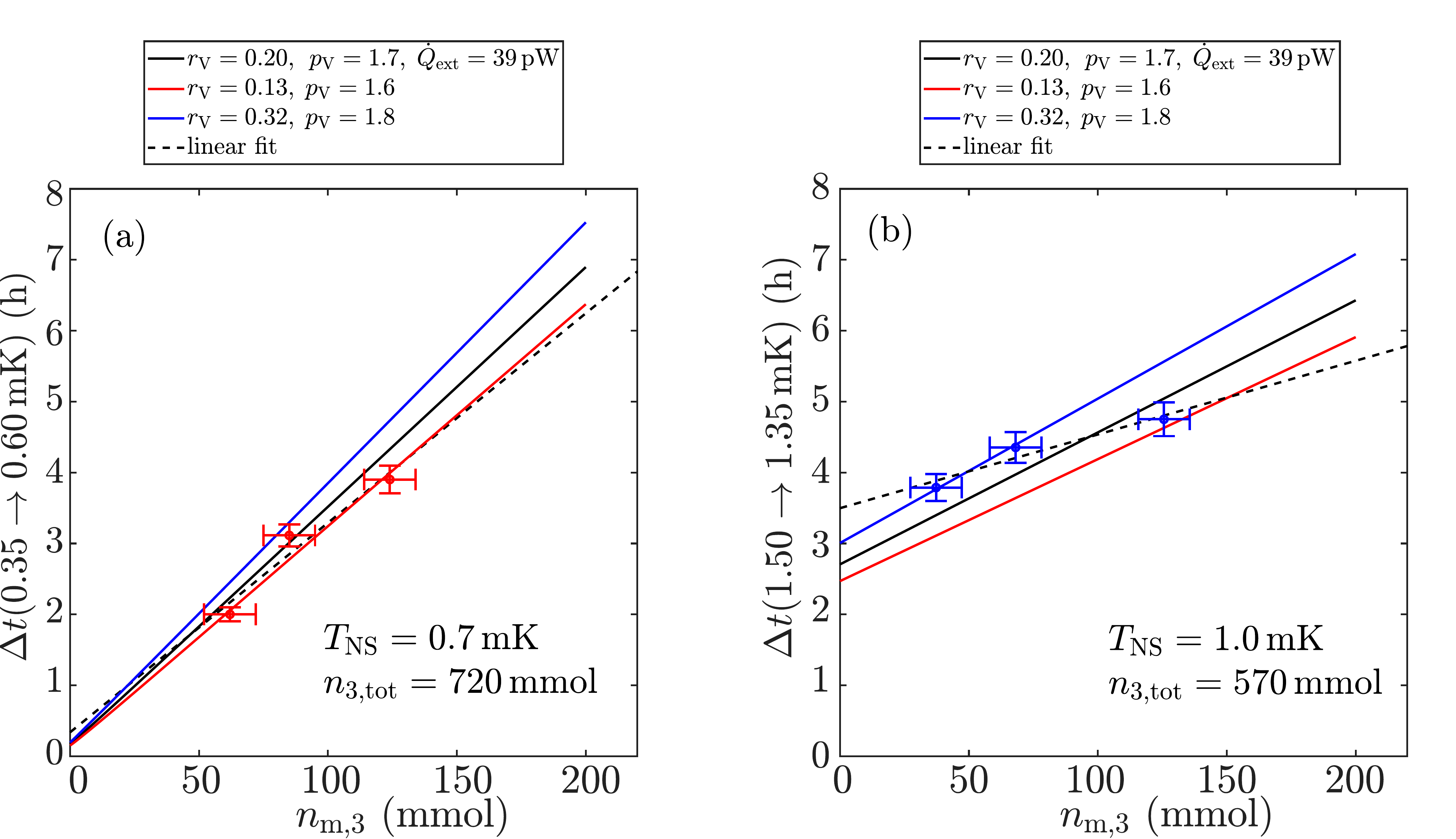}

\caption{(color online) Relaxation time of the cell main volume temperature
from $T_{\mathrm{L}}=0.35\,\mathrm{mK}$ to $0.60\,\mathrm{mK}$ after
melting solid \protect\textsuperscript{4}He periodically (a), and
from $T_{\mathrm{L}}=1.50\,\mathrm{mK}$ to $1.35\,\mathrm{mK}$ after
growing solid \protect\textsuperscript{4}He periodically (b) at different
mixture amounts (datapoints). Solid lines were obtained by modeling
the system with different Kapitza resistance parameters (cf. Fig.~\ref{fig:delta_T-1}b),
while the dashed lines are linear fits to the datapoints, for comparison\label{fig:delta_T-2}}
\end{figure}

The fitting process is not as straightforward as in Section~\ref{subsec:cell-wall}
for the plain cell wall Kapitza coefficients, because the thermal
conductivity of the channel connecting the two cell volumes \cite{Riekki_kappa}
now also plays a more important role, and the heat-exchanger volume temperature can
no longer be assumed to be equal to the nuclear stage temperature.
Thus, we simulated the entire system using equations of Section~\ref{sec:Thermal-model}
and varied the Kapitza coefficients. During the fitting, we first
chose $p_{\mathrm{V}}$, and then adjusted $r_{\mathrm{V}}$ trying
to make computed cell temperature match the measured one. The boundary
condition during the fitting process was that each combination of
$p_{\mathrm{V}}$ and $r_{\mathrm{V}}$ should result a constant Kapitza
resistance value at the upper limit of our range of interest, 10 mK.
This restriction made the fitting process more straightforward, and
it ensured that the computational model kept behaving consistently
above the $T_{c}$. The result of the analysis is shown in Figs.~\ref{fig:delta_T-1}
and \ref{fig:delta_T-2}. 

Figure~\ref{fig:delta_T-1} relates to the measurement of Fig.~\ref{fig:stepwise-growth},
where Fig.~\ref{fig:delta_T-1}a illustrates how changing either
only the exponent $p_{\mathrm{V}}$, or the coefficient $r_{\mathrm{V}}$
(without the 10 mK restriction) affects the simulated cooldown time.
Next, Fig.~\ref{fig:delta_T-1}b shows the fits with adjusted $r_{\mathrm{V}}$
for each $p_{\mathrm{V}}$, demonstrating that $p_{\mathrm{V}}=1.7$
is the most appropriate choice across the whole dataset. Another element
in the analysis is the somewhat variant external heat leak to the
experimental cell, effect of which is illustrated in Fig.~\ref{fig:delta_T-1}c.
Since the computed slope spread with typical range of $\dot{Q}_{\mathrm{ext}}\approx20...60\,\mathrm{pW}$
is of same order as in the fits of Fig.~\ref{fig:delta_T-1}b, we
conclude that anything within $p_{\mathrm{L}}=1.6$ and $1.8$ is
acceptable by adjusting the heat leak accordingly. This provides the
confidence bounds to our fitted parameters: $p_{\mathrm{V}}=\left(1.7\pm0.1\right)$
and $r_{\mathrm{V}}=\left(0.2\pm0.1\right)\,\mathrm{WK}^{-p_{\mathrm{V}}-1}$.

Figure~\ref{fig:delta_T-2} shows how these Kapitza parameters suit
with two more measurements at other temperature spans. The data of
Fig.~\ref{fig:delta_T-2}a was obtained during stepwise melting run,
and Fig.~\ref{fig:delta_T-2}b during another stepwise growth run.
The cooldown rates in Fig.~\ref{fig:delta_T-2}a are reproduced slightly
better with $p_{\mathrm{L}}=1.6$ than with 1.7, which is as good
as the other displayed options for Fig.~\ref{fig:delta_T-2}b. The
temperature range in Fig.~\ref{fig:delta_T-2}a is already quite
close to the saturation limit of our QTF thermometer, and thus is
likely the least reliable of the presented datasets.

\begin{figure}[t]
\center\includegraphics[width=0.75\columnwidth]{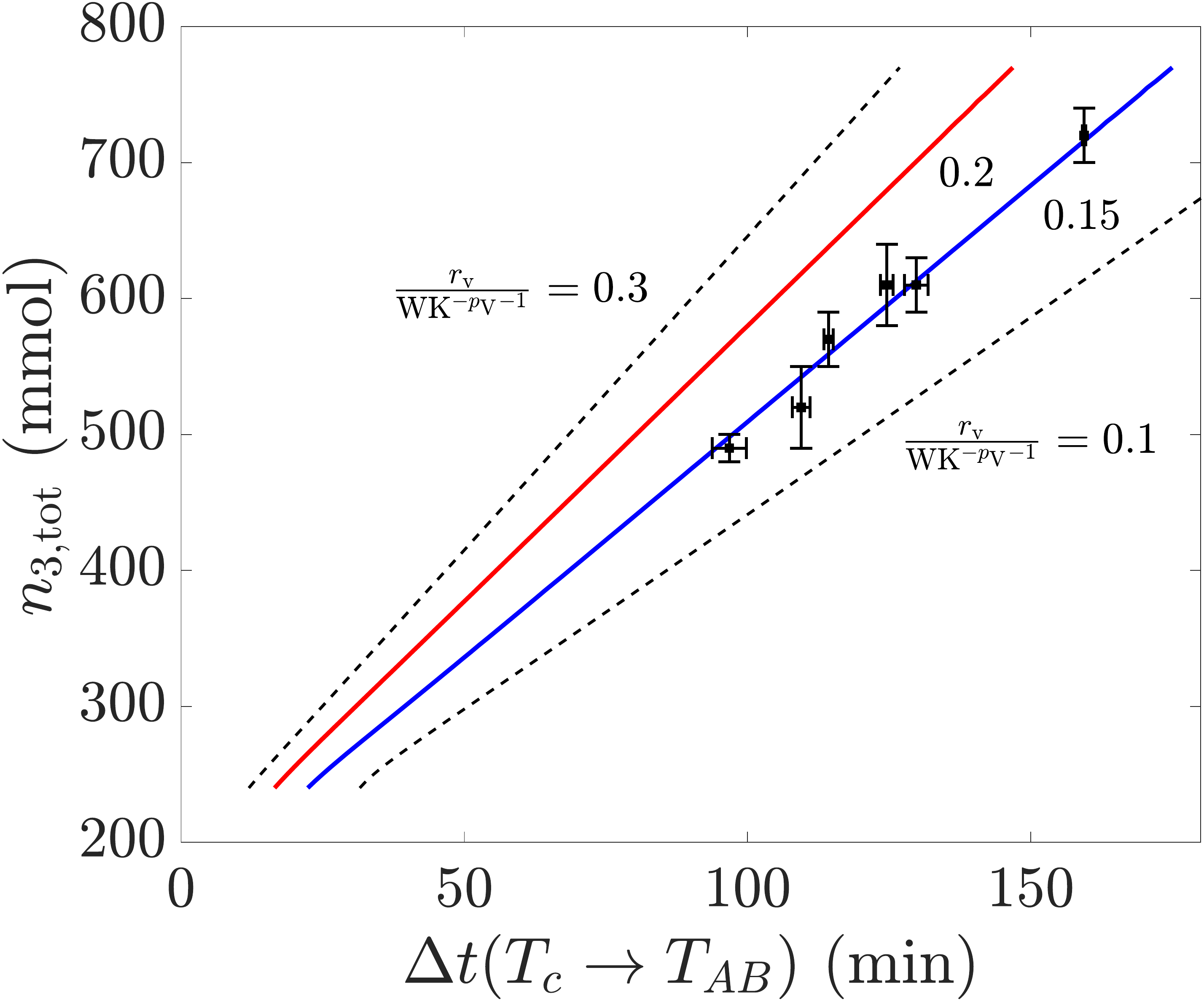}\caption{(color online) Time it took for the cell main volume to cool from
the $T_{c}$ to the $T_{AB}$ versus the total amount of \protect\textsuperscript{3}He
in the experimental cell (main volume + heat-exchanger volume + sinter
+ connecting channel). Various lines are computed $T_{c}\rightarrow T_{AB}$
times, assuming no \protect\textsuperscript{3}He in the mixture phase.
The dashed black lines show the behavior at the upper and lower end
of the $r_{\mathrm{V}}$ confidence bounds, while the red line is
the low temperature fit, and blue the best $r_{\mathrm{V}}$ to the
current dataset with $p_{\mathrm{V}}=1.7$ kept constant \label{fig:TcTab}}
\end{figure}
We should also acknowledge that the sinter Kapitza coefficient values
are likely not the same throughout the entire temperature range. This
was examined by analyzing the cooldown behavior close to the $T_{c}$
by plotting the time it took to cool from the $T_{c}$ to the $T_{AB}$
with varying amounts of \textsuperscript{3}He in the system, illustrated
in Fig.~\ref{fig:TcTab}. Here, superfluid \textsuperscript{3}He
still provides a significant contribution to the total heat capacity
of the system. Each datapoint in Fig.~\ref{fig:TcTab} represents
average $T_{c}\rightarrow T_{AB}$ time, taken over multiple precools
with maximal \textsuperscript{4}He crystal sizes. They are compared
against computed values, calculated by assuming that there is no mixture
in the system, that the nuclear stage is at constant $T_{\mathrm{NS}}=0.5\,\mathrm{mK}$
temperature, and that the background heat leak is constant $\dot{Q}_{\mathrm{ext}}=80\,\mathrm{pW}$.
At these temperatures, the heat leak is small compared to the heat
flow through the sinter and thus did not significantly affect the
computed $T_{c}\rightarrow T_{AB}$ cooling times. As we chose here
to keep the exponent $p_{\mathrm{V}}=1.7$ unchanging, we found the
best correspondence to the data with slightly reduced $r_{\mathrm{V}}=0.15\,\mathrm{WK}^{-p_{\mathrm{V}}-1}$
value. Nevertheless, the adjusted value is still within the limits
of our confidence bounds. Since our region of interest lies mainly
below 1 mK, we chose not to include any temperature dependence in
$r_{\mathrm{V}}$ to our computational model. Rather, we used the
value $r_{\mathrm{V}}=0.2\,\mathrm{WK}^{-p_{\mathrm{V}}-1}$ when
analyzing all our low temperature procedures, and kept an option to
use a slightly smaller value while treating data near the $T_{c}$.
\begin{table}[t]
\center%
\begin{tabular}{|cc||cc|cc|cc|cc|}
\hline 
\multicolumn{2}{|c||}{us} & \multicolumn{2}{c|}{(mix)\cite{Oh1994}} & \multicolumn{2}{c|}{(mix)\cite{Voncken_sinter}} & \multicolumn{2}{c|}{(\textsuperscript{3}He-B)\cite{Voncken_sinter}} & \multicolumn{2}{c|}{(\textsuperscript{3}He-B)\cite{Ahonen1978}}\tabularnewline
\hline 
$p$ & $R_{0}$ & $p$ & $R_{0}$  & $p$ & $R_{0}$  & $p$ & $R_{0}$  & $p$ & $R_{0}$\tabularnewline
\hline 
1.6 & 80 &  &  &  &  & 1 & 700 & 1 & 1100\tabularnewline
$1.7$ & $50$ & 2 & 10..30 & 2 & 6.5 & 2 & 0.5 &  & \tabularnewline
1.8 & 20 &  &  & 3 & 0.0029 & 3 & $0.2\cdot10^{-3}$ &  & \tabularnewline
(2) & (12) &  &  &  &  &  &  &  & \tabularnewline
\hline 
\end{tabular}\caption{Comparison between our sinter Kapitza parameters and the values received
by others in saturated mixture and \protect\textsuperscript{3}He-B.
The unit of $R_{0}$ is $\mathrm{m^{2}K^{\mathit{p}+1}W^{-1}}$. The
values in parentheses are our determined Kapitza exponent rounded
to the closest integer value and $R_{0}$ scaled to correspond it
(see text). \label{tab:parameter-comparison}}
\end{table}

The determined $r_{\mathrm{V}}=\left(0.2\pm0.1\right)\,\mathrm{WK}^{-p_{\mathrm{V}}-1}$,
with $10\,\mathrm{m^{2}}$ sinter surface area, corresponds to the
area-scaled prefactor $R_{0,\mathrm{V}}=\left(50\pm30\right)\,\mathrm{m^{2}}\mathrm{K}^{p_{\mathrm{V}}+1}\mathrm{W^{-1}}$.
To enable comparison with the measurements made by others, we can round our determined Kapitza exponent to the closest integer
value (2), and scale $R_{0}$ to maintain the constant Kapitza resistance
at 10 mK to get $12\,\mathrm{m^{2}K^{3}W^{-1}}$. Oh \emph{et al.} \cite{Oh1994} determined at 1 MPa that the Kapitza resistance
between their sinter and saturated mixture followed exponent $p=2$
with the coefficient $R_{0}$ between $10-30\mathrm{\,m^{2}}\mathrm{K}^{3}\mathrm{W^{-1}}$
depending on the magnetic field of their experiment. Voncken \emph{et al.} \cite{Voncken_sinter},
on the other hand, measured the Kapitza resistance of saturated mixture
and sinter in situation where the phase-separation boundary was within
the sinter, receiving either $p=2$ or 3, with $R_{0}=6.5\mathrm{\,m^{2}}\mathrm{K}^{3}\mathrm{W^{-1}}$
or $0.0029\mathrm{\,m^{2}}\mathrm{K}^{4}\mathrm{W^{-1}}$, respectively.
The comparison is summarized in Table \ref{tab:parameter-comparison}
which also includes the Kapitza resistances in \textsuperscript{3}He-B
from Refs.~\cite{Voncken_sinter} and \cite{Ahonen1978}. Our heat-exchanger
volume is mostly filled by pure \textsuperscript{3}He, but since
there is \textsuperscript{4}He readily available in the system, all
available surfaces are covered by it. This naturally includes the
sinter, which is why our observed Kapitza resistance parameters are
more in line with mixture parameters determined by others than with
the values in \textsuperscript{3}He-B.

\section{Helium isotope proportions\label{sec:Helium-quantities}}

Throughout the experiment, we kept log of the total amount of \textsuperscript{3}He
in the system, and how it was split between the different volumes. Initially,
we had total of 700 mmol of \textsuperscript{3}He, but learned that
about 1 mol was needed to have sufficiently large pure \textsuperscript{3}He
phase both in the cell main volume and in the lower bellows placed
within the dilution refrigerator.

Of the total \textsuperscript{3}He, 200-400 mmol was in the bellows
volume to ensure that the mixture there was always at saturation.
The heat-exchanger and the connecting channel required about 190 mmol
of \textsuperscript{3}He to completely fill the open volume, and,
based on the Kapitza resistance analysis of the previous section,
we assume that the pores of the sinter were completely filled with
saturated mixture. Since we had 11 g of sinter, with density $10.5\cdot10^{3}\,\mathrm{kg/m^{3}}$
and filling factor 0.5, we had 90 mmol of saturated (8.1\%) mixture
in trapped in the sinter, meaning we had additional 7 mmol of \textsuperscript{3}He
stored in the heat-exchanger volume. The rest resided in the main
volume of the experimental cell. It is not clear whether the mixture
trapped in the sinter should be exactly at the bulk saturation concentration,
but we deemed it a reasonable approximation.

The optimal amount of \textsuperscript{3}He in the main volume was
about 400 mmol. Below that, at small \textsuperscript{4}He crystal
sizes, the \textsuperscript{3}He QTF also became immersed in mixture,
thus making it rather useless as a thermometer. Conversely, if there
were a lot more than 400 mmol of \textsuperscript{3}He in the main
volume the solid \textsuperscript{4}He phase could not have been
grown to maximal size, as now the pure \textsuperscript{3}He phase
took so much space.
\begin{figure}
\center\includegraphics[width=0.75\columnwidth]{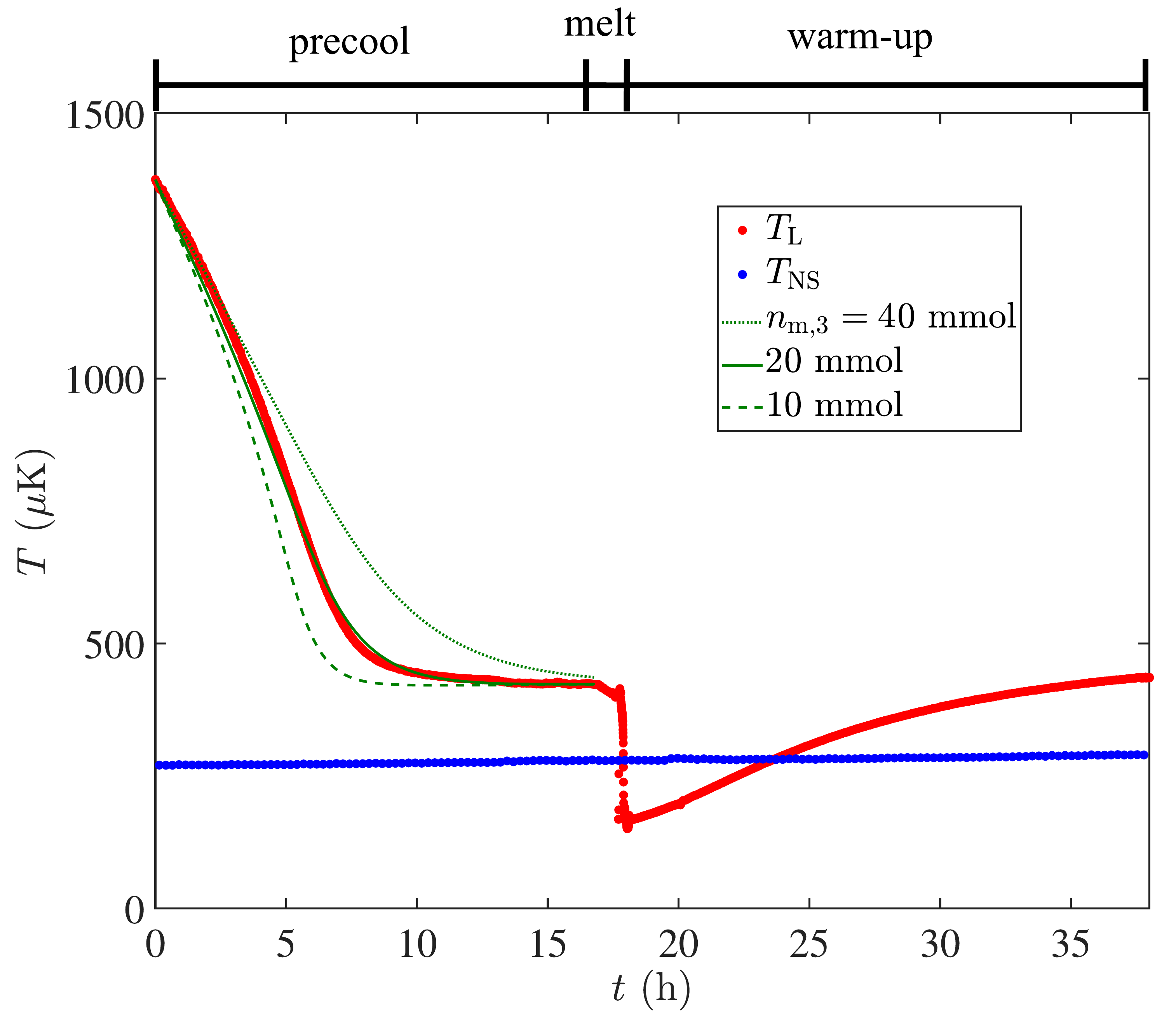}\caption{(color online) Example of data obtained during a successful melting
run, with different stages labeled. The nuclear stage temperature
$T_{\mathrm{NS}}$ (by the PLM) is shown in blue, while the cell main
volume temperature $T_{\mathrm{L}}$ (by the QTF measurement, $f_{32}$) is shown
in red. The green lines are computed $T_{\mathrm{L}}$ with different
amounts of \protect\textsuperscript{3}He in the mixture phase (shown
in the legend), illustrating the sensitivity of the relaxation to
the heat capacity in the system\label{fig:Example-of-data}}

\end{figure}

When the solid phase is present at millikelvin temperatures, the pressure
of the system is fixed at $2.564\,\mathrm{MPa}$, and the molar volumes
of the phases are constant. The size of the solid phase was determined
by tracking the total amount of \textsuperscript{4}He added to (or
removed from) the cell through the superleak line starting from the
nucleation. If \textsuperscript{4}He is transferred at rate $\dot{n}_{4}$,
the solid is changing size at rate $\dot{n}_{s}=10.5\dot{n}_{4}$ \cite{Riekki2019}.
Since the total volume of the cell is known as well, the volume that
is left, after solid \textsuperscript{4}He and pure \textsuperscript{3}He,
can be assumed to be filled by saturated \textsuperscript{3}He\textendash \textsuperscript{4}He
mixture. Below $\sim1.5\,\mathrm{mK}$, when the entropy of pure \textsuperscript{3}He
has become small compared to mixture, we can cross-check the mixture
amount by observing the relaxation of the cell temperature toward
the nuclear stage temperature. In Fig.~\ref{fig:Example-of-data},
we illustrate how even a small assumed change in the mixture amount significantly
alters the computed time constant of the process. When we had sufficiently
undisturbed relaxation period, we could determine the mixture amount
within the accuracy of $5\,\mathrm{mmol}$. If we again take the total
\textsuperscript{3}He amount as given, then the mixture amount fixes
the solid \textsuperscript{4}He amount, enabling us to cross-check
it against the amount determined from the \textsuperscript{4}He flow
measurement. These two were always consistent within about 10\%. 

The relaxation time gives the total heat capacity, and thus entropy
of the system, which is critically important in determining the lowest
temperature obtained in the melting process. This is, of course, true
within the confines of our QTF temperature calibration. If temperatures
change across the board, the heat capacities and the helium amounts
in different phases deduced from them naturally change as well.

\section{Melting the solid\label{subsec:Melting}}

\subsection{Analysis\label{subsec:Analysis}}

To calculate the temperature of the liquid in the main volume of the
cell, in situations where the QTF thermometer had become insensitive,
we need to solve the system of differential equations Eqs.~\eqref{eq:L-heat}-\eqref{eq:V-heat}
for the sinter volume temperature $T_{\mathrm{V}}$, and the main
volume temperature $T_{\mathrm{L}}$ as a function of time. The initial
value for $T_{\mathrm{L}}$ was the reading given by the QTF thermometer,
while the initial $T_{\mathrm{V}}$ value was attained recursively
starting from the mean value between the measured $T_{\mathrm{L}}$
and $T_{\mathrm{NS}}$.

In the following discussion, we focus on 6 precool\textendash melt\textendash warm-up
runs that meet the following criteria: 1) the time between final crystal
growth and start of the melt was sufficiently long so that we could
determine the amount of mixture in the system from the relaxation
time as in Fig.~\ref{fig:Example-of-data}, as well as the pre-melting
background heat leak from $T_{\mathrm{L}}-T_{\mathrm{NS}}$ in the
end, 2) the precooling temperature was low enough to make an attempt
at sub-100 $\mu$K temperatures viable, 3) melting process started
without heat pulse caused by the removal of solid from the cell-side
superleak end, and 4) the follow-up time after the melt was observed
for long enough period to see the saturation temperature of the system
and the post-melting background heat leak (again based on $T_{\mathrm{L}}-T_{\mathrm{NS}}$),
and the nuclear stage temperature was stable during this time. Figure
\ref{fig:Example-of-data} showed an example of a dataset fulfilling
these criteria with different stages of the operation labeled. The
precool stage is only the final precooling period, after the solid
\textsuperscript{4}He was grown to maximal size.

Additionally, as further examples, we show 4 melts done above,
or near, the $T_{c}$, one melting done with the thermal gate operated
as originally intended, one example of a melting performed using the
bellows system, as well as one cyclical melting/growing process showing
an asymmetric behavior between growing and melting the solid.

\subsection{Heat leak during melting\label{subsec:Heat-leak-during}}

\begin{figure}[p]
\vspace*{\smallskipamount}
\hspace{0.2\columnwidth}\subfloat{\noindent \centering{}\includegraphics[bb=0bp 0bp 659bp 55bp,width=0.75\columnwidth]{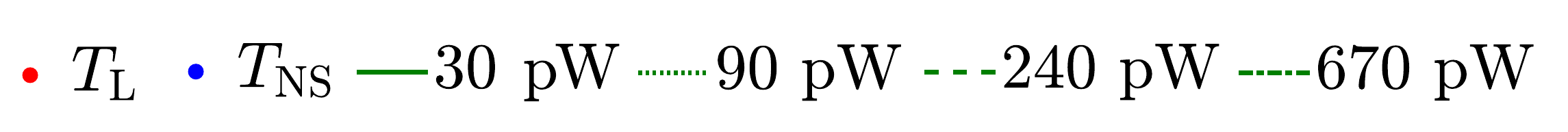}}\vspace*{-0.02\textheight}

\subfloat{\includegraphics[width=0.5\columnwidth]{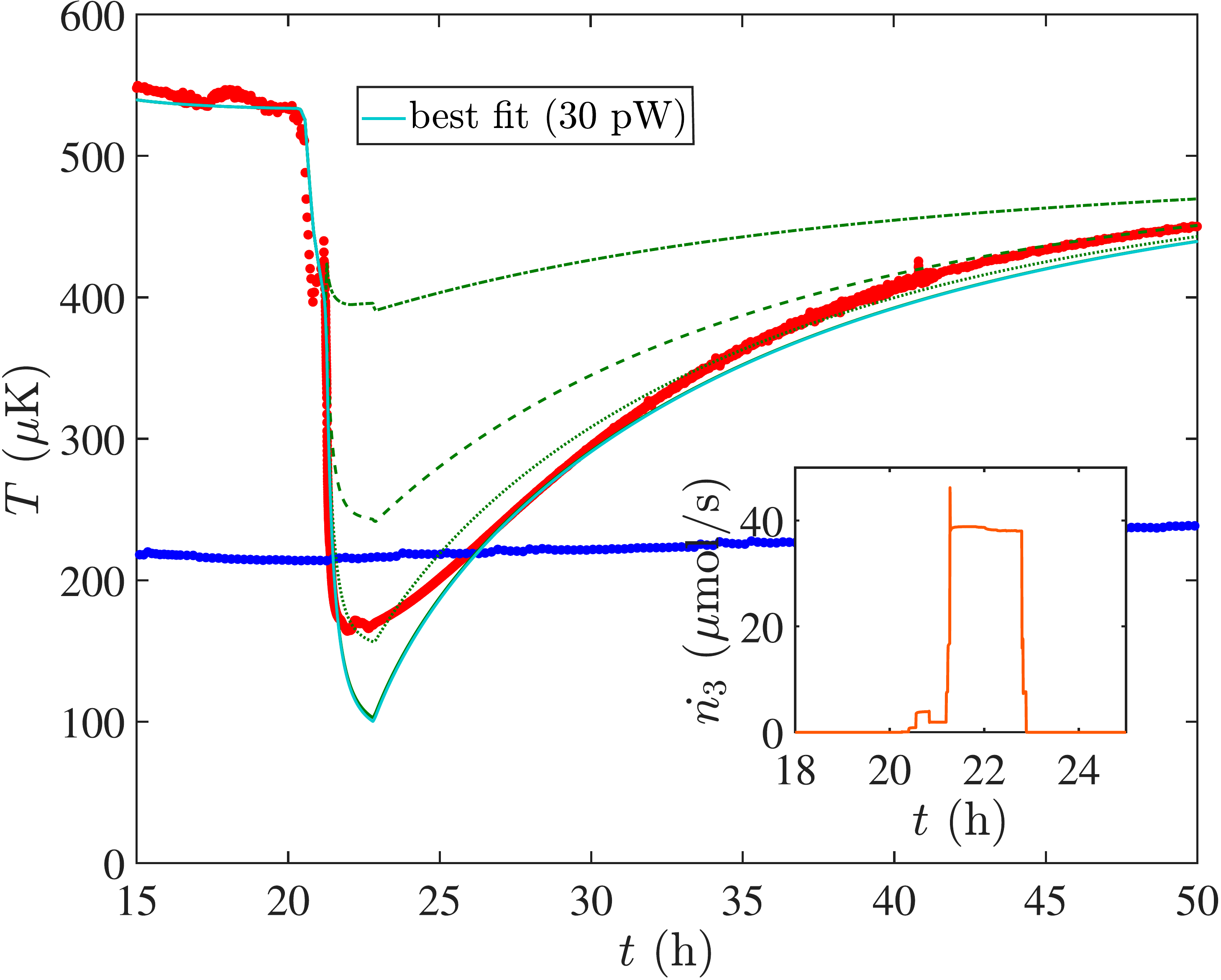}}\quad{}\subfloat{\includegraphics[width=0.5\columnwidth]{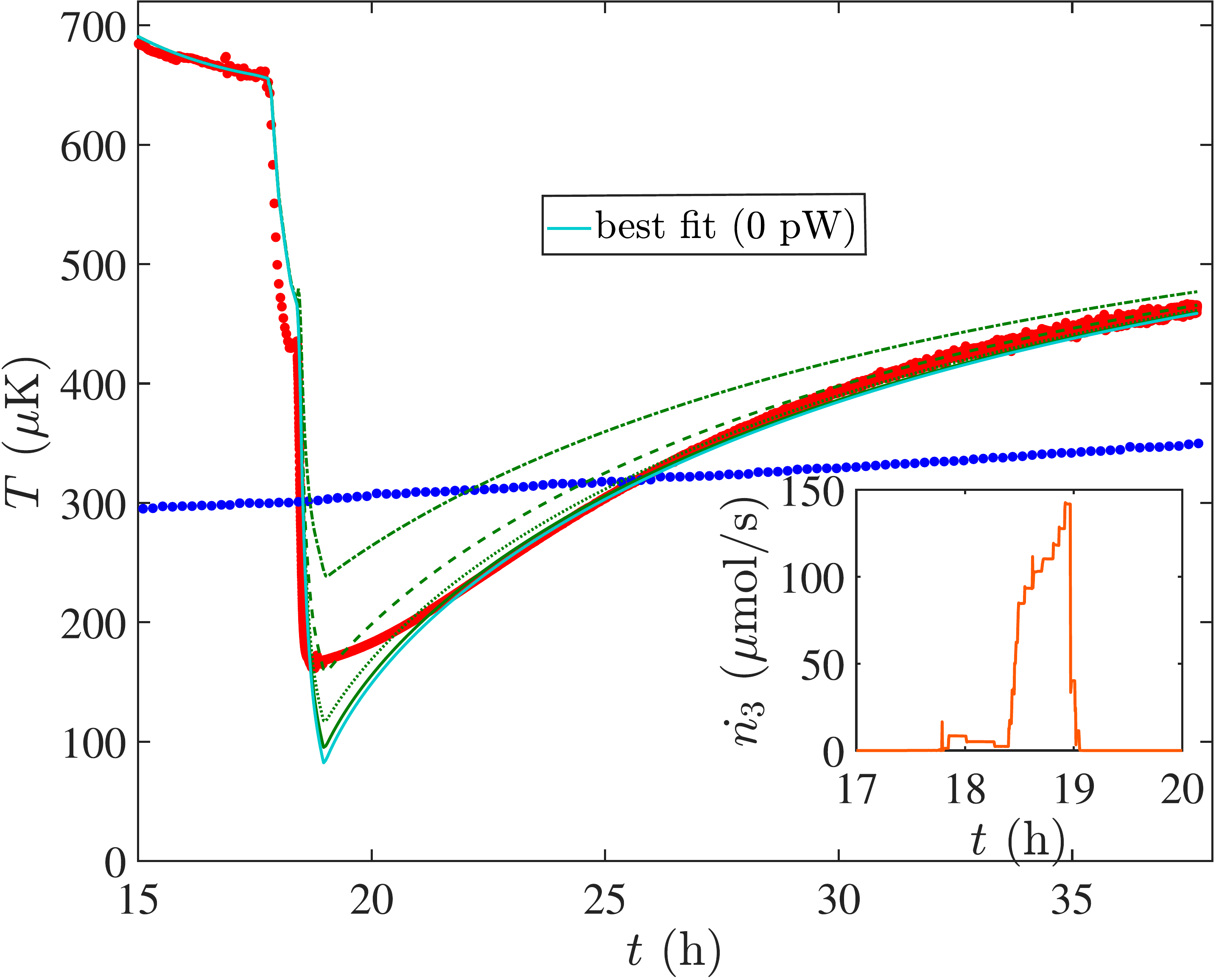}}\vspace*{-0.02\textheight}

\qquad{}%
\begin{minipage}[t]{0.5\columnwidth}%
\noindent \begin{flushleft}
(a) $\left(610\pm20\right)\,\mathrm{mmol}$, $\left(30\pm5\right)\,\mathrm{mmol}$,
$\left(2920/170\pm60\right)\,\mathrm{mmol}$, $50/40\,\mathrm{pW}$
\par\end{flushleft}%
\end{minipage}\enskip{}\quad{}%
\begin{minipage}[t]{0.5\columnwidth}%
\noindent \begin{flushleft}
(b) $\left(610\pm20\right)\,\mathrm{mmol}$, $\left(30\pm5\right)\,\mathrm{mmol}$,
$\left(2940/170\pm60\right)\,\mathrm{mmol}$, $78/43\,\mathrm{pW}$
\par\end{flushleft}%
\end{minipage}

\vspace*{\smallskipamount}
\subfloat{\includegraphics[width=0.5\columnwidth]{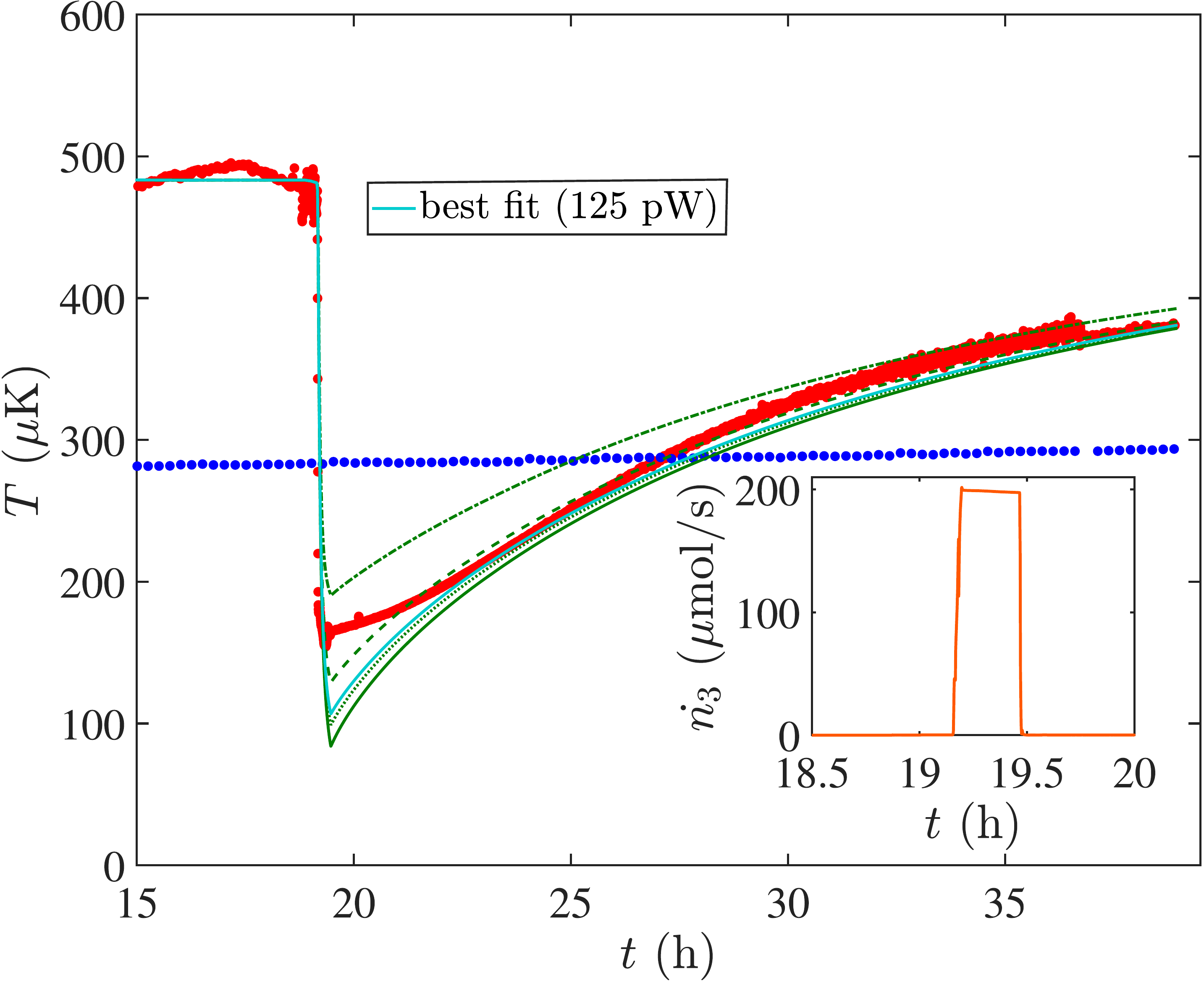}}\quad{}\subfloat{\includegraphics[width=0.5\columnwidth]{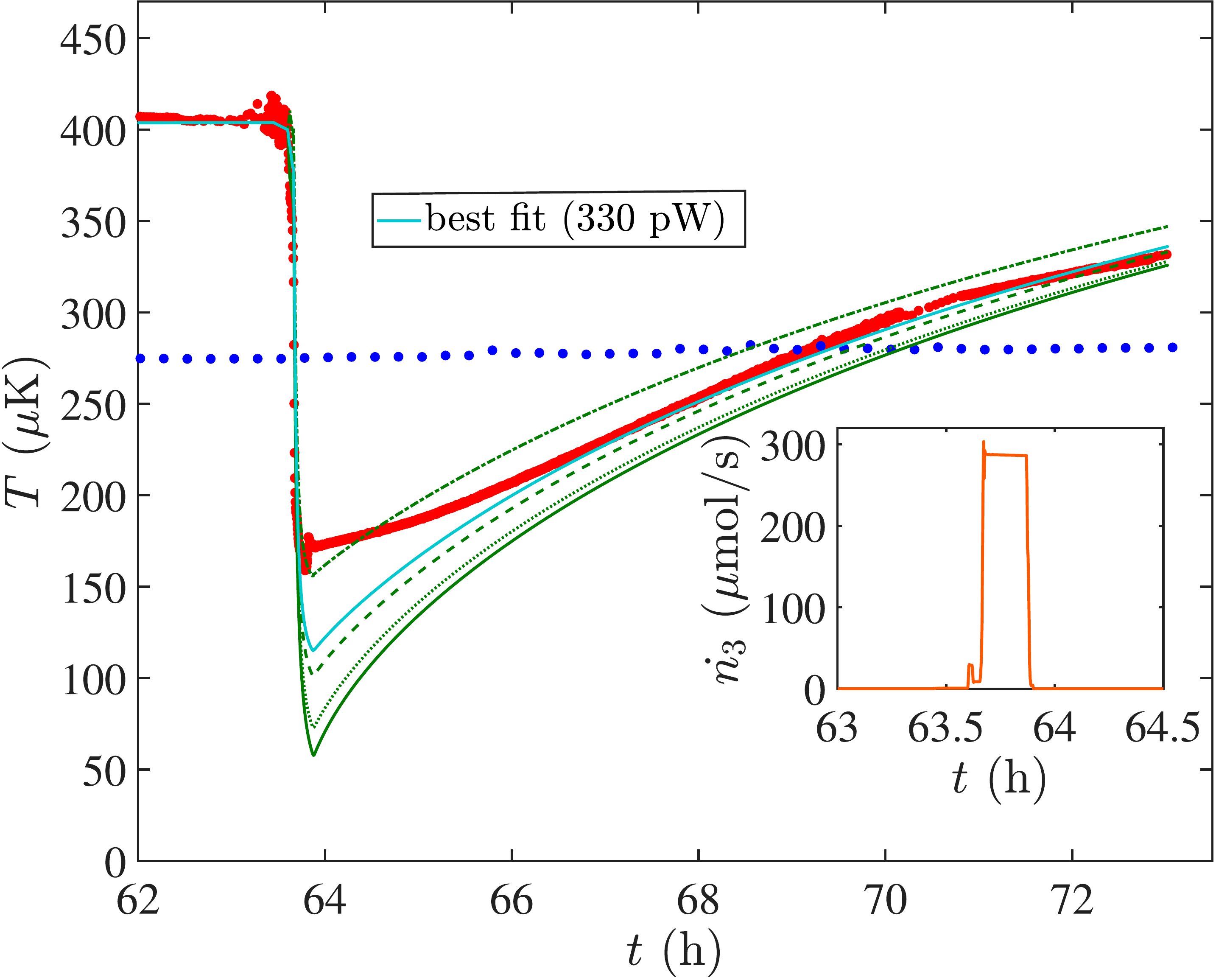}}\vspace*{-0.02\textheight}

\qquad{}%
\begin{minipage}[t]{0.5\columnwidth}%
\noindent \begin{flushleft}
(c) $\left(570\pm20\right)\,\mathrm{mmol}$, $\left(39\pm5\right)\,\mathrm{mmol}$,
$\left(2890/210\pm40\right)\,\mathrm{mmol}$, $33/26\,\mathrm{pW}$
\par\end{flushleft}%
\end{minipage}\enskip{}\quad{}%
\begin{minipage}[t]{0.5\columnwidth}%
\noindent \begin{flushleft}
(d) $\left(570\pm20\right)\,\mathrm{mmol}$, $\left(30\pm5\right)\,\mathrm{mmol}$,
$\left(3030/270\pm60\right)\,\mathrm{mmol}$, $17/37\,\mathrm{pW}$
\par\end{flushleft}%
\end{minipage}\vspace*{\smallskipamount}
\subfloat{\includegraphics[width=0.5\columnwidth]{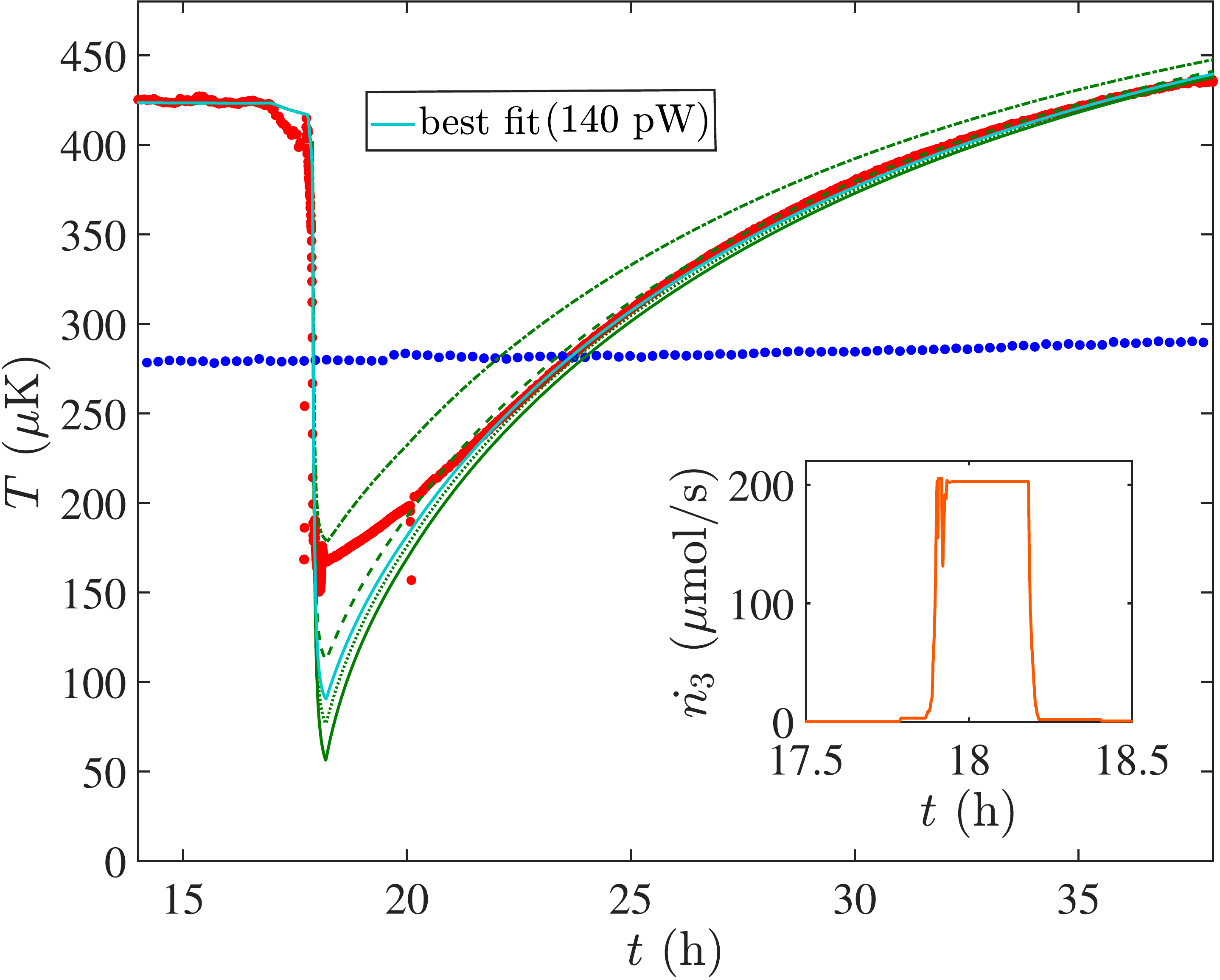}}\quad{}\subfloat{\includegraphics[width=0.5\columnwidth]{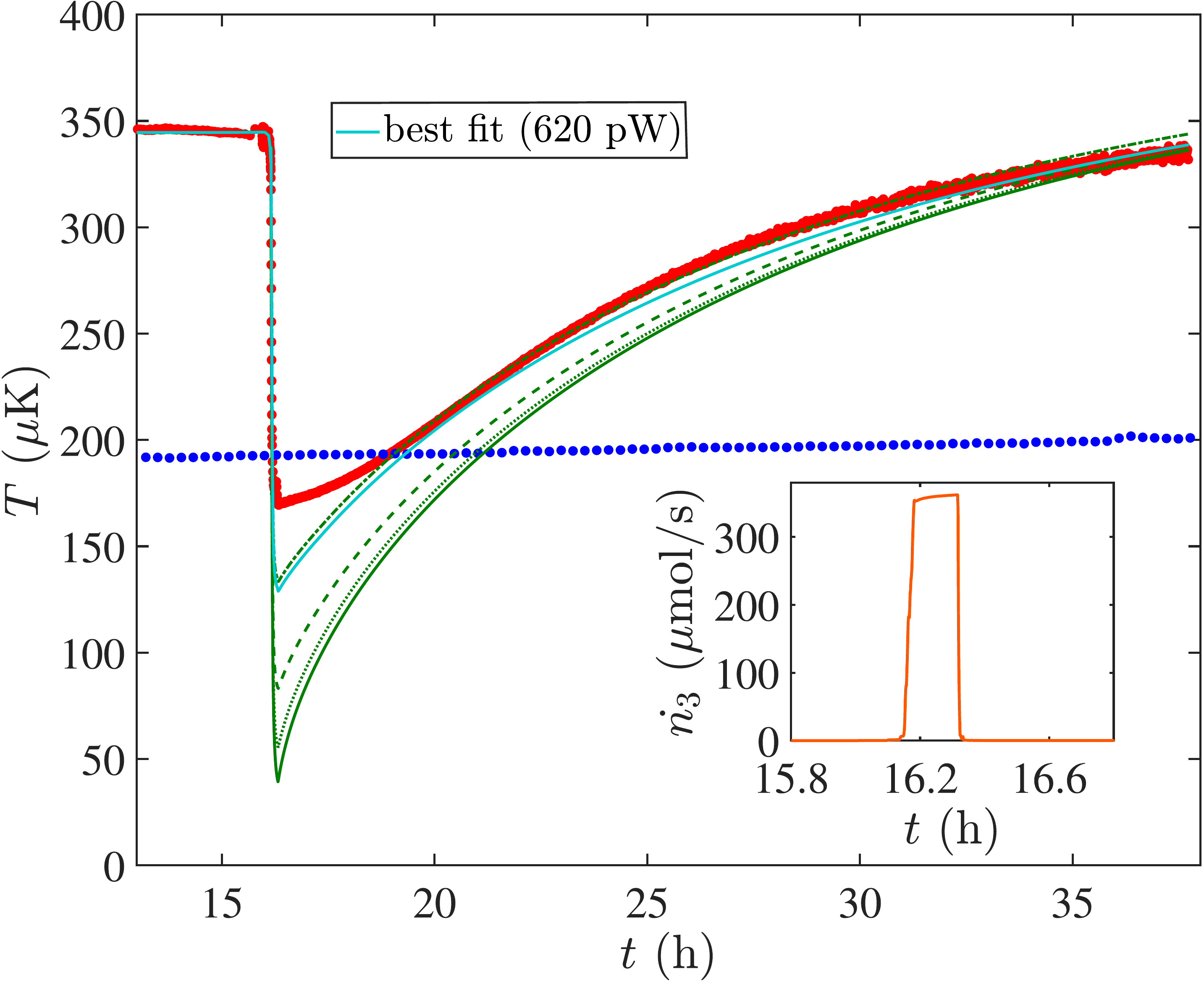}}\vspace*{-0.02\textheight}

\qquad{}%
\begin{minipage}[t]{0.5\columnwidth}%
\noindent \begin{flushleft}
(e) $\left(570\pm20\right)\,\mathrm{mmol}$, $\left(20\pm4\right)\,\mathrm{mmol}$,
$\left(3150/430\pm50\right)\,\mathrm{mmol}$, $18/39\,\mathrm{pW}$
\par\end{flushleft}%
\end{minipage}\enskip{}\quad{}%
\begin{minipage}[t]{0.5\columnwidth}%
\noindent \begin{flushleft}
(f) $\left(720\pm20\right)\,\mathrm{mmol}$, $\left(17\pm4\right)\,\mathrm{mmol}$,
$\left(3020/560\pm40\right)\,\mathrm{mmol}$, $14/20\,\mathrm{pW}$
\par\end{flushleft}%
\end{minipage}

\caption{(color online) Measured nuclear stage temperature $T_{\mathrm{NS}}$
and measured cell main volume temperature $T_{\mathrm{L}}$ as a function
of time, with computed $T_{\mathrm{L}}$ at various heat leaks during
melting shown in green. For each subfigure, the cyan line indicates
the best fit to the post-melting warm-up period, while the inset shows
the \protect\textsuperscript{3}He phase-transfer rate during the
melt. Below each subfigure: total \protect\textsuperscript{3}He in
the system, amount of \protect\textsuperscript{3}He in the mixture
phase before melting, solid \protect\textsuperscript{4}He before/after
melting, and background heat leak before/after melting. $t=0$ is
the time when the final solid growth was finished (full relaxation
shown in Fig.~\ref{fig:melt_mintemps})\label{fig:melt_heatleaks}}
\end{figure}
Heat leak, along with the melting rate and the amount of mixture prior
to melting, are the most important quantities in determining the final
temperature. The Kapitza resistance of the plain cell wall is so high
at these temperatures that heat flow through it is practically zero.
The heat flow through the sinter of the heat-exchanger volume is still
notable, but the thermal conductivity of the connecting channel becomes
so small that the main volume of the cell becomes effectively decoupled
from the heat-exchanger volume as the melting is carried out.

The background heat leak $\dot{Q}_{\mathrm{ext}}$ will have two different
values: the value before melting determined from the difference between
$T_{\mathrm{L}}$ and $T_{\mathrm{NS}}$ at the end of the precool,
and the value after melting deduced from it
long after the melting is done. The ratio of these two background
heat leaks varied from 0.5 to 1.8.
\begin{figure}[b]
\center\includegraphics[width=0.75\columnwidth]{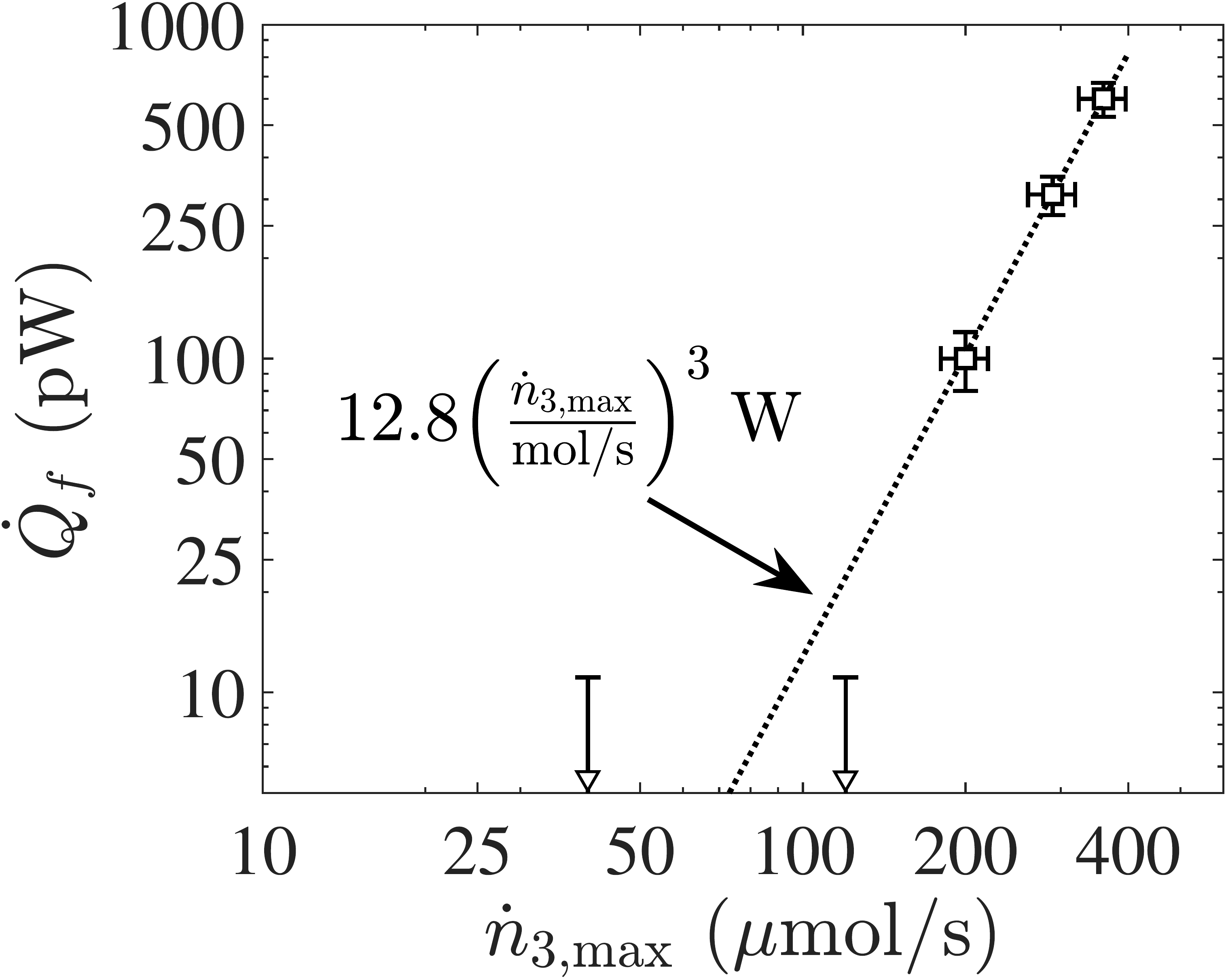}\caption{Heat leak $\dot{Q}_{f}$ as a function of the maximum \protect\textsuperscript{3}He
phase-transfer rate $\dot{n}_{3}$ determined from Fig.~\ref{fig:melt_heatleaks}\label{fig:Viscous-heat-leak}}

\end{figure}

To deduce the relation between the melting rate $\dot{n}_{3}$ and
the heat leak $\dot{Q}_{f}$ in Eq.~\eqref{eq:L-heat}, the assumed
value for $\dot{Q}_{f}$ was allowed to vary between 0 and 700 pW
to find the best correspondence with the experimental $T_{\mathrm{L}}$
data. The resulting collection of computed $T_{\mathrm{L}}$ data
are shown in Fig.~\ref{fig:melt_heatleaks}, with the six melts
fulfilling the criteria discussed in Section \ref{subsec:Analysis}.
The full solution to Eqs.~\eqref{eq:L-heat}-\eqref{eq:V-heat} also
includes the heat-exchanger volume temperature $T_{\mathrm{V}}$,
but we omitted it from the figures for clarity. We sought $\dot{Q}_{f}$
value that would make the post-melting warm-up rate match with the
QTF measured behavior. As said, at lowest temperatures, the QTF response
was saturated and only around $300\,\mathrm{\mu K}$ the reading became
reasonably reliable once again. The criterion for the ``best fit''
was that the computed $T_{\mathrm{L}}$ would not cross the measured
value during the warm-up, but would approach it asymptotically at
the earliest possible time.

When $\dot{n}_{3}$ is below about $150\,\mathrm{\mu mol/s}$ no additional
heat leak during the melt is resolved; in fact the heat leak can even
appear to be less than the after-melting value. Then above $200\,\mathrm{\mu mol/s}$
the heat leak is rapidly increased ending up to more than $600\,\mathrm{pW}$
with the highest phase-transfer rate of $360\,\mathrm{\mu mol/s}$
in Fig.~\ref{fig:melt_heatleaks}e. The highest phase-transfer rate
the superleak could sustain by \textsuperscript{4}He extraction was
about $500\,\mathrm{\mu mol/s}$, but the test run using that was
not performed under appropriate conditions to be included in this
analysis.

By subtracting the post-melting heat leak value from the ``best fit''
value during the melt, we get the heat leak identified as $\dot{Q}_{f}$.
Figure \ref{fig:Viscous-heat-leak} shows $\dot{Q}_{f}$ as a function
of the \textsuperscript{3}He phase-transfer rate $\dot{n}_{3}$,
where the data falls on a third power dependence. The indicated uncertainty
was $10\%$ in $\dot{n}_{3}$, and $10\,\mathrm{pW}+10\%$ in $\dot{Q}_{f}$.
Only the parts of the error bars of the low melting rate data are
visible on the logarithmic scale, while the measurements from Fig.~\ref{fig:melt_heatleaks}c
and \ref{fig:melt_heatleaks}e fall right on top of each other. The
reason for the data to follow $\dot{n}_{3}^{3}$ dependence is not
understood. If the origin of this heat leak were viscous losses, it
should follow $\dot{n}_{3}^{2}$ behavior instead. 

\begin{figure}[t]
\includegraphics[width=1\columnwidth]{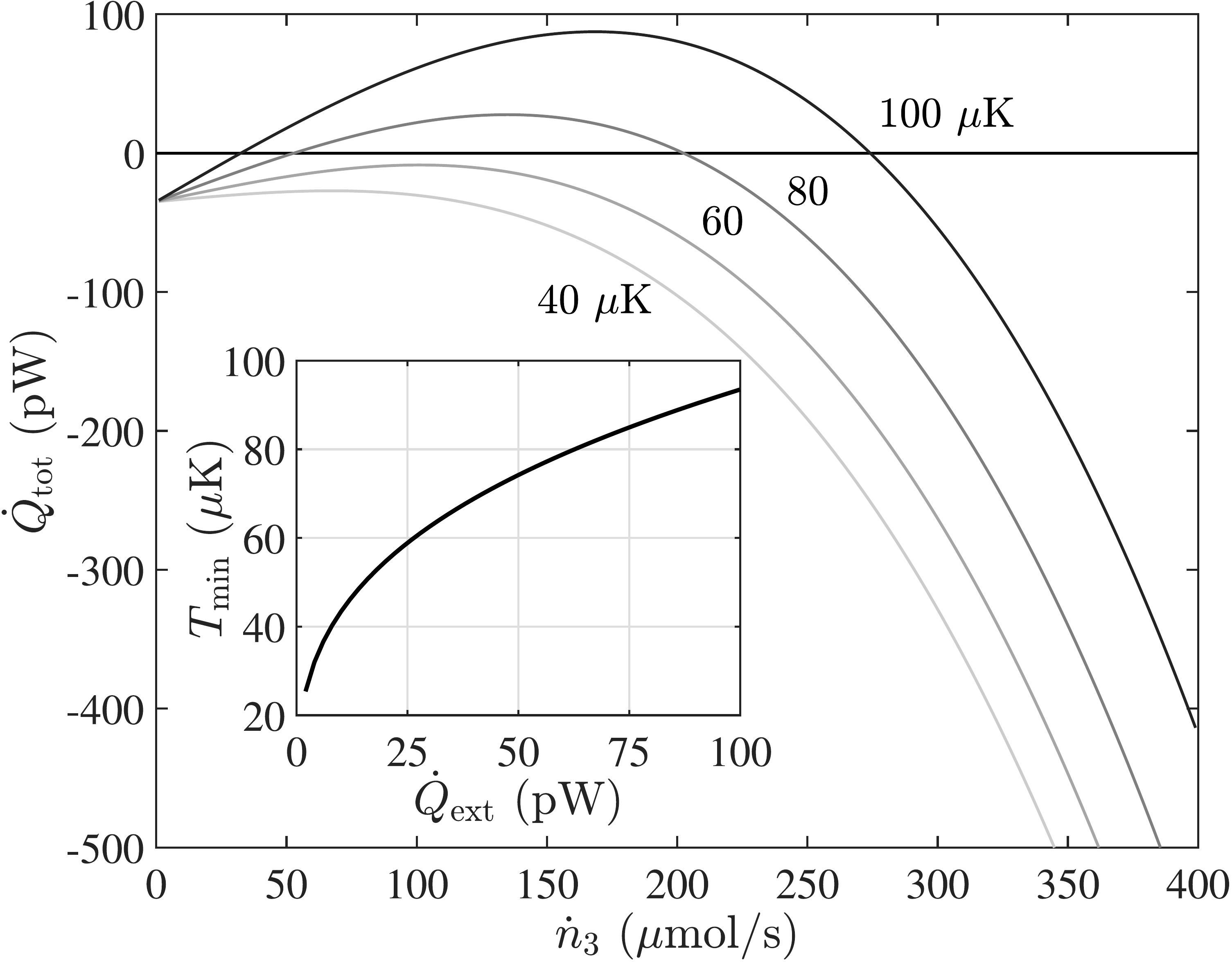}

\caption{Heat balance in the system $\dot{Q}_{\mathrm{tot}}=\dot{Q}_{\mathrm{melt}}-\dot{Q}_{f}-\dot{Q}_{\mathrm{ext}}$
at several temperatures as the function of the \protect\textsuperscript{3}He
phase-transfer rate. Inset shows the minimum achievable temperature
as a function of the background heat leak at the optimal melting rate
given by Eq.~\eqref{eq:n3dot-optimal}\label{fig:calc-mintemps}}
\end{figure}

Now that we have evaluated the heat leaks, we can calculate the total
heat load to the system at ultra-low temperatures $\dot{Q}_{\mathrm{tot}}=\dot{Q}_{\mathrm{melt}}-\dot{Q}_{f}-\dot{Q}_{\mathrm{ext}}$
(from the right side of Eq.~\eqref{eq:L-heat}). We ignore the heat
flow through the plain cell wall $\dot{Q}_{\mathrm{direct}}$ due
to its massive Kapitza resistance, and $\dot{Q}_{\mathrm{tube}}$
as the superfluid \textsuperscript{3}He thermal conductivity in the
connecting channel is already rather small. The graph in Fig.~\ref{fig:calc-mintemps}
is drawn with $\dot{Q}_{\mathrm{ext}}=35\,\mathrm{pW}$, the average
of the post-melting heat leak values from Fig.~\ref{fig:melt_heatleaks},
while
\begin{equation}
\dot{Q}_{f}=12.8\left(\frac{\dot{n}_{3}}{\mathrm{mol/s}}\right)^{3}\,\mathrm{W}\label{eq:Qeta}
\end{equation}
is based on Fig.~\ref{fig:Viscous-heat-leak}, and we have $\dot{Q}_{\mathrm{melt}}=109\,\mathrm{\frac{J}{mol\,K^{2}}}\dot{n}_{3}T^{2}$
 \cite{Riekki2019}. Net positive values correspond to cooling in the
system, and give the lowest achievable temperature around $65\,\mathrm{\mu K}$
with $110\,\mathrm{\mu mol/s}$ \textsuperscript{3}He phase-transfer
rate. 
Below that rate, the cooling from the melting/mixing process is not
enough to overcome the background heat leak, while above it the losses
due to the flow rate become inefficiently large. At the optimal \textsuperscript{3}He
phase-transfer rate $\dot{n}_{\mathrm{3,opt}}$

\begin{equation}
\dot{n}_{\mathrm{3,opt}}=\left(\frac{0.048\dot{Q}_{\mathrm{ext}}}{1.22}\right)^{1/3}\label{eq:n3dot-optimal}
\end{equation}
the minimum temperature $T_{\mathrm{min}}$ is given by

\begin{equation}
T_{\mathrm{min}}=\sqrt{\frac{12.8\dot{n}_{3,\mathrm{opt}}^{3}+\dot{Q}_{\mathrm{ext}}}{109\dot{n}_{3,\mathrm{opt}}}},\label{eq:Tmin}
\end{equation}
and is shown in the inset of Fig.~\ref{fig:calc-mintemps} as a function
of the background heat leak $\dot{Q}_{\mathrm{ext}}$. As a side note,
the \textsuperscript{3}He phase-transfer rate $\dot{n}_{3}$ and
\textsuperscript{4}He extraction rate through the superleak $\dot{n}_{4}$
are related by $\dot{n}_{3}=0.84\dot{n}_{4}$ \cite{Riekki2019}.

Unfortunately, the optimal set of conditions with regards to the total
\textsuperscript{3}He amount, low precooling temperature and residual
heat leak, as well as optimal melting procedure never met in the actual
experiment. Instead, in attempts to compensate for the background
heat leak, most of our melts gravitated towards using as high rates
as reasonably possible, which now, in retrospect, after full analysis
of the system, was not the optimal approach. As Fig.~\ref{fig:calc-mintemps}
clearly demonstrates, $\dot{n}_{3}$ of range $100-150\,\mathrm{\mu mol/s}$
would have been the most beneficial.
\begin{figure}[p]
\vspace*{\smallskipamount}
\hspace{0.25\columnwidth}\subfloat{\includegraphics[bb=0bp 0bp 659bp 55bp,width=0.6\columnwidth]{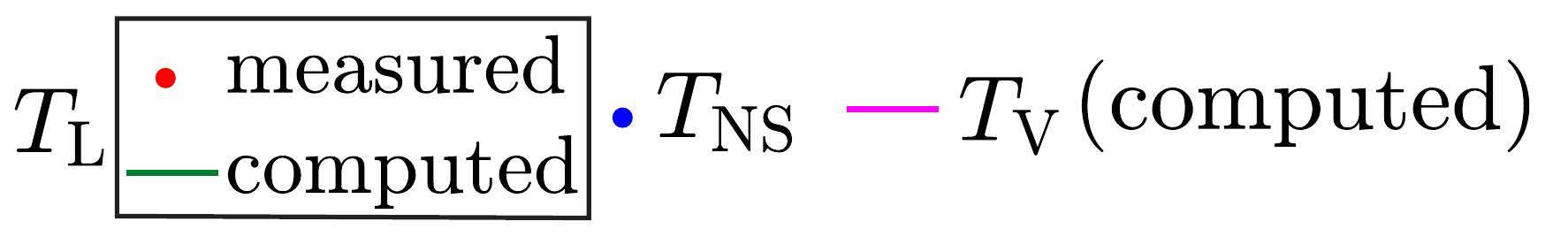}}\vspace*{-0.02\textheight}

\subfloat{\includegraphics[width=0.5\columnwidth]{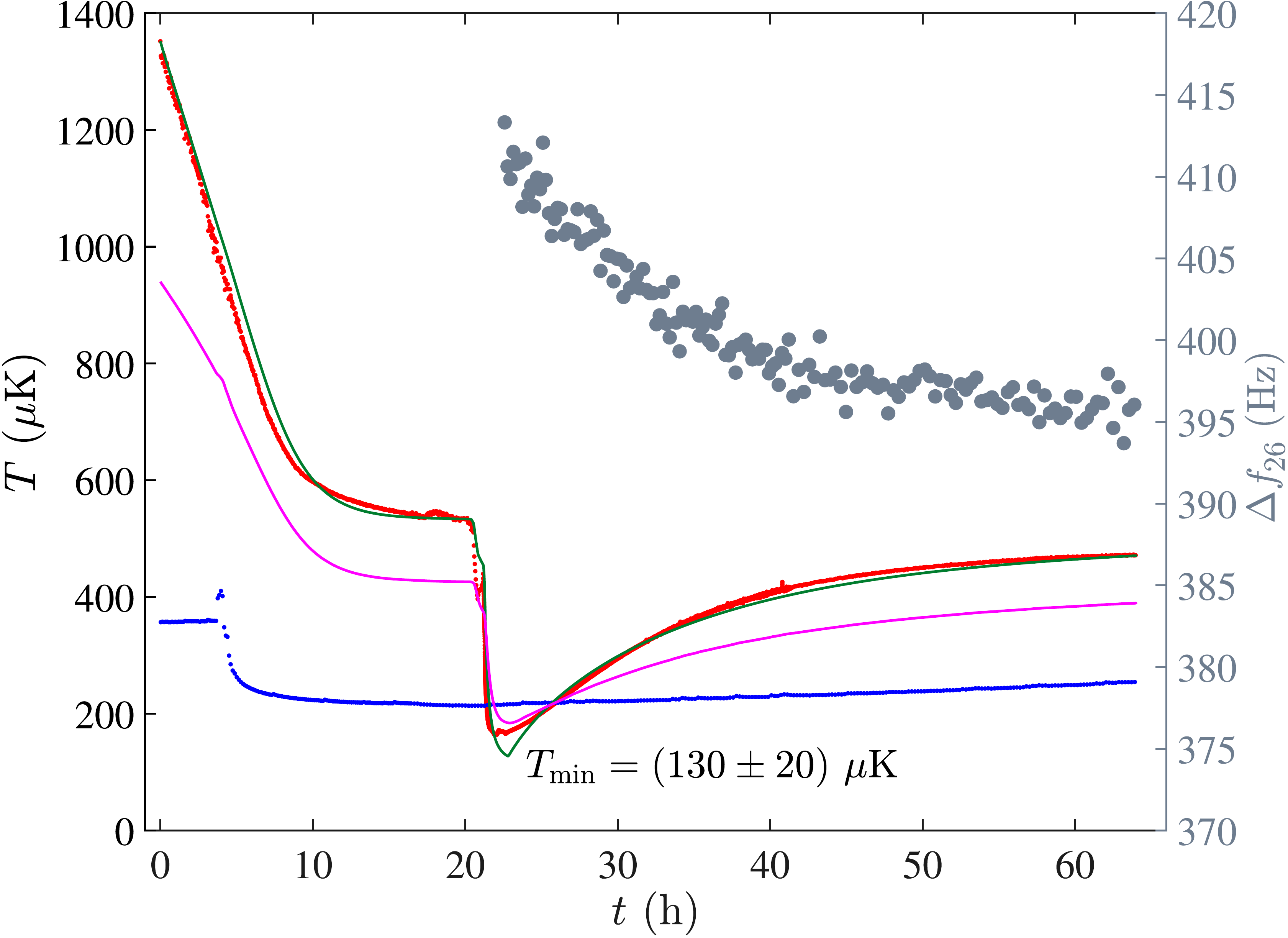}}\quad{}\subfloat{\includegraphics[width=0.5\columnwidth]{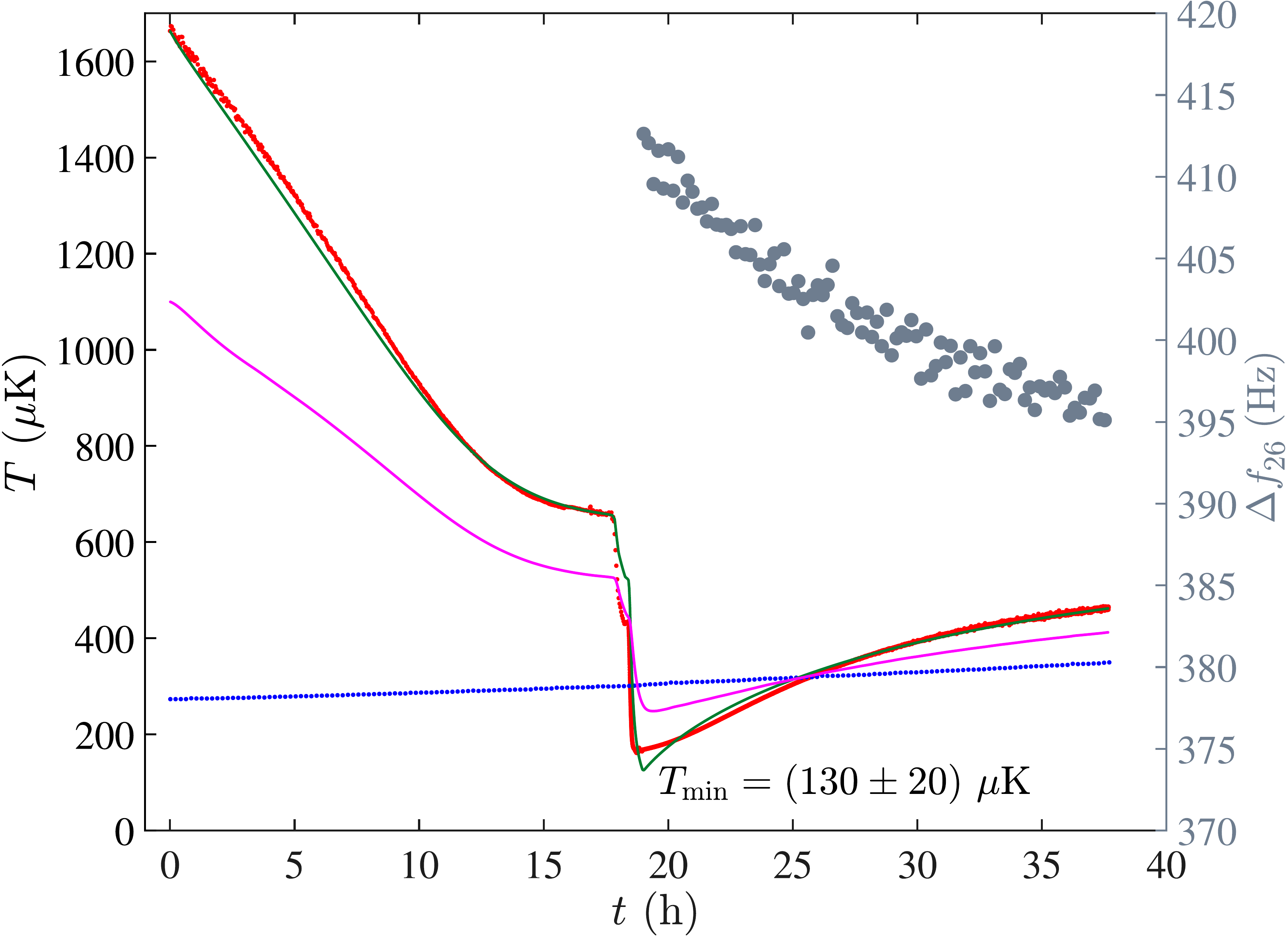}}\vspace*{-0.02\textheight}

\enskip{}%
\begin{minipage}[t]{0.5\columnwidth}%
\noindent \begin{flushleft}
(a) $\left(610\pm20\right)\,\mathrm{mmol}$, $\left(30\pm5\right)\,\mathrm{mmol}$,
$\left(2920/170\pm60\right)\,\mathrm{mmol}$, $50/40\,\mathrm{pW}$,
$40\,\mathrm{\mu mol/s}$
\par\end{flushleft}%
\end{minipage}\enskip{}%
\begin{minipage}[t]{0.5\columnwidth}%
\noindent \begin{flushleft}
(b) $\left(610\pm20\right)\,\mathrm{mmol}$, $\left(30\pm5\right)\,\mathrm{mmol}$,
$\left(2940/170\pm60\right)\,\mathrm{mmol}$, $78/43\,\mathrm{pW}$,
$120\,\mathrm{\mu mol/s}$
\par\end{flushleft}%
\end{minipage}

\vspace*{\smallskipamount}
\subfloat{\includegraphics[width=0.5\columnwidth]{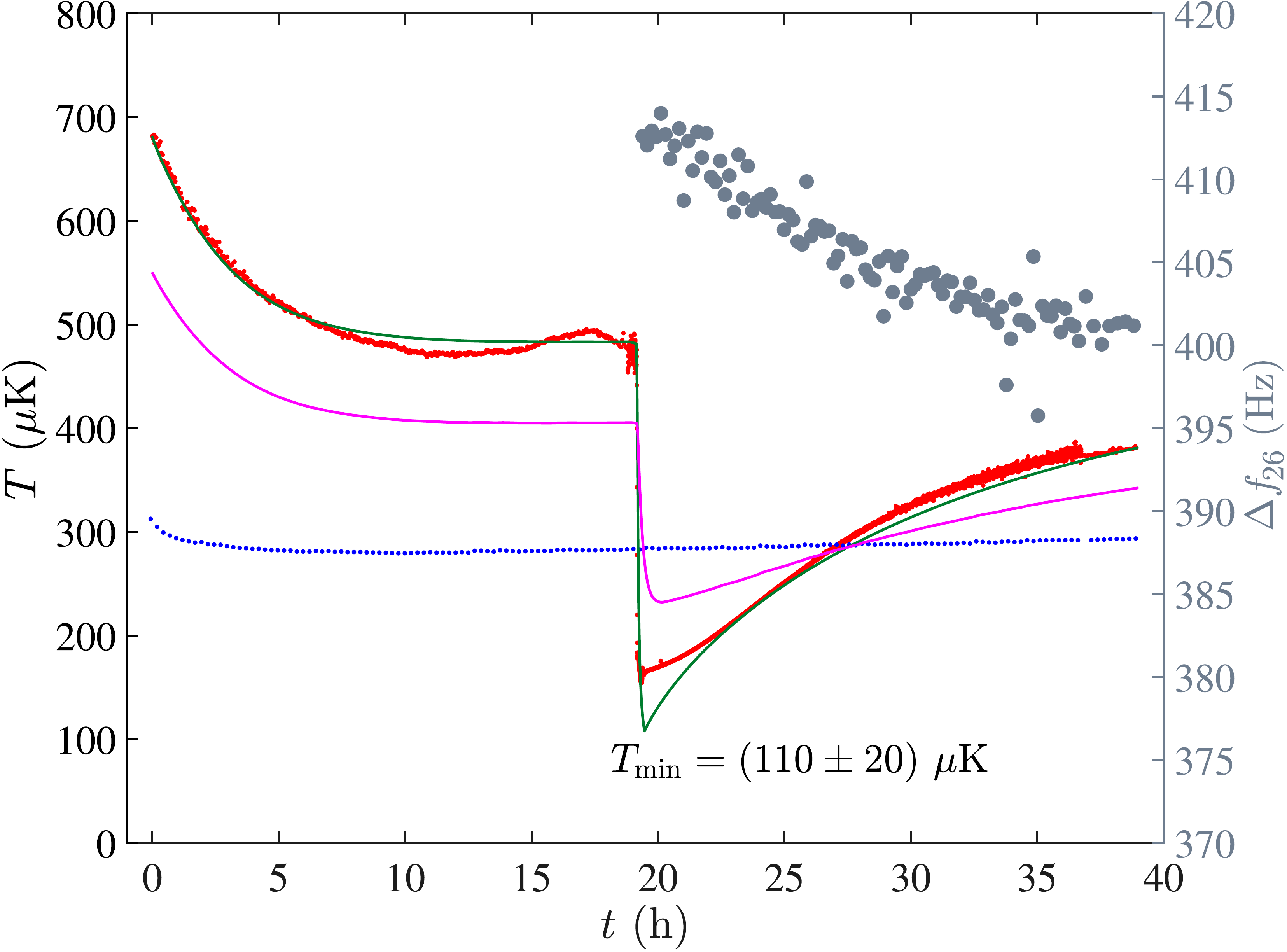}}\quad{}\subfloat{\includegraphics[width=0.5\columnwidth]{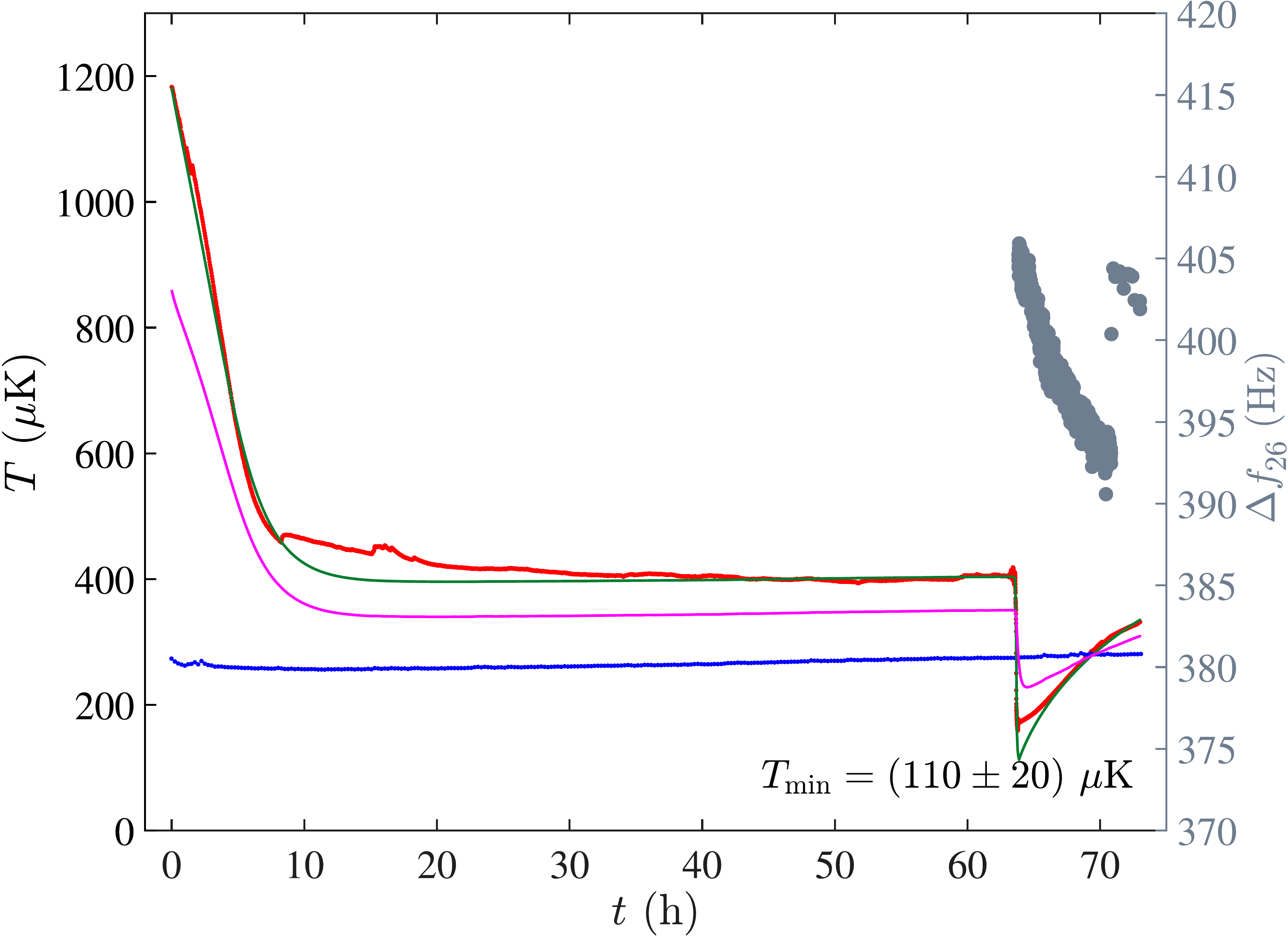}}\vspace*{-0.02\textheight}

\enskip{}%
\begin{minipage}[t]{0.5\columnwidth}%
\noindent \begin{flushleft}
(c) $\left(570\pm20\right)\,\mathrm{mmol}$, $\left(39\pm5\right)\,\mathrm{mmol}$,
$\left(2890/210\pm40\right)\,\mathrm{mmol}$, $33/26\,\mathrm{pW}$,
$200\,\mathrm{\mu mol/s}$
\par\end{flushleft}%
\end{minipage}\enskip{}%
\begin{minipage}[t]{0.5\columnwidth}%
\noindent \begin{flushleft}
(d) $\left(570\pm20\right)\,\mathrm{mmol}$, $\left(30\pm5\right)\,\mathrm{mmol}$,
$\left(3030/270\pm60\right)\,\mathrm{mmol}$, $17/37\,\mathrm{pW}$,
$290\,\mathrm{\mu mol/s}$ 
\par\end{flushleft}%
\end{minipage}\vspace*{\smallskipamount}
\subfloat{\includegraphics[width=0.5\columnwidth]{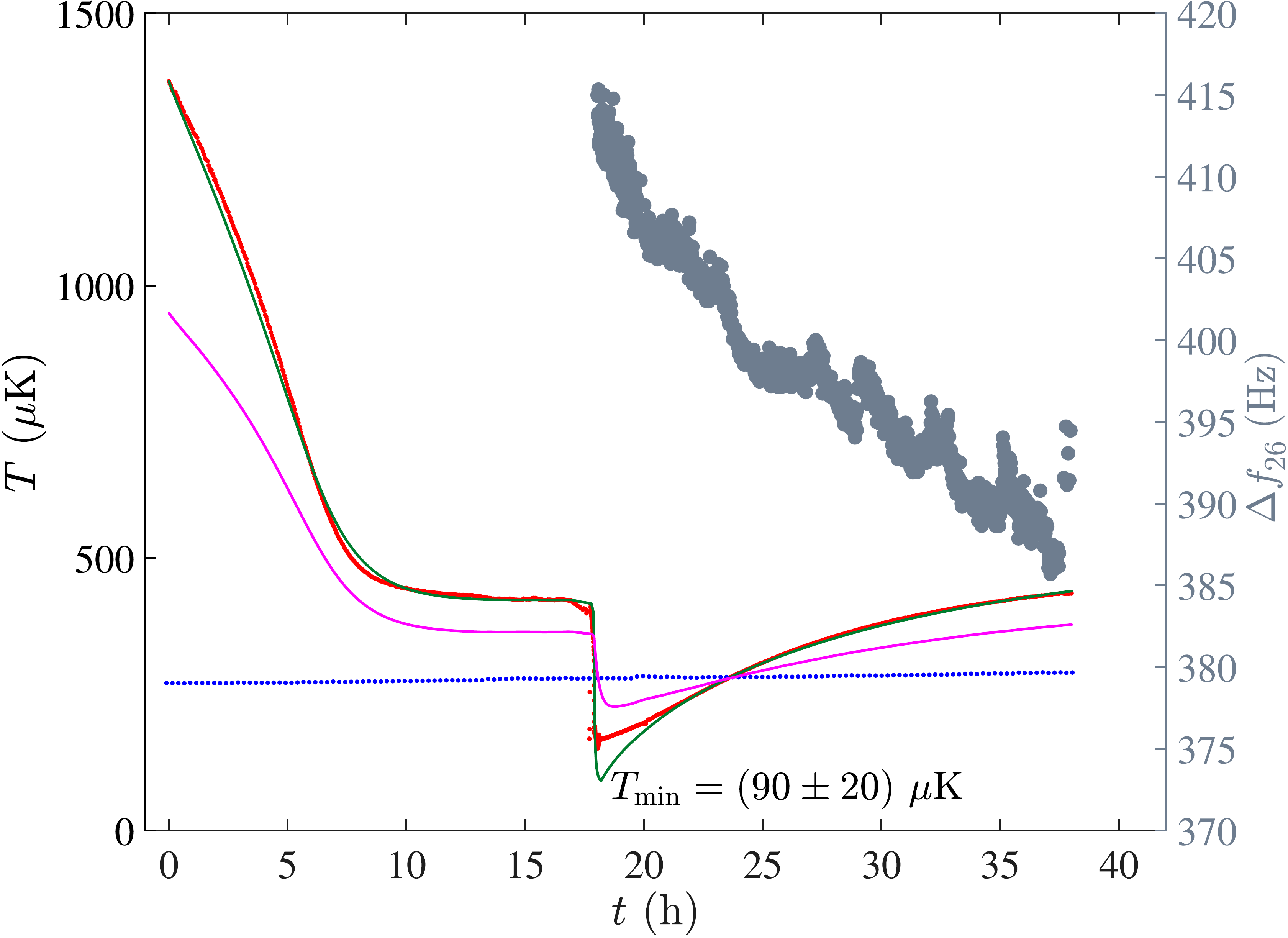}}\quad{}\subfloat{\includegraphics[width=0.5\columnwidth]{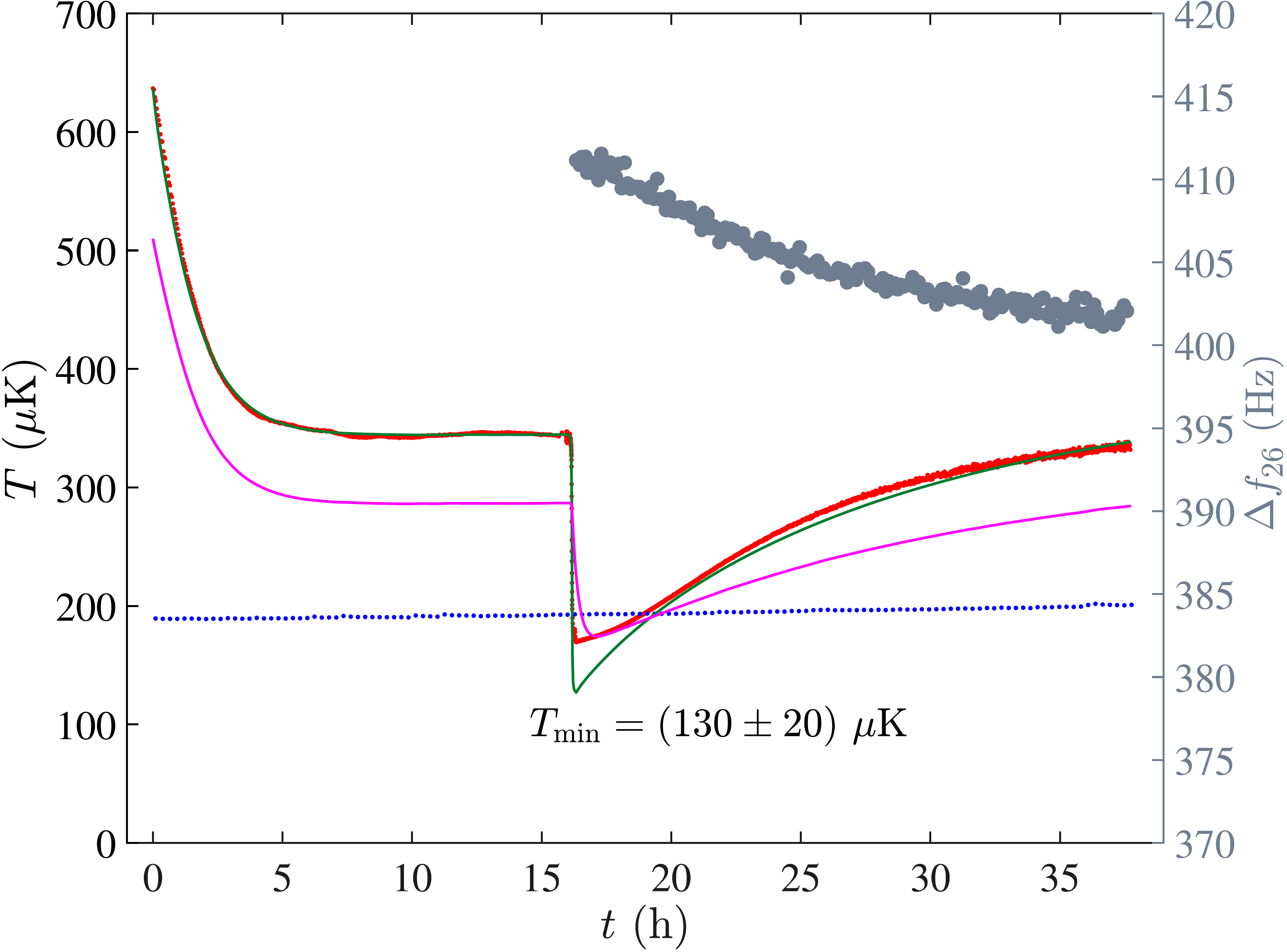}}\vspace*{-0.02\textheight}

\enskip{}%
\begin{minipage}[t]{0.5\columnwidth}%
\noindent \begin{flushleft}
(e) $\left(570\pm20\right)\,\mathrm{mmol}$, $\left(20\pm4\right)\,\mathrm{mmol}$,
$\left(3150/430\pm50\right)\,\mathrm{mmol}$, $18/39\,\mathrm{pW}$,
$200\,\mathrm{\mu mol/s}$
\par\end{flushleft}%
\end{minipage}\enskip{}%
\begin{minipage}[t]{0.5\columnwidth}%
\noindent \begin{flushleft}
(f) $\left(720\pm20\right)\,\mathrm{mmol}$, $\left(17\pm4\right)\,\mathrm{mmol}$,
$\left(3020/560\pm40\right)\,\mathrm{mmol}$, $14/20\,\mathrm{pW}$,
$360\,\mathrm{\mu mol/s}$
\par\end{flushleft}%
\end{minipage}

\caption{(color online) Left y-axis: Measured nuclear stage temperature $T_{\mathrm{NS}}$
and measured cell main volume temperature $T_{\mathrm{L}}$ along
with computed $T_{\mathrm{L}}$ and computed heat-exchanger volume
temperature $T_{\mathrm{V}}$. The lowest computed temperature is
also written out. Right y-axis: resonance width of the mixture QTF
as it emerges from the solid \protect\textsuperscript{4}He. At $t=0$
the solid growth was stopped. Below each subfigure: total \protect\textsuperscript{3}He
in the system, amount of \protect\textsuperscript{3}He in the mixture
phase before melting, solid \protect\textsuperscript{4}He before/after
melting, background heat leak before/after melting, and the mean \protect\textsuperscript{3}He
phase-transfer rate $\dot{n}_{3}$ (cf. Fig.~\ref{fig:melt_heatleaks})
\label{fig:melt_mintemps}}
\end{figure}

\subsection{Lowest temperatures\label{subsec:Lowest-temperatures}}

Having now all needed parameters, we can proceed to calculate the
lowest temperatures obtained in the actual melts. These are shown
in Fig.~\ref{fig:melt_mintemps}, where the subfigures correspond
to the melts in Fig.~\ref{fig:melt_heatleaks}. This time we present
the data starting from the time when the final solid \textsuperscript{4}He
growth was completed. Furthermore, the figure shows the calculated
heat-exchanger volume temperature $T_{\mathrm{V}}$, and behavior
of the mixture phase QTF as it emerges from solid at about the midpoint
of the melt.

Comparing the lowest temperature achieved in Fig.~\ref{fig:melt_mintemps}a,
with low melting rates to Fig.~\ref{fig:melt_mintemps}d and \ref{fig:melt_mintemps}f,
at high rates, we see that the increased phase-transfer rate did not
result in a lower temperature, even with notably better precooling
conditions due to the increased heat leak $\dot{Q}_{f}$. The lowest
temperature we obtain is $\left(90\pm20\right)\,\mathrm{\mu K\approx}\frac{T_{c}}{\left(29\pm5\right)}$
in Fig.~\ref{fig:melt_mintemps}e with the phase-transfer rate of
about $200\,\mathrm{\mu mol/s}$. The confidence bounds in the temperature
include the uncertainties in the helium amounts in different phases,
in the temperature calibration of QTF, heat leaks, melting rate, and
in the thermal transfer parameters of the system. Out of these, the
initial amount of mixture and the heat leak were the most significant.

The response of the mixture fork as it emerges from the solid during
the melt, and thus when the temperatures are at their lowest, is shown
on the right y-axes of Fig.~\ref{fig:melt_mintemps}. The points
displayed are five-point averages from the measured values. In Figs.~\ref{fig:melt_mintemps}d-e,
the mixture QTF was measured mostly in the tracking mode \cite{Pentti_Rysti_Salmela},
resulting in more datapoints. But as the jump in \ref{fig:melt_mintemps}d
illustrates, as we switch from tracking to full-spectrum sweeps, the
tracking parameters determined several days earlier, when the QTF
was still out from solid, are no longer dependable. This is caused
predominantly by drifting background signal beneath the resonance
response. During the other runs, we used full sweeps to provide more
reliable data, and between Figs.~\ref{fig:melt_mintemps}a-c and
Fig.~\ref{fig:melt_mintemps}f, we decreased the number of sampling
frequencies per sweep to improve data gathering rate.
\begin{figure}[t]
\includegraphics[width=1\columnwidth]{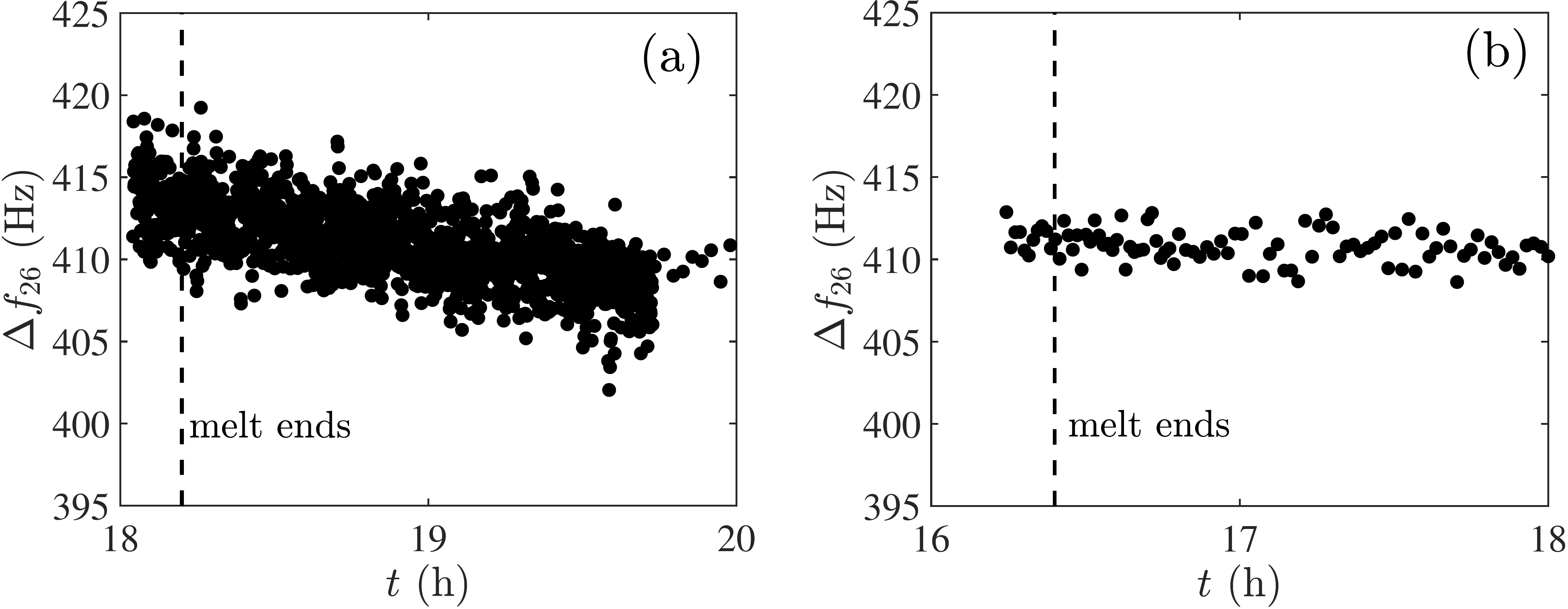}

\caption{Close-up view of the mixture QTF width at the lowest temperatures.
(a) corresponds to the measurement of Fig.~\ref{fig:melt_mintemps}e,
and (b) of Fig.~\ref{fig:melt_mintemps}f\label{fig:f26-zoom}}

\end{figure}

Figure \ref{fig:f26-zoom} takes a close-up look of the mixture QTF
data from Figs.~\ref{fig:melt_mintemps}e-f, except that the data
presented now is not averaged. In Fig.~\ref{fig:f26-zoom}a the data
was obtained by the tracking mode, while in Fig.~\ref{fig:f26-zoom}b
it was received from narrow-span frequency sweeps. The QTF response
shows no indication of the superfluid phase in mixture. When the QTF
is released from solid and becomes measurable, the width stays approximately
constant during the melt, starting to change only after it is over.
Initially, the QTF is in its saturation regime (cf. Fig.~\ref{fig:Resonace-width-freq})
and does not respond to changes in temperature. But after the melt,
when temperature has increased enough, some sensitivity is recovered. The slope in Fig.~\ref{fig:f26-zoom}a may have been affected by the mentioned drift in the signal background during tracking.
The unexciting response is in agreement with the determined lowest
temperatures, as the mixture superfluid transition is expected to
occur only at around $40\,\mathrm{\mu K}$ \cite{Effective_3He_interactions}.
To reach that, the background heat leak should be below 8 pW, as given
by Eqs.~\eqref{eq:n3dot-optimal} and \eqref{eq:Tmin}, and the optimal
phase-transfer rate would be about $70\,\mathrm{\mu mol/s}$.
\begin{figure}[b]
\center\includegraphics[width=0.75\columnwidth]{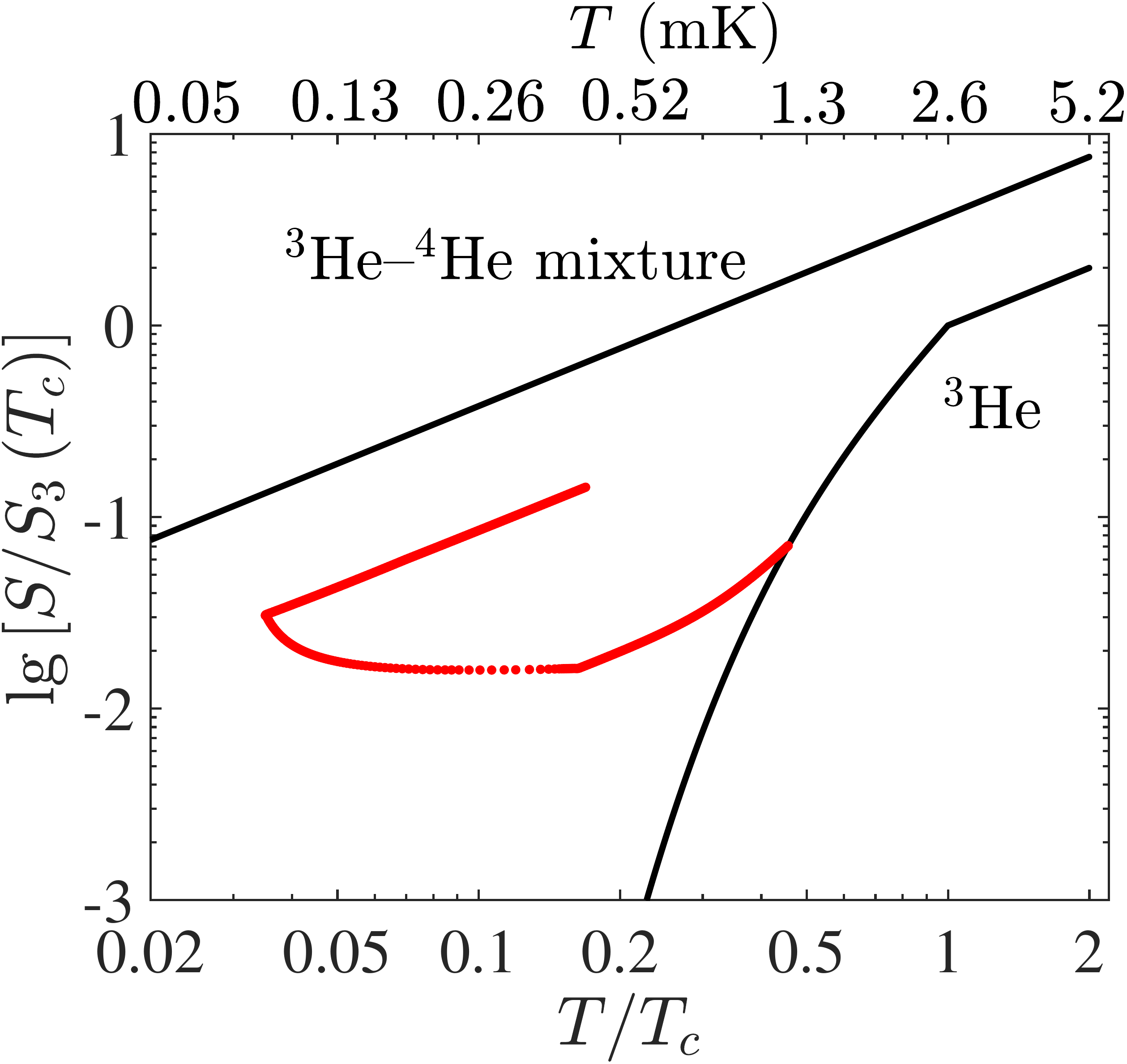}\caption{(color online) Entropy during the precool\textendash melt\textendash warm-up
cycle (cf. Fig.~\ref{fig:melt_mintemps}e and \ref{fig:Example-of-data})
with the entropies of pure \protect\textsuperscript{3}He and saturated
\protect\textsuperscript{3}He\textendash \protect\textsuperscript{4}He
mixture \cite{Riekki2019}, for reference. Entropy values are scaled
by its value at the pure \protect\textsuperscript{3}He $T_{c}$\label{fig:Entropy}}
\end{figure}

In Fig.~\ref{fig:Entropy}, we show the entropy of the system during
an experimental run. In this example, we have used the computed temperature
from Fig.~\ref{fig:melt_mintemps}e. During precool, the entropy
deviates from the pure \textsuperscript{3}He entropy, as there is
finite amount of mixture present, and its $\propto T$ proportional
entropy is the main contributor to the total entropy of the system
below $0.5T_{c}\approx1.3\,\mathrm{mK}$ \cite{Riekki2019}. As the
melt is started, the entropy initially follows the adiabatic behavior
going horizontally away from the pure \textsuperscript{3}He curve
towards the mixture curve. But, below $0.05T_{c}$, the heat leaks
force the system to stay at an elevated temperature. At the end of
the melt, there is still pure \textsuperscript{3}He phase present,
which means that the actual entropy curve stops short of the mixture-only
curve, but goes parallel with it during the warm-up period.
\begin{figure}[t]
\center\includegraphics[width=0.75\columnwidth]{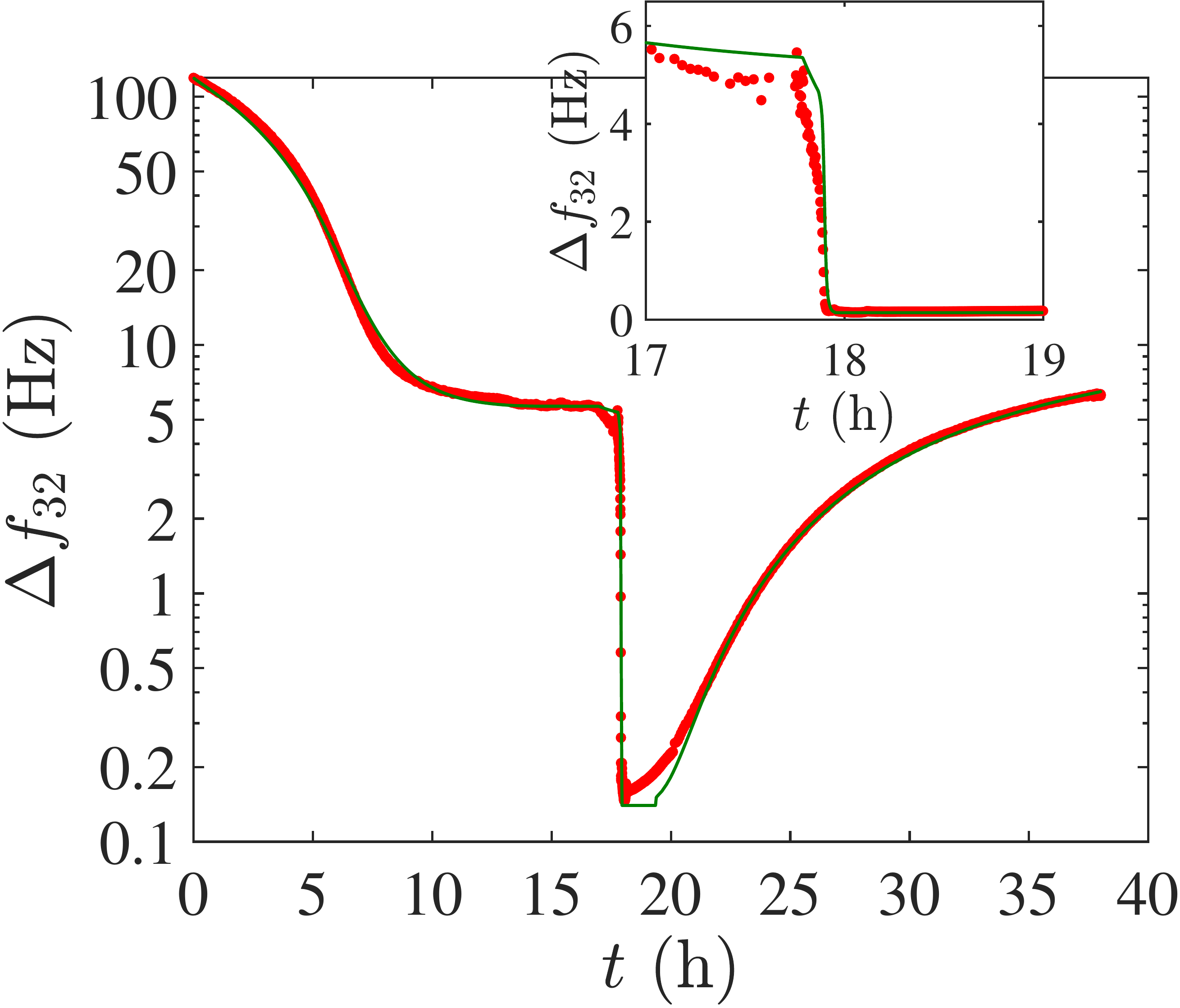}

\caption{(color online) Example of resonance width of the pure \protect\textsuperscript{3}He
QTF during a melting procedure, compared against width calculated
backwards from the computed temperature. The data here is the same
as in Figs.~\ref{fig:Example-of-data}, and \ref{fig:melt_mintemps}e\label{fig:Resonance-width} }
\end{figure}

Finally, to provide further cross-check between the computed and measured
temperature, Fig.~\ref{fig:Resonance-width} compares the measured
resonance width of the pure \textsuperscript{3}He QTF to the value
converted from the simulated cell main volume temperature (cf. Fig.~\ref{fig:melt_mintemps}e)
using the calibration from Section \ref{subsec:Quartz-tuning-fork}.
Note, that the residual width in the calibration was assumed to be
140 mHz which results the flat region in the computed values at the
lowest temperatures. The maximum difference between the measured and
calculated width is of order 1 Hz (20\%), which occurred at the start
of the melt. This can be due to the \textsuperscript{4}He film covering
the QTF shifting, since we disturb the equilibrium by initiating \textsuperscript{4}He
flow through the superleak. During the other melts discussed in Figs.~\ref{fig:melt_heatleaks}
and \ref{fig:melt_mintemps} the maximum difference is of the same
order.

\subsection{Melts at higher temperatures\label{subsec:Melts-at-higher-Ts}}

\begin{figure}
\vspace*{\smallskipamount}
\hspace{0.25\columnwidth}\subfloat{\includegraphics[bb=0bp 0bp 659bp 55bp,width=0.6\columnwidth]{Fig16_legend}}\vspace*{-0.02\textheight}

\subfloat{\includegraphics[width=0.5\columnwidth]{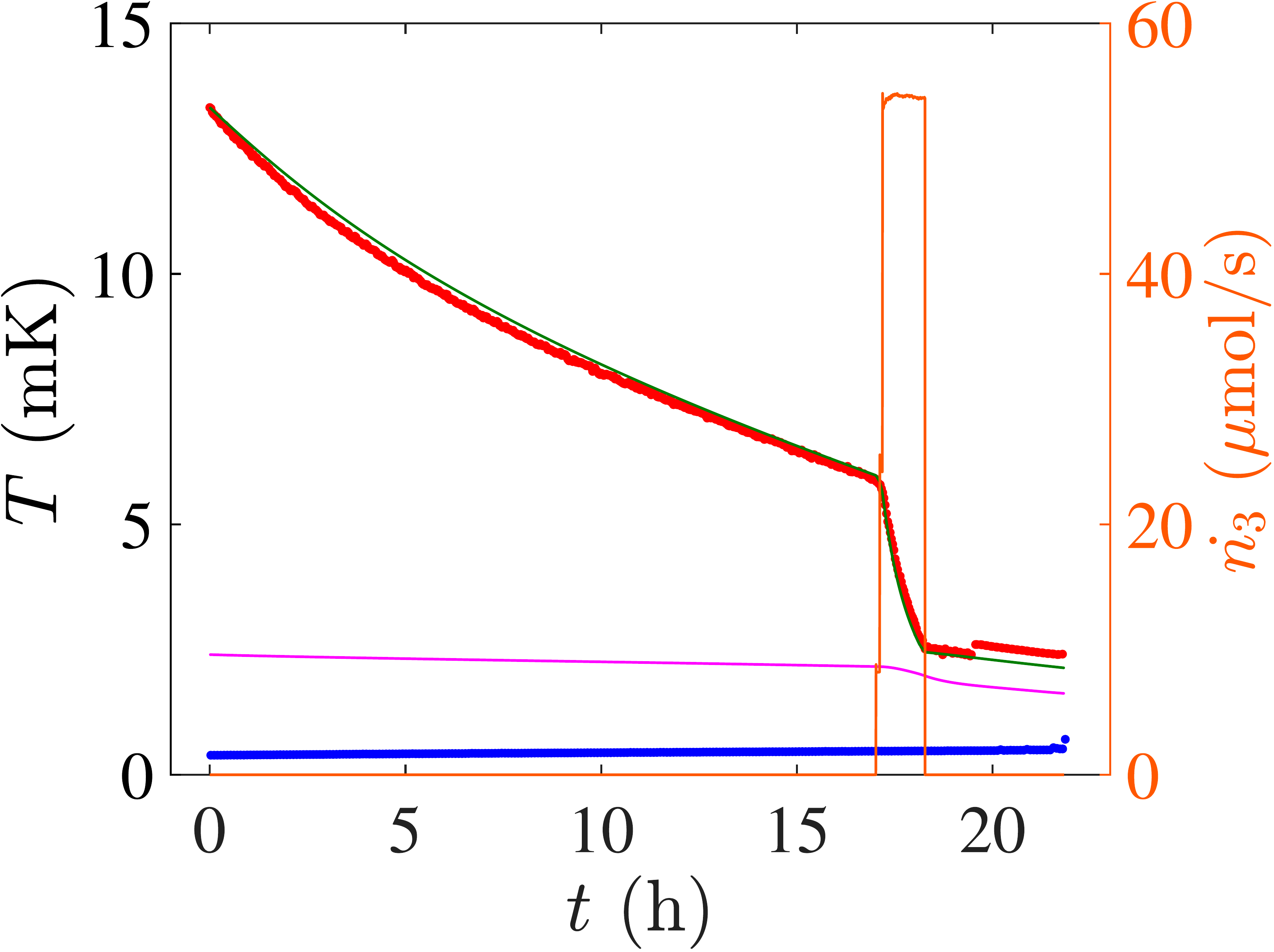}}\quad{}\subfloat{,\includegraphics[width=0.5\columnwidth]{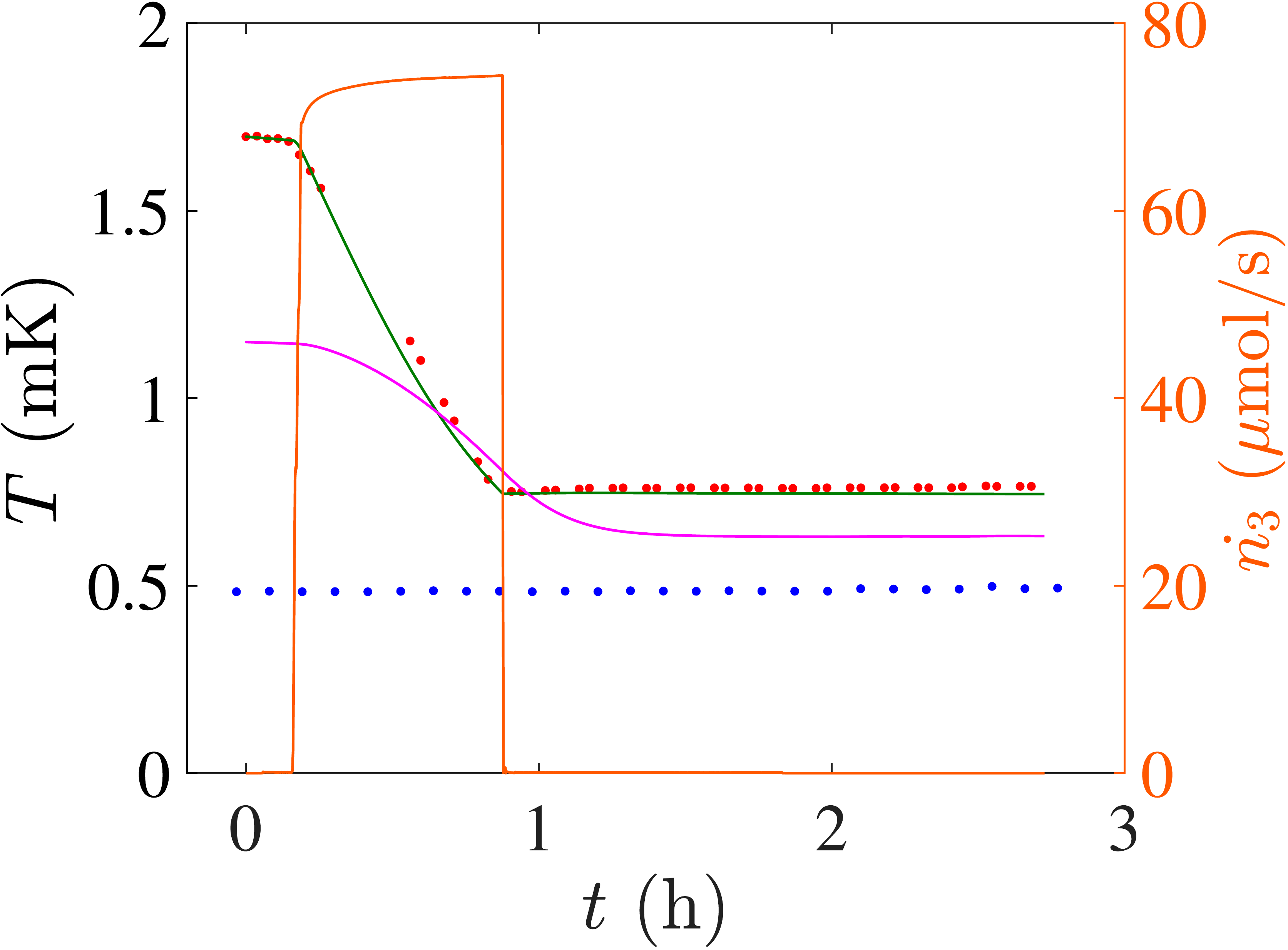}}\vspace*{-0.02\textheight}

\enskip{}%
\begin{minipage}[t]{0.5\columnwidth}%
\noindent \begin{flushleft}
(a) $\left(610\pm20\right)\,\mathrm{mmol}$, $\left(20\pm10\right)\,\mathrm{mmol}$,
$\left(3100/380\pm60\right)\,\mathrm{mmol}$
\par\end{flushleft}%
\end{minipage}\enskip{}%
\begin{minipage}[t]{0.5\columnwidth}%
\noindent \begin{flushleft}
(b) $\left(720\pm20\right)\,\mathrm{mmol}$, $\left(65\pm10\right)\,\mathrm{mmol}$,
$\left(2340/10\pm40\right)\,\mathrm{mmol}$
\par\end{flushleft}%
\end{minipage}

\vspace*{\smallskipamount}
\subfloat{\includegraphics[width=0.5\columnwidth]{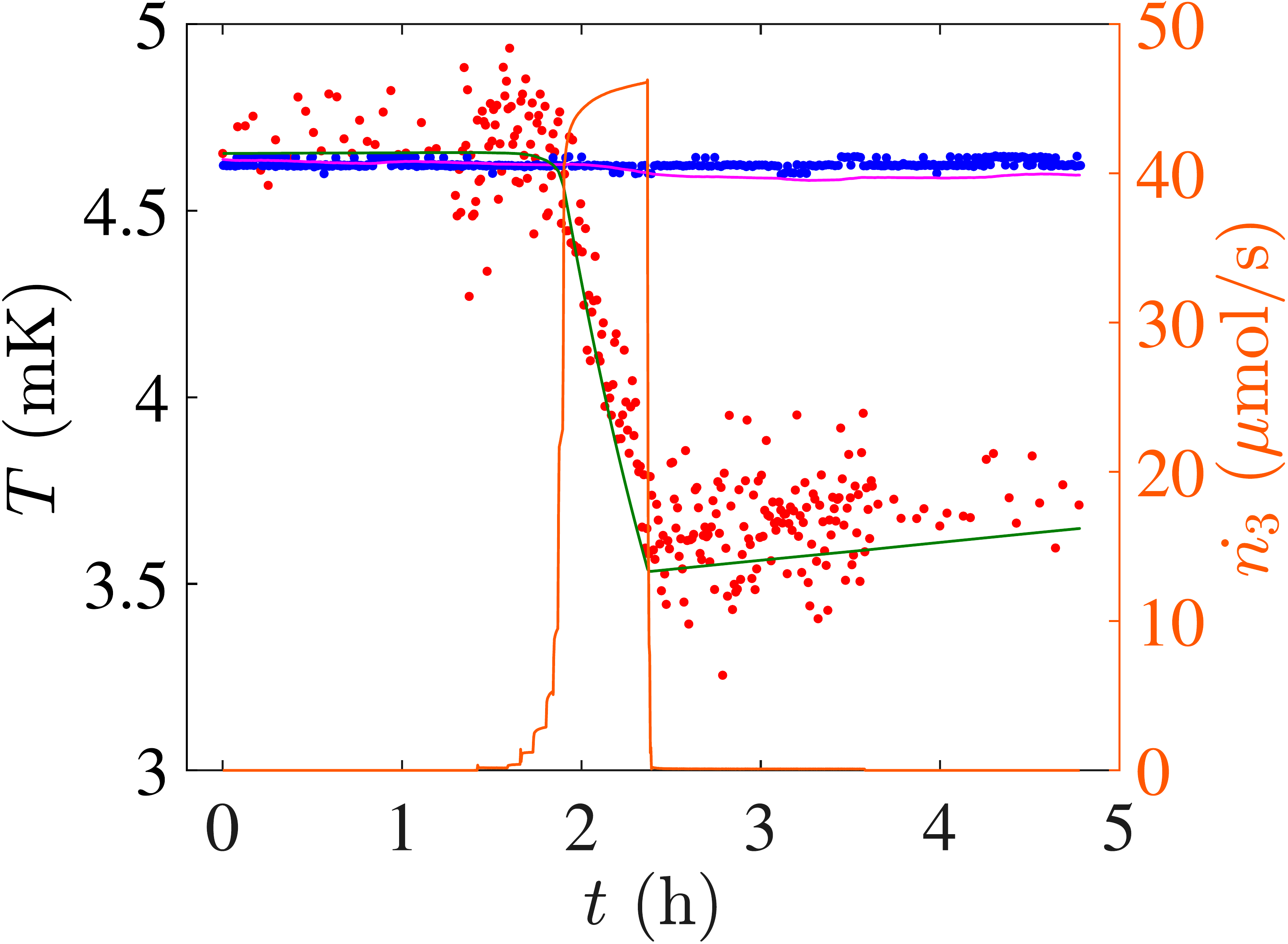}}\quad{}\subfloat{\includegraphics[width=0.5\columnwidth]{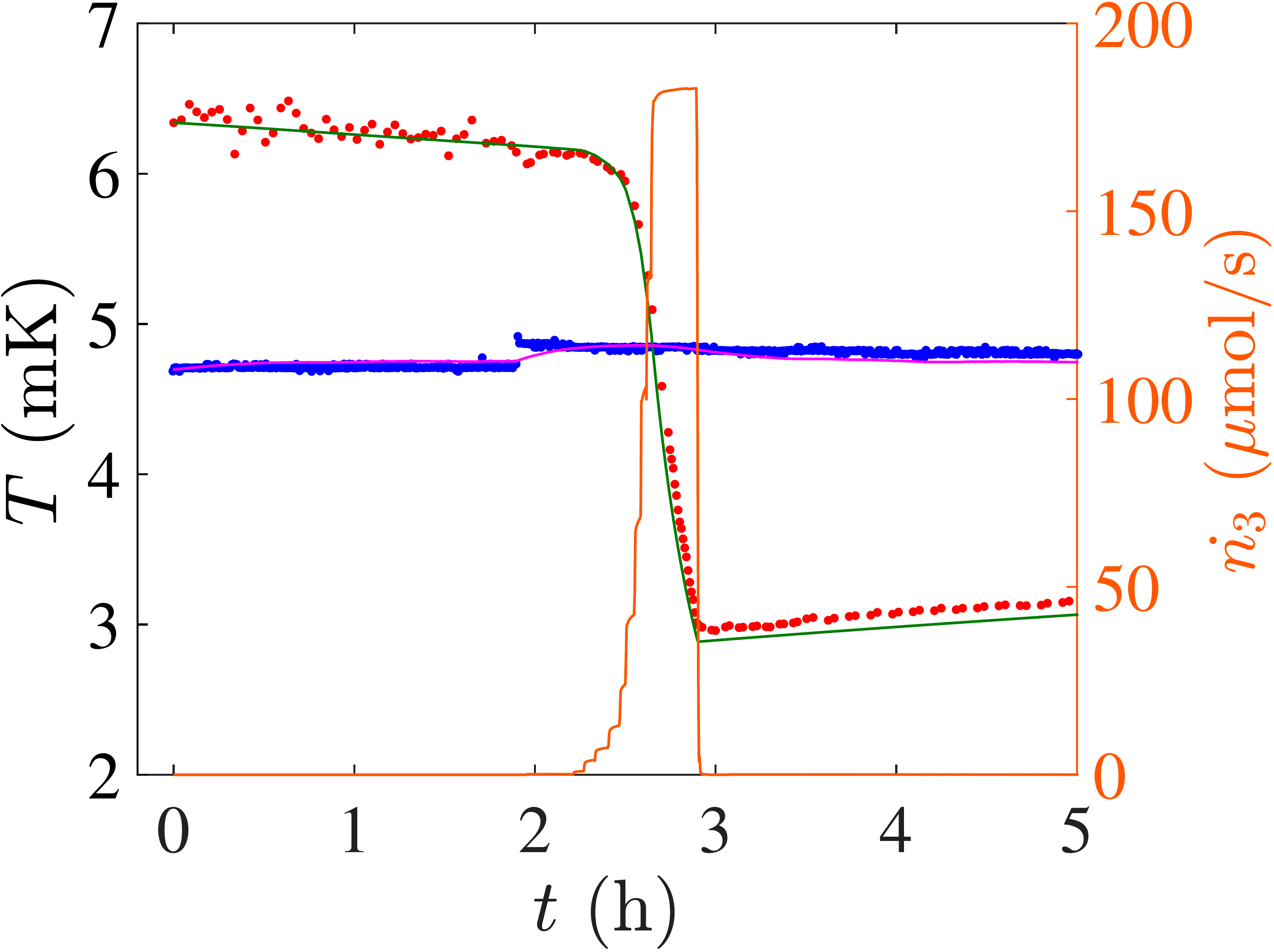}}\vspace*{-0.02\textheight}

\enskip{}%
\begin{minipage}[t]{0.5\columnwidth}%
\noindent \begin{flushleft}
(c) $\left(610\pm20\right)\,\mathrm{mmol}$, $\left(110\pm10\right)\,\mathrm{mmol}$,
$\left(1920/900\pm60\right)\,\mathrm{mmol}$
\par\end{flushleft}%
\end{minipage}\enskip{}%
\begin{minipage}[t]{0.5\columnwidth}%
\noindent \begin{flushleft}
(d) $\left(610\pm20\right)\,\mathrm{mmol}$, $\left(50\pm10\right)\,\mathrm{mmol}$,
$\left(2700/30\pm60\right)\,\mathrm{mmol}$
\par\end{flushleft}%
\end{minipage}\vspace*{-0.01\textheight}

\caption{(color online) Left y-axis: measured nuclear stage temperature $T_{\mathrm{NS}}$,
measured cell main volume temperature $T_{\mathrm{L}}$, with computed
$T_{\mathrm{L}}$, and computed heat-exchanger volume temperature $T_{\mathrm{V}}$. Right y-axis: \protect\textsuperscript{3}He phase-transfer
rate. Below each subfigure: total \protect\textsuperscript{3}He in
the system, amount of \protect\textsuperscript{3}He in the mixture
phase before melting, and solid \protect\textsuperscript{4}He before/after
melting. Background heat leak was kept constant 80 pW (negligible
at these temperatures)\label{fig:highT}}
\end{figure}
Figure \ref{fig:highT} illustrates the correspondence between measured
and computed temperatures in melts where the QTF-thermometer was still
completely in reliable reading range. At these temperatures, the background
heat leak $\dot{Q}_{\mathrm{ext}}$ of order tens of picowatts is
practically irrelevant, so we simply chose to use largest value from
Fig.~\ref{fig:melt_mintemps}. In some cases we needed to adjust
the amount of solid before the melting from our logbook values, but
the adjustment was always 10\% or less. Also, the quality of the QTF
measurement varies in these examples, because we tried different measurement
methods in preparation for lower temperature operations. For example,
in Fig.~\ref{fig:highT}b, we used alternating full spectrum sweeps
between \textsuperscript{3}He QTF and the mixture QTF, and tracking
mode with as low excitation as possible in Fig.~\ref{fig:highT}c.
In Fig.~\ref{fig:highT}d, we used the method that we found to work
best: full sweeps right until melting, tracking mode during melt,
and then full sweeps again after. All across the board, we could reproduce
the measured $T_{\mathrm{L}}$ within reasonable accuracy. 

\subsubsection{Thermal gate operation\label{subsec:Thermal-gate-operation}}

\begin{figure}[b]
\includegraphics[width=1\columnwidth]{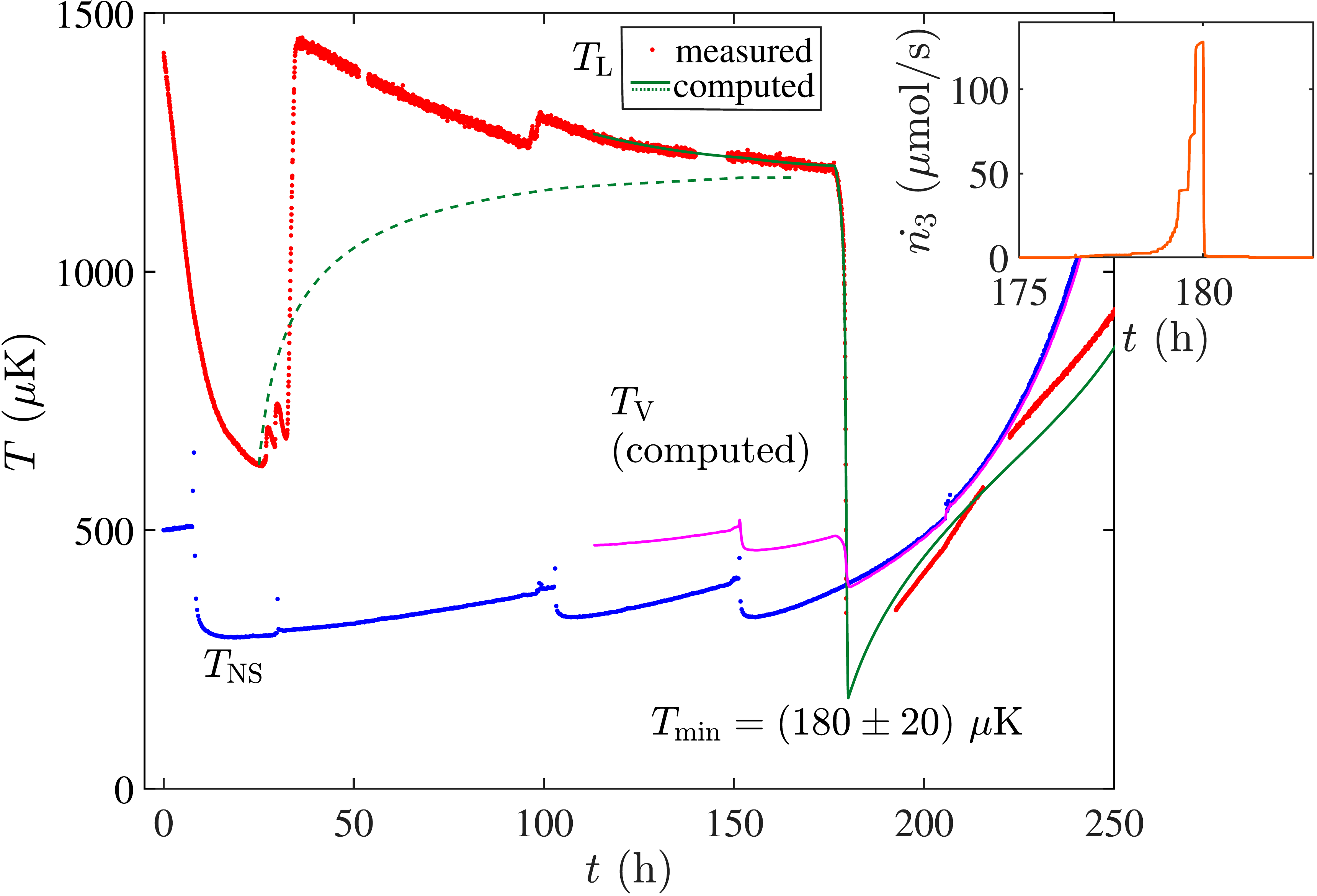}\caption{(color online) Measured nuclear stage temperature $T_{\mathrm{NS}}$,
measured cell main volume temperature $T_{\mathrm{L}}$, with computed
$T_{\mathrm{L}}$, and computed heat-exchanger volume temperature $T_{\mathrm{V}}$ during thermal gate operation. Dashed green line shows
the simulated temperature behavior at the closing of the thermal gate
assuming no additional heating during the operation, in clear conflict
with the measured behavior. The solid green line shows a computation
that was carried out from the point where the thermal gate was fully
closed (TG at $\sim0.3\,\mathrm{MPa}$). Total \protect\textsuperscript{3}He
$\left(610\pm20\right)\,\mathrm{mmol}$, \protect\textsuperscript{3}He
in mixture before the melt $\left(19\pm5\right)\,\mathrm{mmol}$,
the amount of solid at the beginning/in the end $\left(3110/250\pm60\right)\,\mathrm{mmol}$,
and heat leak before/after melt $47/27\,\mathrm{pW}$. Inset shows
the \protect\textsuperscript{3}He phase-transfer rate during the
melt\label{fig:TG-operation}}
\end{figure}
As the experiment was planned, the thermal gate was supposed to be
used at the end of the precool to isolate the main volume of the cell
from the heat-exchanger volume in effort to minimize all heat leaks
during the melting process (cf. Section \ref{subsec:Thermal-gate}).
We learned, however, that the gate did not work as intended, which
is demonstrated in Fig.~\ref{fig:TG-operation}. When the gate is
open, from 0 to 25 h, the precooling proceeds towards sub-500 $\mu$K
temperature, as usual. But as soon as we start to close the gate by
increasing pressure in its bellows, heating spikes appear in the main
cell volume that become more severe as the gate is further shut off.
We attribute this behavior to friction that occurs when the bellows
operated stainless-steel ball is pressed against its saddle, either
due to vibrations, or due to imperfect alignment. Furthermore, the
volume of the TG bellows is small, and there were two lines connected
to it; one was an ordinary capillary, while the other was a superleak
line. This could make it possible for a sound mode to oscillate between
the lines and cause mechanical vibrations in the system. In attempt
to eliminate this possibility, we filled the TG bellows with \textsuperscript{3}He\textendash \textsuperscript{4}He
mixture, to have the normal \textsuperscript{3}He component to dampen
the modes, but to no avail.

The second shortcoming of the thermal gate idea was that its operation
would only prevent heat leaks arriving to the heat-exchanger volume
($\dot{Q}_{\mathrm{extV}}$ in Eq.~\eqref{eq:V-heat}) from reaching
the main volume, but would do nothing to the direct heat leaks there
($\dot{Q}_{\mathrm{ext}}$ in Eq.~\eqref{eq:L-heat}). The dashed
green line in Fig.~\ref{fig:TG-operation} illustrates the simulated
temperature behavior under the assumption of no extra heating caused
by closing of the gate. It is clear that the behavior calculated this
way does not represent the actual course of events, as the temperature
does not jump up enough. On the other hand, the value of the new equilibrium
temperature is explained well by our computational model, as closing
the thermal gate basically removes the $\dot{Q}_{\mathrm{tube}}$
contribution to Eq.~\eqref{eq:L-heat} (in our calculations we reduced
the conductivity of the channel in the closed state by a factor 15).

Closing the gate thus only removes a small contribution to total heat
load of the main volume, an advantage lost because of the large extra
heating caused by the operation of the gate.

\subsubsection{Melting with the bellows system\label{subsec:Melting-bellows}}

\begin{figure}
\includegraphics[width=1\columnwidth]{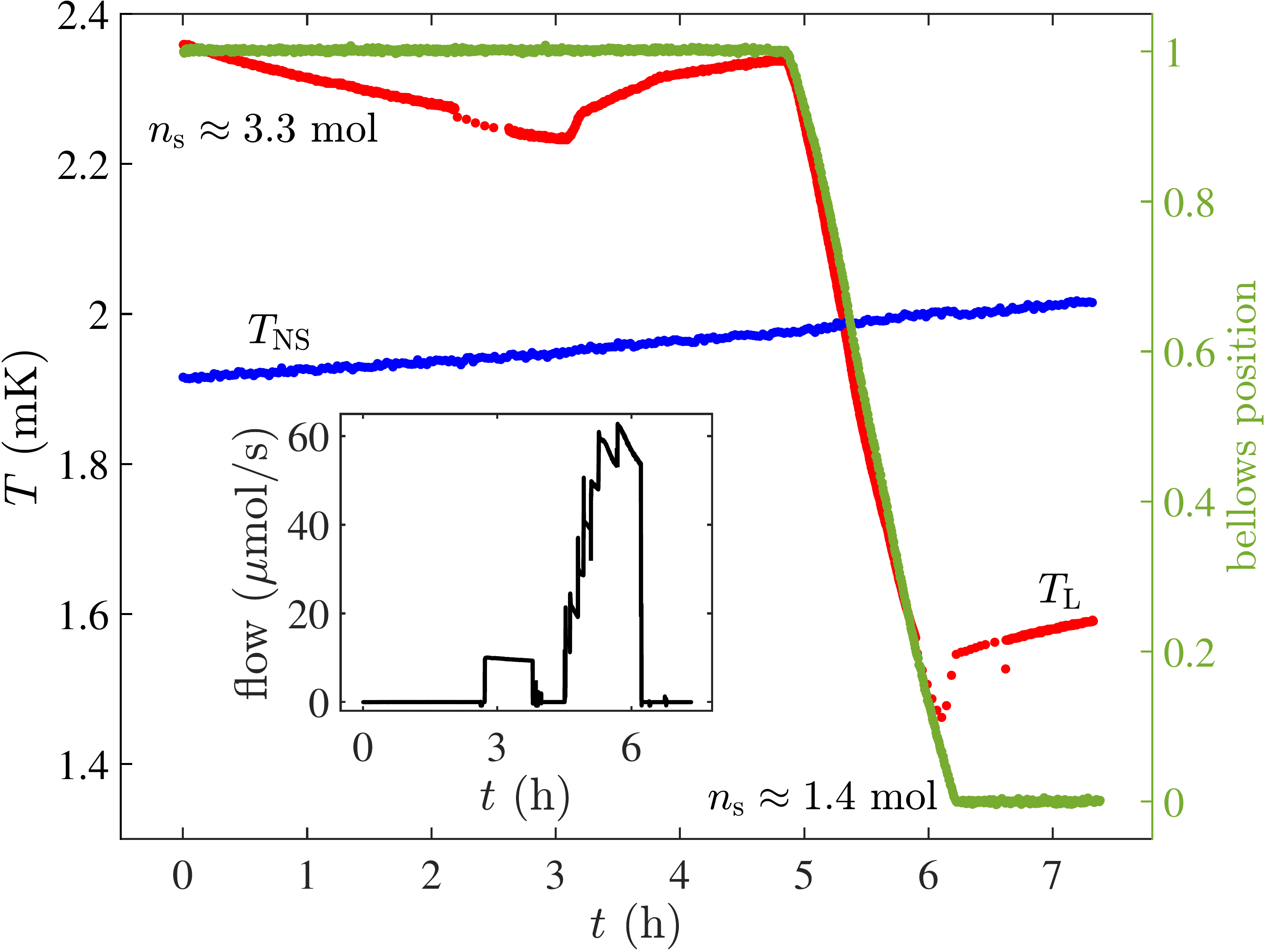}\caption{(color online) Left y-axis: measured nuclear stage temperature $T_{\mathrm{NS}}$,
and measured cell main volume temperature $T_{\mathrm{L}}$ and during
a melting performed by the bellows system. Right y-axis: the position
of the bellows in units where 1 is fully extended and 0 fully retracted.
Inset shows the \protect\textsuperscript{4}He extraction rate from
the upper bellows \label{fig:Bellows}}
\end{figure}
An alternative method to carry out the melting procedure was to use
the bellows system placed within the dilution refrigerator of the
cryostat, as described in Section \ref{subsec:Bellows-system}. The idea
behind it was that the bellows would separate the melting cell from
anything above the dilution refrigerator temperature ($\sim10\,\mathrm{mK}$),
as now the 1 K end of the superleak line could be blocked with solid
\textsuperscript{4}He, as it was no longer needed. However, we encountered
several problems that made the utilization of the bellows challenging,
which are illustrated in Fig.~\ref{fig:Bellows}.

The first problem was that the bellows could not be moved enough to
accommodate melting of the entire \textsuperscript{4}He crystal.
The maximum initial solid amount was usually slightly over 3 mol,
and even by fully using the range of the bellows we could melt only
60\% of it. Secondly, when the bellows was near the limits of its
movement range, it caused heating in the main volume of the experimental
cell. In Fig.~\ref{fig:Bellows}, when we began to pump the upper
bellows near 3 h mark, the cell main volume immediately showed warming up, even if the bellows was not even moving. As the bellows eventually
started to move, the solid was melted and the liquid in the cell cooled
down as intended. When the bellows reached the other end of its range,
another heat pulse in $T_{\mathrm{L}}$ was observed. Since the heating
was already a problem at 2 mK range, at sub-1 mK, as the cooling power
of the melting process decreases rapidly, its effect would be even
more detrimental. So, in order to avoid extra heating, we would have
to keep the bellows from reaching either end of its movement capacity,
but that would make the first problem even worse, i.e., we could melt
even less of the total solid. The third, and final problem was that
by using the bellows we sacrifice the ability to determine the amount
of solid accurately, as then we cannot use the flow measurement to
determine the extracted amount of \textsuperscript{4}He. We can,
of course, convert the movement of the bellows to melted solid amount,
but that would introduce more uncertainty in the estimations.

In conclusion, the drawbacks of the bellows operation outweighed the
potential advantages, which is why we focused our efforts to perform
the melts by pumping the superleak from room temperature.

\subsubsection{Extra heating during solid growth\label{subsec:Extra-heating-during-growth}}

\begin{figure}[b]
\center\includegraphics[width=1\columnwidth]{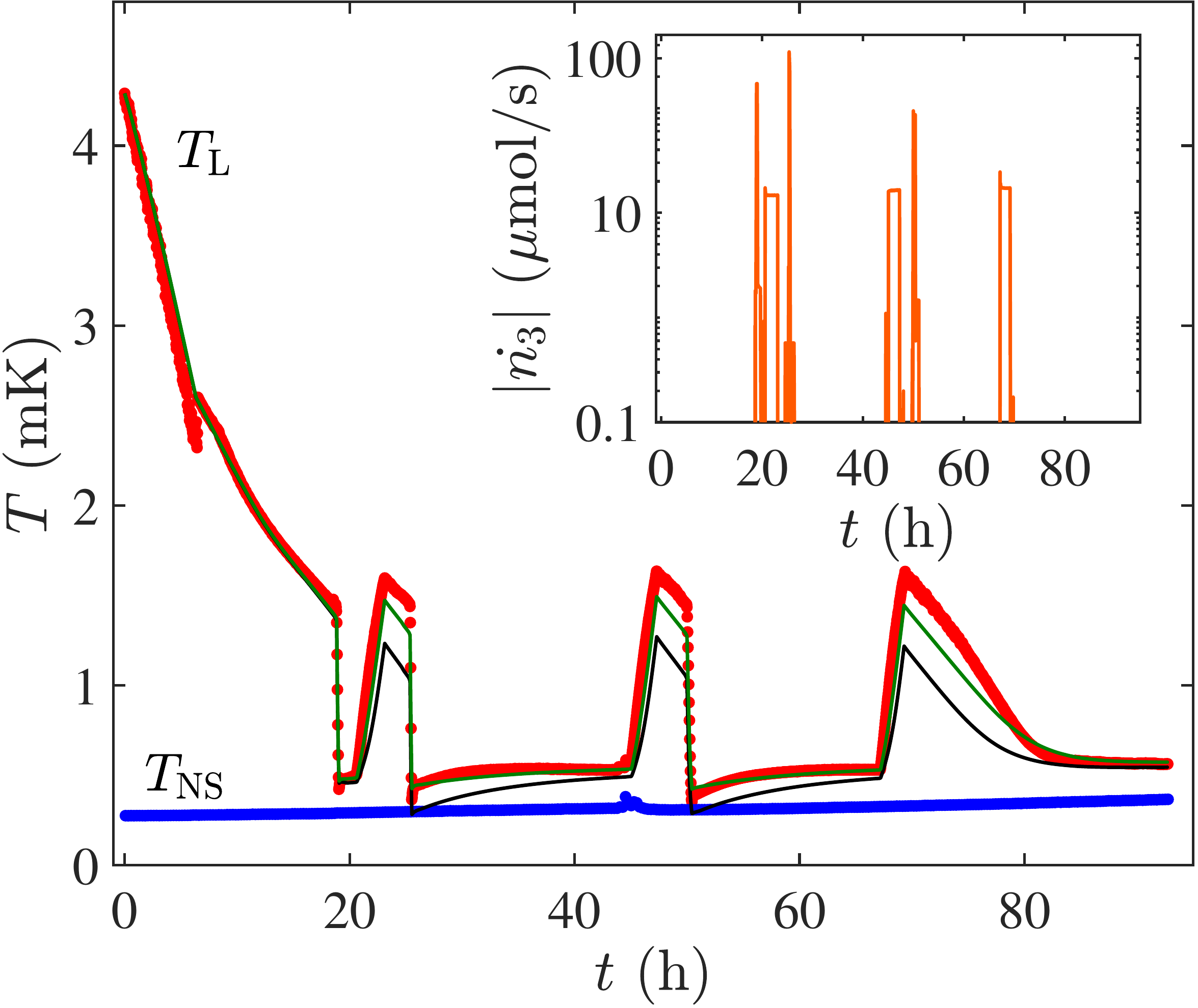}\caption{(color online) Measured cell main volume temperature $T_{\mathrm{L}}$
and nuclear stage temperature $T_{\mathrm{NS}}$ during three growing-melting
cycles. Green line is the computed $T_{\mathrm{L}}$ with extra 300
pW heating added during the growth phase, while the black line is
computed without the extra heating. Inset shows the absolute value
of the \protect\textsuperscript{3}He phase-transfer rate\label{fig:Growth}}
\end{figure}
As we have described in the previous sections, our computational model
can reproduce precools and melts throughout the entire temperature
range. But during the growths of the solid \textsuperscript{4}He,
it requires further tuning. As an illustrative example of this, in
Fig.~\ref{fig:Growth} we have first a precool from 4 mK to 1.5 mK,
followed by a partial melt bringing temperature to about 0.5 mK. Almost
immediately after, we grew the solid phase back to the initial amount,
then observed the precool for a short while before melting the solid
once again. After recording the post-melting warm-up period, the sequence
was repeated. Then, after the third growth, we chose to observe the
post-growth precool period longer instead. The solid amount before
the melts was about 2.9 mol, and 1.2 mol at their ends.

Our initial attempt to replicate the measured temperature with the
computational model resulted the black line in Fig.~\ref{fig:Growth}.
The background heat leak was 50 pW, determined from the post-melting
relaxation, and as the flow-dependent heat leak we used $\dot{Q}_{f}$,
from Eq.~\eqref{eq:Qeta}. During growths, the latent heat from the
transfer of \textsuperscript{3}He from mixture to pure phase plus
$\dot{Q}_{f}$ are not sufficient to cause $T_{\mathrm{L}}$ to increase
enough to meet the measured values. On the other hand, during the
melts, the computed temperature falls below the measured values.

To make the temperatures match, we found it necessary to add more
heating to the growth periods. This turned out to be a balancing
act, as extra heating during growths caused the calculated temperature
at the end of the melt to be increased as well (due to the increased
starting temperature, naturally). We could not make both the initial
and final temperatures match perfectly, but the best attempt is shown
as the green line in Fig.~\ref{fig:Growth}, where additional $300\,\mathrm{pW}$
of heating is added to the growth periods. The additional heating
seemed not to relate to the growth rate, as did $\dot{Q}_{f}$, discussed
in Section \ref{subsec:Heat-leak-during}. But rather, it appeared
to depend on the crystal size \textemdash{} when there was just a
small amount of solid, this additional heating was larger than when
the crystal was big. This suggests that surface energies are involved,
although the detailed account for this effect remains an open question.
Treiner \cite{Treiner1993} found, using the theoretical framework
of Ref.~\cite{Pavloff1991}, that a single \textsuperscript{3}He
atom introduced to a partly solidified \textsuperscript{4}He system
would preferably bind on the first liquid \textsuperscript{4}He layer
on the surface of the solid \textsuperscript{4}He phase. How this
applies to a system, where plenty of \textsuperscript{3}He is available
needs further considerations, but if there is excess \textsuperscript{3}He
on the solid surface, that might play a role in the observed asymmetric
behavior between growing and melting the crystal.
\begin{figure}[p]
\subfloat{\includegraphics[width=0.5\columnwidth]{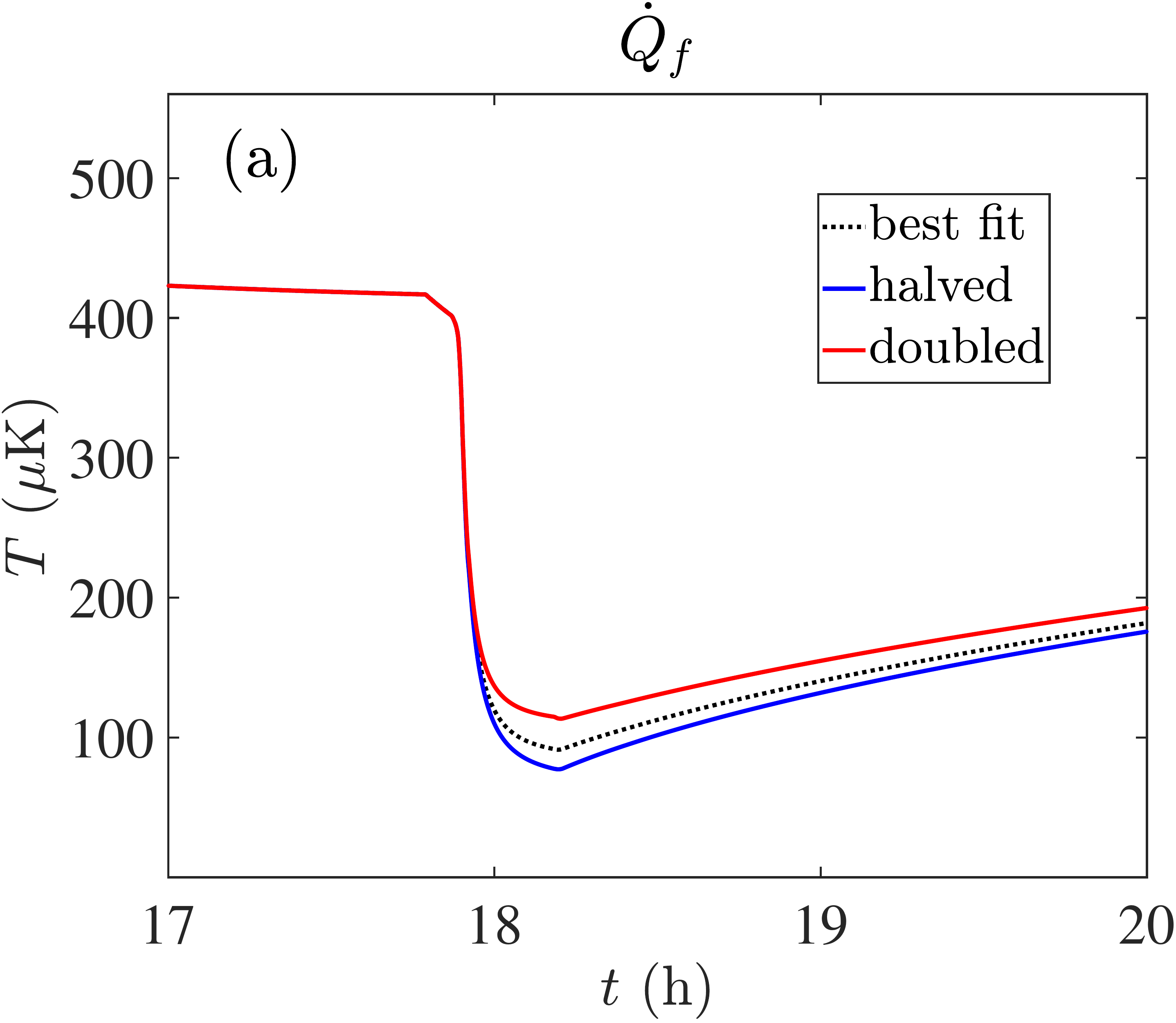}}\quad{}\subfloat{\includegraphics[width=0.5\columnwidth]{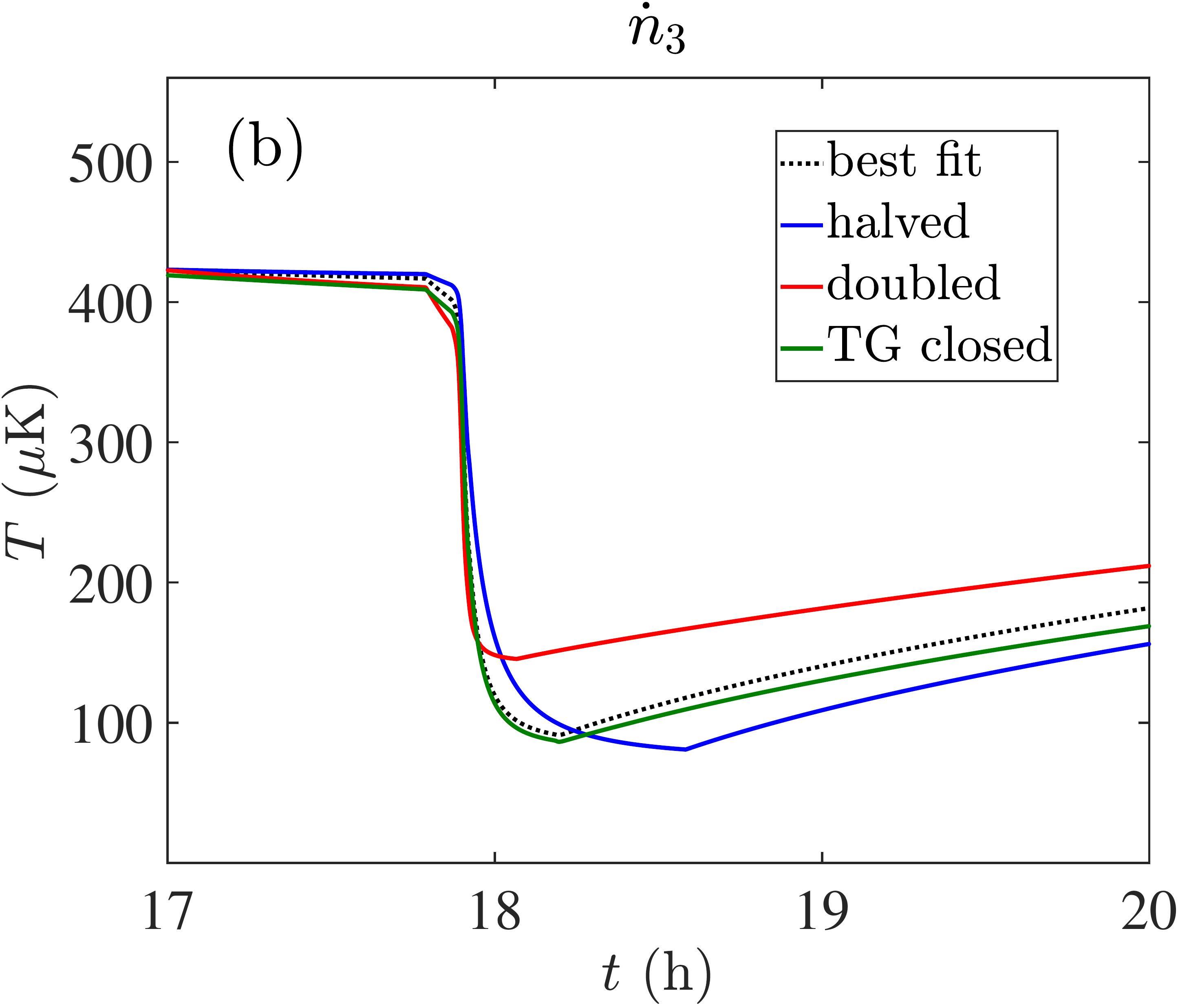}}

\subfloat{\includegraphics[width=0.5\columnwidth]{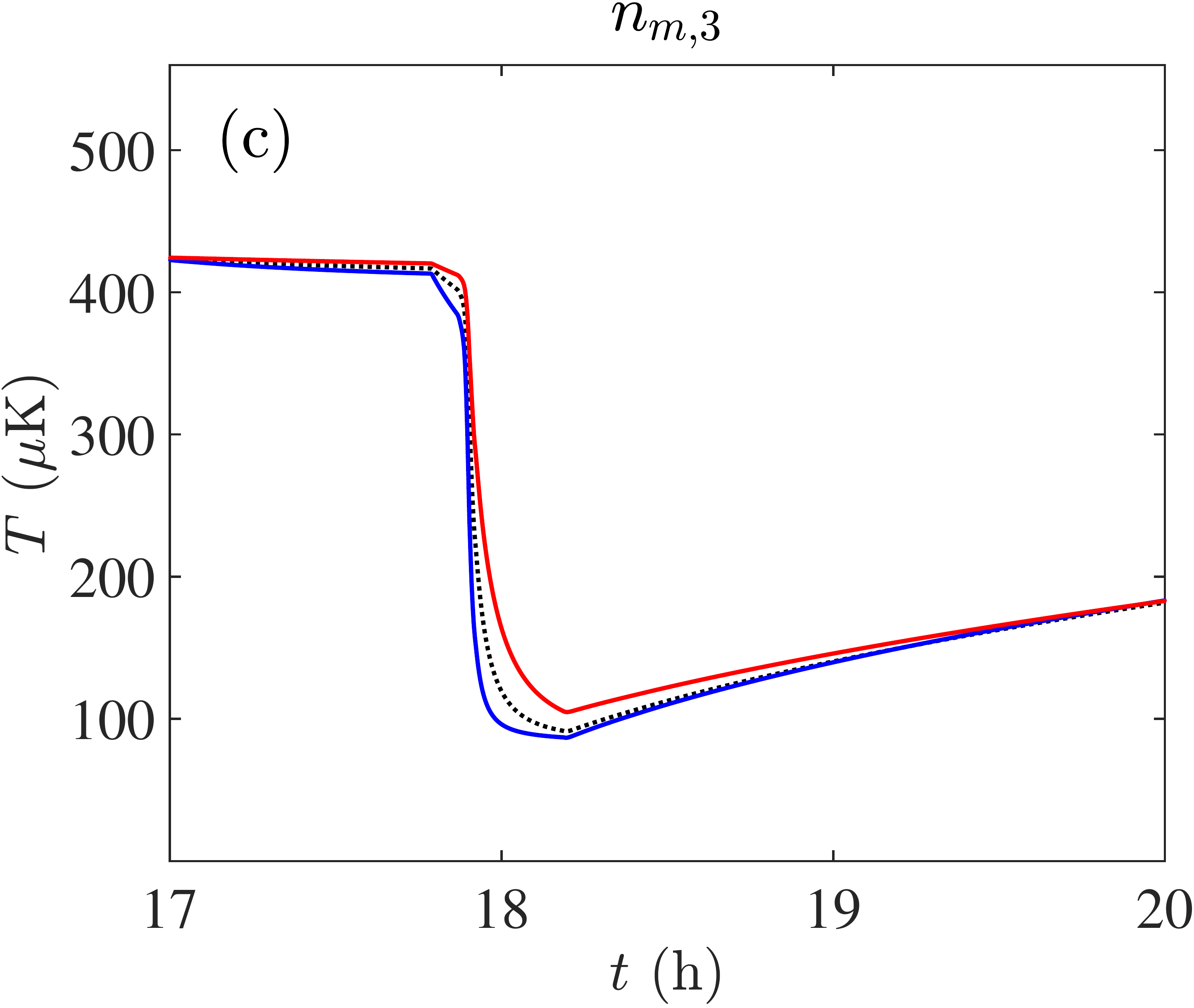}}\quad{}\subfloat{\includegraphics[width=0.5\columnwidth]{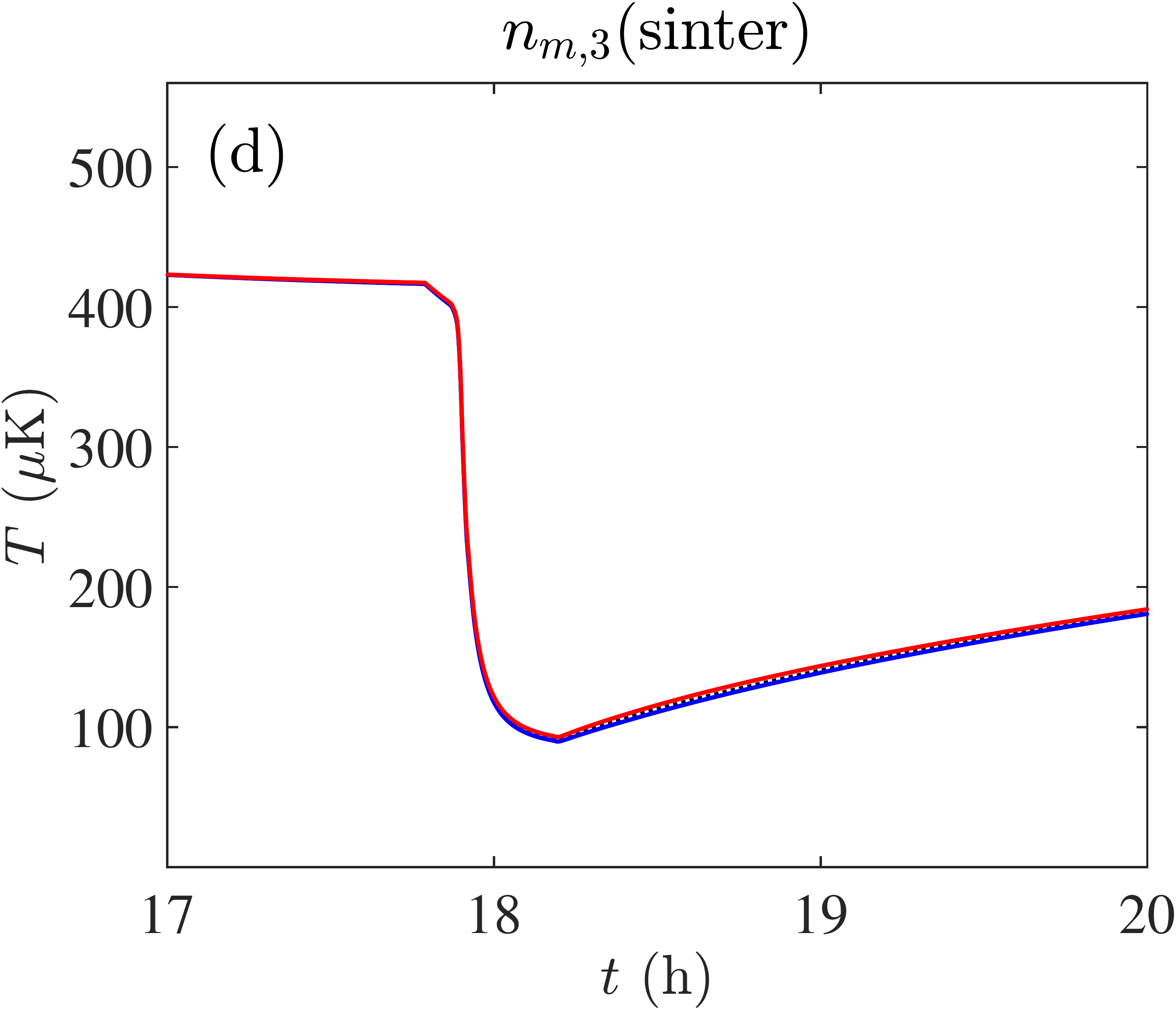}}

\enskip{}\subfloat{\includegraphics[width=0.5\columnwidth]{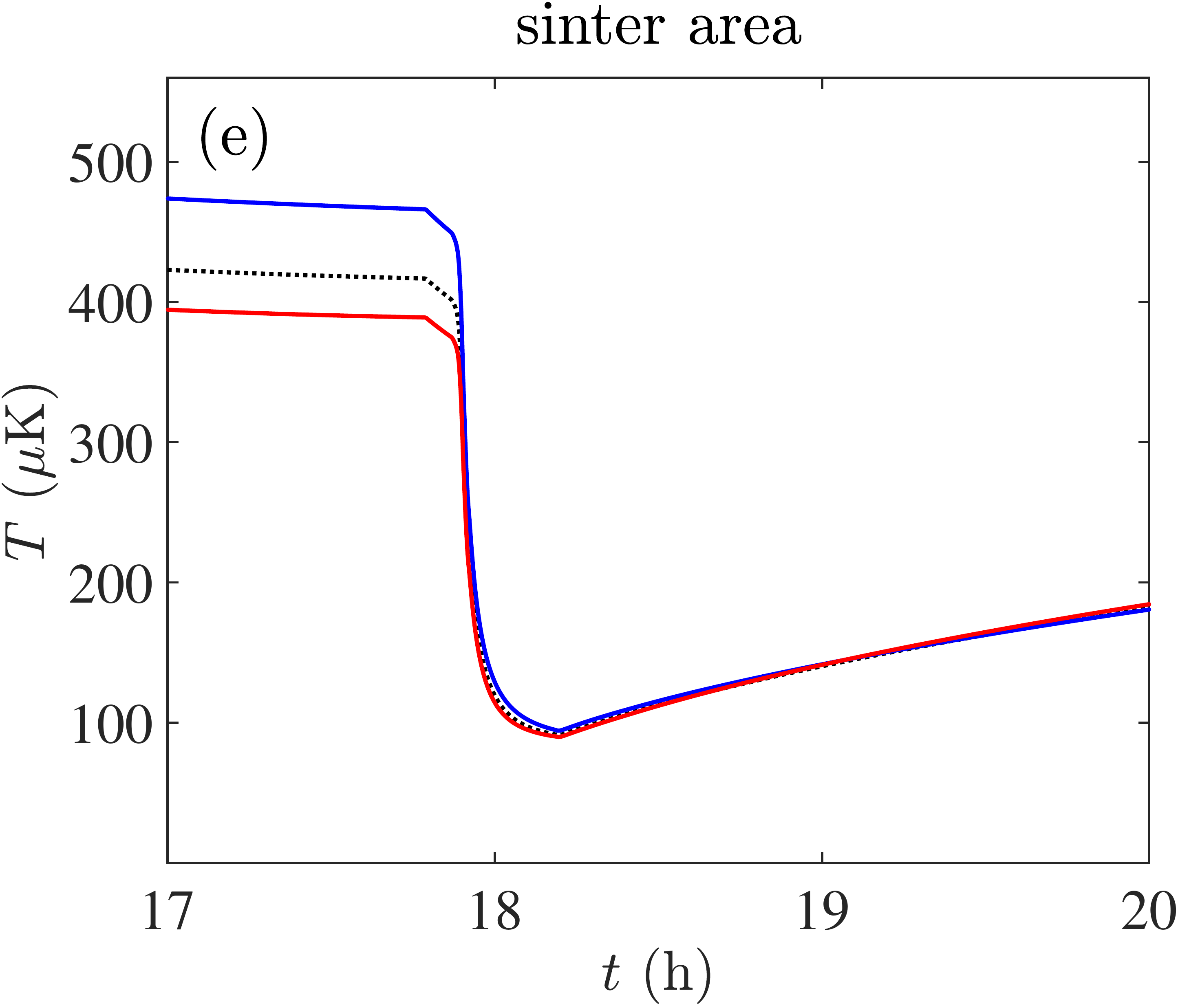}}\quad{}\subfloat{\includegraphics[width=0.5\columnwidth]{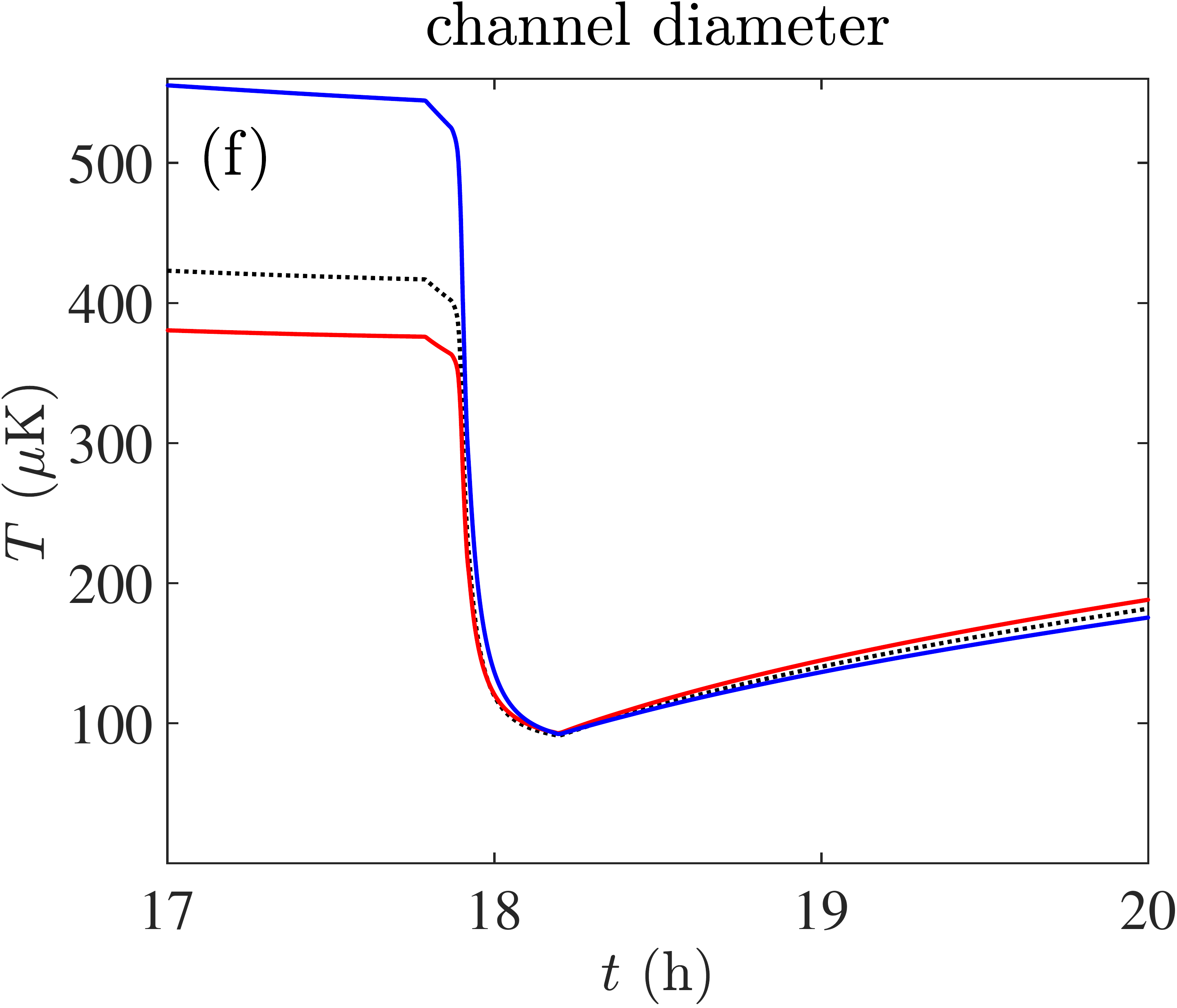}}

\caption{(color online) Simulated melting behavior with altered parameters
compared to the calculation of Fig.~\ref{fig:melt_mintemps}e. Red
lines correspond to two times larger parameter value, while in blue
lines the value is halved. (a) Flow-rate dependent heat leak $\dot{Q}_{f}$,
(b) \protect\textsuperscript{3}He phase-transfer rate, plus the green
line shows the melting if the thermal gate was closed at the beginning
of the melt (without any additional heating), (c) the amount of \protect\textsuperscript{3}He
in the mixture phase before the melt, (d) the amount of \protect\textsuperscript{3}He
trapped as mixture in the pores of the sinter in the heat-exchanger
volume, (e) surface area of the sinter (effectively $2r_{\mathrm{V}}$
and $0.5r_{\mathrm{V}}$ of Section \ref{subsec:sinter}), and (f)
diameter of the channel connecting the cell main volume and the heat-exchanger
volume\label{fig:vary-params}}
\end{figure}

\subsection{Simulations with modified parameters\label{subsec:Simulations-altered-params}}

Finally, we take a look at how modifying certain experimental details
would affect the computed melting behavior. We focus on the melt presented
in Fig.~\ref{fig:melt_mintemps}e, as it resulted the lowest calculated
temperature so far. The main source of heat was the flow-dependent
heat leak $\dot{Q}_{f}$. In Fig.~\ref{fig:vary-params}a, we have
halved or doubled the numerical prefactor in $\dot{Q}_{f}=12.8\left(\frac{\dot{n}_{3}}{\mathrm{\mu mol/s}}\right)^{3}$,
resulting in approximately $10\,\mathrm{\mu K}$ decrease, or $20\,\mathrm{\mu K}$
increase in the lowest temperature, respectively. Next, in Fig.~\ref{fig:vary-params}b,
we have kept the original numerical prefactor, but modified the phase-transfer
rate $\dot{n}_{3}$. The plot was constructed keeping the final solid
amount constant, which is why the double-rate melt takes less time
than the original one, and opposite for the half-rate melt. Doubling
$\dot{n}_{3}$ has dramatic adverse effect to the ultimate temperature,
as the flow-dependent heat leak rapidly increases. In fact, having
halved the rate would have resulted in about $10\,\mathrm{\mu K}$
lower temperature. The additional line in Fig.~\ref{fig:vary-params}b
demonstrates the melting behavior under the assumption that the thermal
gate could be closed right before initiating the melt without additional
heating. Since the majority of the heat leak ($\dot{Q}_{f}+\dot{Q}_{\mathrm{ext}}$)
arrives to the main volume of the cell, closing off the heat-exchanger
volume does not significantly decrease the total heat load to the
melting process, and the lowest temperature is thus almost unchanged.
Then, Fig.~\ref{fig:vary-params}c shows that even decreasing the
amount of initial mixture does not help to bring the final temperature
down as long as the relatively large $\dot{Q}_{f}$ is in place. Increasing
mixture, on the other hand, would further hamper the performance,
because then the before-melting state would contain additional entropy.

Figures \ref{fig:vary-params}d-f demonstrate parameters that do not
have much effect on the ultimate temperature. Since the amount of
mixture in the heat-exchanger volume is small (7 mmol), and the volume
is separated from the main volume by the poor thermal conductivity
of the connecting channel, its alteration results in only minuscule
effect at the lowest temperatures. In the last two subfigures, we
alter the thermal connection between the nuclear stage and the experimental
cell by changing the amount of sinter (Fig.~\ref{fig:vary-params}e),
or modifying the dimensions of the connecting channel (Fig.~\ref{fig:vary-params}f).
Better thermal contact results in decreased precooling temperature,
as expected. But, during the melt, the heat flow from nuclear stage
to the cell is also increased resulting in almost no net change in
the lowest achievable temperature. The same outcome, but in reversed
order, is true for reduced thermal contact.

Thus, the most critical aspect in improving the setup is to take care
of the heat leaks. The most important is the background external heat
leak $\dot{Q}_{\mathrm{ext}}$, and not the flow-dependent heat leak
$\dot{Q}_{f}$, even if it currently was the most significant contribution
to the total heating. If we were able to significantly reduce, or
even completely remove, $\dot{Q}_{\mathrm{ext}}$, we could melt the
crystal at low enough rate for $\dot{Q}_{f}$ to have less effect.
Since at present, $\dot{Q}_{\mathrm{ext}}$ was still significant,
we were led to combat it by increasing the melting rate, which ended
up causing further heating problems. In reality we can never completely
get rid of $\dot{Q}_{\mathrm{ext}}$, so the solution is two-pronged;
reduce the background heat leak as much as possible, while at the
same time improve the superleak performance in hopes to reduce $\dot{Q}_{f}$.
After that, if the heat leaks were under control, the next step is
to reduce the amount of mixture in the state before the melt. Improving
the precooling temperature comes in question only if there is a reliable,
non-heating, way to isolate the heat-exchanger volume from the main
volume.

\section{Conclusions and suggestions for future improvements\label{sec:Conclusions}}

We studied a novel cooling method that operates with mixture of \textsuperscript{3}He
and \textsuperscript{4}He, at the \textsuperscript{4}He crystallization
pressure $2.564\,\mathrm{MPa}$ \cite{Pentti_Thermometry} at the \textsuperscript{3}He
saturation molar concentration $8.1\%$ \cite{Pentti_etal_solubility} motivated
by the search for the coveted superfluid transition of \textsuperscript{3}He
in dilute \textsuperscript{3}He\textendash \textsuperscript{4}He
mixture \cite{Effective_3He_interactions}. The heart of the experimental
setup consisted of a large melting cell ($77\,\mathrm{cm^{3}}$),
and a separate sinter-filled heat-exchanger volume ($5\,\mathrm{cm^{3}}$
and $10\,\mathrm{m^{2}}$ surface area), with a pressure operated
thermal gate in between.

In the method, the isotopes are first separated by growing solid \textsuperscript{4}He
in the experimental cell, followed by a precool with an adiabatic
nuclear demagnetization refrigerator to below $0.5\,\mathrm{mK}$.
Then, the solid is melted, allowing \textsuperscript{3}He and \textsuperscript{4}He
to mix in a heat absorbing process \cite{Riekki2019}. The final temperature
depends on the initial entropy of the system, and on the heat leak
during the melting process. 

The major challenge in the data-analysis was that the quartz tuning
fork resonators (QTFs) that were used for thermometry became insensitive
at the lowest temperatures. Thus, we had to resolve in constructing
a computational model of the system to determine the achieved temperatures.
We showed that the Kapitza resistance of the plain cell wall followed
$R_{K}=R_{0}/\left(AT^{p}\right)$ with the exponent $p_{\mathrm{L}}=\left(2.6\pm0.2\right)$,
and coefficient $R_{0,\mathrm{L}}=\left(0.17\pm0.05\right)\,\mathrm{m^{2}}\mathrm{K}^{p_{\mathrm{L}}+1}\mathrm{W^{-1}}$,
while for the sinter the corresponding parameters were $p_{\mathrm{V}}=\left(1.7\pm0.1\right)$
and $R_{0,\mathrm{V}}=\left(50\pm30\right)\,\mathrm{m^{2}}\mathrm{K}^{p_{\mathrm{V}}+1}\mathrm{W^{-1}}$.

The main factor limiting the lowest temperatures turned out to be
the heat leak to the experimental cell during the melting procedure.
We learned that it constituted of two parts: generic background heat
leak, and the flow-rate dependent (and thus melting-rate dependent)
contribution, of which the latter was most dominant at the melting
rates employed. We concluded that the optimal \textsuperscript{3}He
phase-transfer rate would have been around $100..150\,\mathrm{\mu mol/s}$
(corresponding to $120..180\,\mathrm{\mu mol/s}$ \textsuperscript{4}He
extraction rate), and that during the experiment, we often used somewhat
too large values. The lowest temperature obtained was $\left(90\pm20\right)\,\mathrm{\mu K\approx}\frac{T_{c}}{\left(29\pm5\right)}$
($T_{c}=2.6\,\mathrm{mK}$ \cite{Pentti_Thermometry}) with about $200\,\mathrm{\mu mol/s}$
phase-transfer rate, which would still correspond to record-low temperature
obtained in \textsuperscript{3}He\textendash \textsuperscript{4}He
mixture, and in pure \textsuperscript{3}He as well. But, we did not
observe any indication of \textsuperscript{3}He superfluidity in
the \textsuperscript{3}He\textendash \textsuperscript{4}He mixture
phase.

We analyzed how changing different parameters would affect the performance
of the setup, and came to a conclusion that to reduce the final temperature
further the most essential thing is to reduce the heat load to the
cell during the melting. To do that, we suggest to simplify the experimental
setup further by removing the thermal gate completely. As we demonstrated,
it did not live up to its purpose in reducing the heat leak to the
main volume of the cell, since the majority of the background heat
load was going directly to the main volume of the experimental cell,
rather than was coming from the nuclear stage through the sinter.
We showed, in fact, that the decreasing thermal conductivity of the
connecting channel during the melt alone is sufficient to decouple
the main volume and the heat-exchanger volume sufficiently from each
other.

Another component to simplify would be the bellows system, since it
did not provide a valid method to carry out the melting due to the
excessive heating its movement caused, and since its range of motion
did not allow us to melt the entire crystal. In place of the complicated
bellows system, we should simply have a buffer volume at the dilution
refrigerator temperature ($10\,\mathrm{mK}$) to isolate the low-temperature
section of the superleak from the high-temperature one. It may be
beneficial to add a second buffer volume to the nuclear stage temperature
($0.5\,\mathrm{mK}$) to isolate the melting cell from anything above
the \textsuperscript{3}He $T_{c}$. This would mean that the superleak
line would then be made of three parts instead of two, adding a layer
of complexity to the setup. But since the goal is to get rid of as
much of the heat leak as possible, that option should be explored.
As another note, the high-temperature end of the superleak should
have a better thermalization to the still temperature ($0.7\,\mathrm{K}$)
to enable us to block it with solid \textsuperscript{4}He more easily.

If the heat leak can be brought under control, the secondary aspect
to take care of is minimizing the amount of mixture at the state before
the melting, because the final temperature can be reduced in proportion
to the reduction in the initial entropy. We could place the cell-side end
of the superleak higher to enable us to grow more crystal, and reduce
the volume of the pure \textsuperscript{3}He phase. We need only
enough \textsuperscript{3}He in the main volume to saturate all the
\textsuperscript{4}He released from solid, plus the amount required
to keep the \textsuperscript{3}He QTF always in the pure phase. We
could also move the mixture QTF higher in the main volume to have
it become measurable sooner, as currently it was available only after
half of the solid was already melted. An additional QTF could be installed
to the heat-exchanger volume to monitor its temperature, which would
enable us to gather more information on the thermal conductivity of
the connecting channel.

In general, the thermometry is challenging when performing experiments
at such ultra-low temperatures. A small step forward could be to have
a \textsuperscript{3}He QTF with smaller dimensions that would improve
the sensitivity to fluid damping. However, it appeared that the residual
width is not solely caused by the intrinsic properties of the QTF,
but by the \textsuperscript{4}He film covering it \cite{Riekki2019a},
which presumably cannot be remedied by altering the QTF size. Even
in the best case, a new oscillator would only postpone the QTF thermometry
problem, as any mechanical oscillator will become insensitive near $100\,\mathrm{\mu K}$.
It is out of the question to measure the liquid temperature by a PLM
thermometer, because it is too slow to follow the rapid changes of
the fluid temperature during melts. A plausible technique would be
to utilize the quadratic temperature dependence
of the crystallization pressure in mixture \cite{Pentti_Thermometry,Rysti2014,Sebedash2006}.
This would require a pressure gauge with about 1 mPa resolution at 2.5 MPa
pressure, placing extreme demands on the stability and readout of
the arrangement. Finally, another solution would be to use a magnon
BEC thermometer that could, in principle, be used down to much lower
temperatures than mechanical oscillators \cite{Autti2012,Heikkinen2013,Pete_thesis}.
To set it up, the experimental cell would need to have a non-metallic
section to house the magnon sample and the NMR coils, introducing
more complexity, and possibly heat leak sources, to the setup.

Regarding the thermal model, we also mentioned the peculiar asymmetric
behavior between growing and melting the solid. We had to assume increased
heating during growing periods to make the computations match with
measured data. It tells us that some element is still missing from
our model that warrants further study.

To reach the target range of temperatures $<40\,\mathrm{\mu K}$,
the background heat leak should be suppressed to below 10 pW, which
does not appear as an entirely impossible task.
\begin{acknowledgements}
This work was supported by the Jenny and Antti Wihuri Foundation Grant
No. 00180313, and it utilized the facilities provided by Aalto University
at OtaNano - Low Temperature Laboratory. The authors thank J. Rysti
and J. T. Mäkinen for insightful discussions.
\end{acknowledgements}

\section*{\textemdash \textemdash \textemdash \textemdash \textemdash \textemdash \textemdash{}}

\bibliographystyle{spphys}

\end{document}